\documentclass[fleqn,usenatbib]{mnras}

\usepackage{newtxtext,newtxmath}
\usepackage[T1]{fontenc}

\DeclareRobustCommand{\VAN}[3]{#2}
\let\VANthebibliography\thebibliography
\def\thebibliography{\DeclareRobustCommand{\VAN}[3]{##3}\VANthebibliography}

\usepackage{graphicx}	% Including figure files
\usepackage{amsmath}	% Advanced maths commands
\def\hho{H$_2$O~}

\def\gg{$\gamma_6$~}
\def\Tef{$T_{\rm eff}$~}

\def\usep\Vsini{$v$ sin $i$~}
\def\Vt{$V_{\rm t}$~}

\def\ts{$\tau_s$~}

\def\Tef{$T_{\rm eff}$~}

\def\usep\Vsini{$v$ sin $i$~}
\def\Vt{$V_{\rm t}$~}

\title[A blue depression in the spectra of M dwarfs]{A blue depression in the optical spectra of M dwarfs}

\author[H.R.A. Jones et al.]{
Hugh R.A. Jones,$^{1}$\thanks{E-mail: h.r.a.jones@herts.ac.uk}
Yakiv Pavlenko,$^{2,1}$
Yuri Lyubchik,$^{2}$
Mike Bessell,$^{3}$
Nicole Allard$^{4,5}$
and David J. Pinfield$^{1}$
\\
$^{1}$Centre for Astrophysics Research, University of Hertfordshire, Hatfield, Hertfordshire, AL10 9AB, UK\\
$^{2}$Main Astronomical Observatory of National Academy of Sciences of Ukraine, Golosiiv, 252650 Kyiv-22, Ukraine\\
$^{3}$Mount Stromlo and Siding Spring Observatories, Institute of Advanced Studies, The Australian National University, Weston Creek P.O., ACT 2611, Australia\\
$^{4}$GEPI, Observatoire de Paris,  Universit\'e\ PSL, UMR 8111, CNRS, 61, Avenue de l’Observatoire, F-75014 Paris, France\\
$^{5}$Sorbonne Universit\'e\, CNRS, UMR7095, Institut d'Astrophysique de Paris, 98bis Boulevard Arago, F-75014 Paris, France\\
}

\date{Accepted 2023 April 10. Received 2023 April 5; in original form 2022 November 10}

\pubyear{}

\begin{document}
\label{firstpage}
\pagerange{\pageref{firstpage}--\pageref{lastpage}}
\maketitle

\begin{abstract}
A blue depression is found in the spectra of M dwarfs from 4000 to 4500\AA. This depression shows an increase toward lower temperatures though is particularly sensitive to gravity and metallicity. It is the single strongest and most sensitive feature in the optical spectra of M dwarfs. The depression appears as centred on the neutral calcium resonance line at 4227\AA~and leads to nearby features being weaker by about two orders of magnitude than predicted. We consider a variety of possible causes for the depression including temperature, gravity, metallicity, dust, damping constants and atmospheric stratification. We also consider relevant molecular opacities which might be the cause identifying AlH, SiH, and NaH in the spectral region. However, none of these solutions are satisfactory. In the absence of a more accurate determination of the  broadening of the calcium line perturbed by molecular hydrogen, we find a promising empirical fit using a modified Lorentzian line profile for the calcium resonance line. Such fits provide a simplistic line-broadening description for this calcium resonance line and potentially other un-modelled resonance lines in cool high pressure atmospheres. Thus we claim the most plausible cause of the blue depression in the optical spectra of M dwarfs is a lack of appropriate treatment of line broadening for atomic calcium. The broad wings of the calcium resonance line develop at temperatures below about 4000K and are analogous to the neutral sodium and potassium features which dominate the red optical spectra of L dwarfs. 
\end{abstract}

\begin{keywords}
stars -- abundance -- late type
\end{keywords}

%%%%%%%%%%%%%%%%%%%%%%%%%%%%%%%%%%%%%%%%%%%%%%%%%%

%%%%%%%%%%%%%%%%% BODY OF PAPER %%%%%%%%%%%%%%%%%%

\section{Introduction}

The ongoing large-scale efforts to determine the content of stellar neighbourhood have made it clear that the dominant spectral type in the stellar neighbourhood is the M dwarfs. For example, within the 10 parsec Gaia sample 61\% are M dwarfs and more than half of these are within the range M3V to M5V \citep{reyle2021}. The spectra of M dwarfs present a relatively monotonic and well modelled sequence  even for subdwarfs (e.g., \citet{lodieu}). Most recent work has been focussed on observations at longer wavelengths. In particular, in the infrared region where more flux is available and new spectral features might be identified. In addition such investigations are beneficial since comparisons with newly discovered cooler objects might readily be made. This can be crucial for an overall understanding of dwarfs where it is important to have a uniform spectral typing system across the  range of spectral types where pressure-broadened atomic features and molecules dominate.  

Despite new infrared diagnostics, many of the key features for M dwarfs are found at optical wavelengths, e.g., (1) lithium (6708\AA) can be a direct indicator of mass (e.g., \citet{martin2022}). (2) H alpha (6563\AA) emission and sodium (8192\AA) absorption can provide a strong indicator for age (e.g., \citet{kiman20,kiman21,schlieder2012}), (3) calcium hydride can provide metallicity (e.g., 6900\AA, \citet{g97}). A range of other useful spectral indices have also been developed by careful inspection of spectra for spectral typing (e.g, \citet{kirkpatrick1991,martin1998}) and also with larger samples of M dwarfs (e.g., \citet{allers,almendrosabad}). As large scale fibre instruments at optical wavelengths become increasing common (e.g., HERMES\citep{buder}, WEAVE\citep{dalton}, 4MOST\citep{dejong}, LAMOST\citep{wang}), large scale databases of M dwarf optical spectra will become increasingly ubiquitous and thus it is important to be able to utilise such spectra to determine reliable properties for M dwarfs.

Although, the blue optical region is a classical region for spectroscopic analysis of stars it has been neglected for M dwarfs. There are observational and modelling difficulties. The observational ones are the intrinsic lack of flux. As the flux peak of stars moves further toward the red the flux drops dramatically. From a modelling perspective, the optical spectra of M dwarfs are impacted by a range of molecular opacities and for the cooler examples also by atomic resonance lines from Na and K. The measurement and/or calculation of billions of transitions associated with the many degrees of freedom of molecular transitions is complex as is the reliable calculation of line broadening at high pressures with a number of different species. For objects with terrestrial-like temperatures some of the work might be verified by comparison with terrestrial spectra. However, for the hotter temperatures associated with M dwarfs, measurements and calculations need to be complete for a range of different species and made to the molecular dissociation energy with numerous transitions involving high energy ro-vibrational states. Thus M dwarfs can provide stringent tests of line identifications and opacities, e.g., \citep{pav22}. Discrepancies between observations and models in the blue optical will lead to errors in accounting for the overall spectral energy distribution for M dwarfs and thus determinations of their overall properties.

Over recent years there has been a substantial improvement in the molecular line lists relevant to M dwarfs. The EXOMOL group\footnote{https://www.exomol.com} have produced many relevant line lists, in particular, the latest TiO line list \citep{McKemmish19} is important across much of the optical spectra for M dwarfs. However, TiO bands became weak or even disappear at 4500\AA, so at shorter wavelength we should have the chance to see deeper layers of the comparatively hot photosphere, where numerous atomic lines form. Blueward of the TiO bandhead at 4500\AA, synthetic spectra do not provide a good match observations with observed lines being much weaker than expected (\citet{pavl17,pav22}. In this paper, we consider the evidence and potential explanations for the depression in the blue optical spectra of M dwarfs in Section 2. Our various attempts to model and explain this depression are presented in Section 3 and discussed in Section 4.

\begin{table}
\centering
\caption{Values of spectral type (SpT), Gaia G band minus 2MASS K band ($G-K$), absolute K magnitude using Gaia parallax ($M_{\rm K}$) and approximate metallicity for objects presented. Data source and/or references for spectral type and metallicity are given in the final column: B07 (\citet{sdss}), K19 (\citet{atlas}), P03 (\citet{pinfield2003}), G17 (\citet{G17}),  M18 (\citet{M18}), S19 (\citet{S19}), S22 (\citet{S22}), L19 (\citet{lodieu}). For the case of GJ551 we note that most recent assessments of the metallicity of GJ551 find solar or higher, e.g., \citet{S22} find [Fe/H]=0.259 and so we adopt the \citet{M18} value of 0.23 based on coevality with alpha cen A and B which is allowed by dynamics \citep{FJ18}.}
\begin{tabular}{llllll}
\hline
Object & SpT & $G-K$ & $M_{\rm K}$ & [M/H] & Data/Reference\\ 
\hline
NLTT5022 & - & 2.58 &  6.19 & -2.0 & this paper\\
NLTT31967 & - & 2.86 &  6.58 & -2.4& this paper\\
GJ411 & M2V & 3.21 &  6.31& -0.09 & P03/S19\\
LHS1691 & usdM2 & 3.30 &  8.14 & -1.8 & K19\\
GJ109 & M3V & 3.51 &  6.53  & -0.06 & this paper / S19\\
LHS2674 & sdM4 & 3.42 &  8.60 & -0.57 & K19\\
GJ699 & M4V & 3.67 & 8.21 & -0.15 & P03/S19\\
GJ402 & M4V & 3.92 &  7.16 & -0.07 & K19/S22\\
GJ299 & M4.5V & 3.74 &  8.51 & -0.17& this paper / S19 \\
TWA33 & M5.5 & 4.25 &  5.67 & 0.0& this paper / G17 \\
GJ551 & M5.5V & 4.60 &  8.81 & 0.23 & this paper / M18\\
LHS2096 &  esdM5.5 & 3.28 &  9.42 & -1.25 & K19\\
GJ406 & M6V & 4.95 & 9.18 & 0.07& this paper / S22\\
\hline
\end{tabular} 
\label{table}
\end{table}

%useful for quick conversions out of simbad https://www.1728.org/magntudj.htm

\begin{figure}
\centering
\includegraphics[width=\columnwidth]{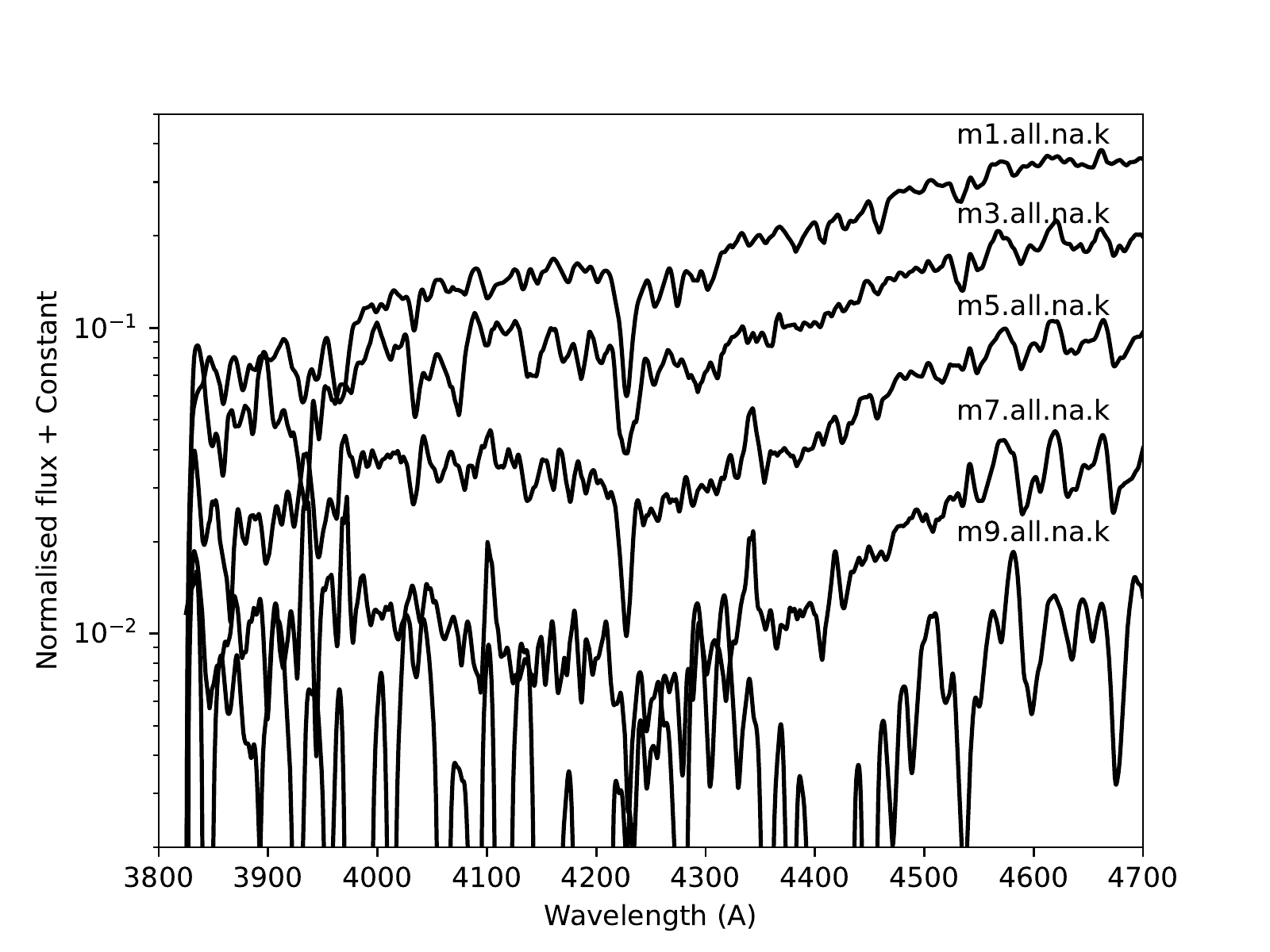}
\includegraphics[width=\columnwidth]{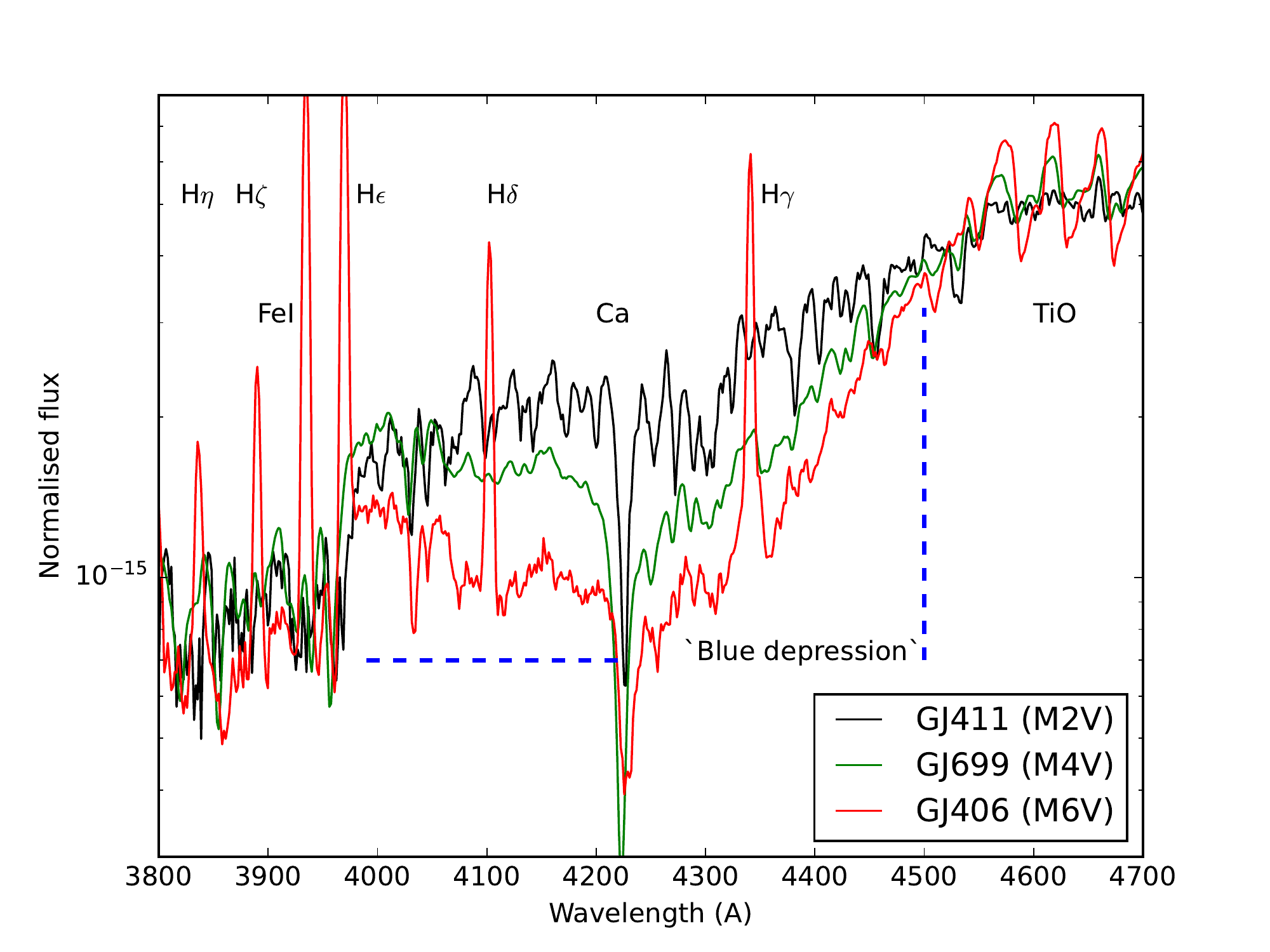}
\caption{The upper plot shows a spectral sequence from M1 to M9 using SDSS averaged spectra from \citet{sdss}. The core of the calcium line does not change much across different M spectral types although there is a broad dip from around 4000 to 4500\AA~which increases in strength to later spectral types. The SDSS averaged M9 dwarf spectrum is of low signal-to-noise particularly in the blue and has been smoothed. The lower plot approximately marks the blue depression using a blue dashed line and shows ISIS spectra overplotted so as emphasise the spectral region of the blue depression feature for the nearby dwarfs GJ411 (M2), GJ699 (M4) and GJ406 (M6). The emission lines in GJ406 are hydrogen lines and are consistent with its relatively young age and the slow decline in activity for late type M dwarfs.}
\label{bluedepression}
\end{figure}

\begin{figure}
\centering
\includegraphics[width=\columnwidth]{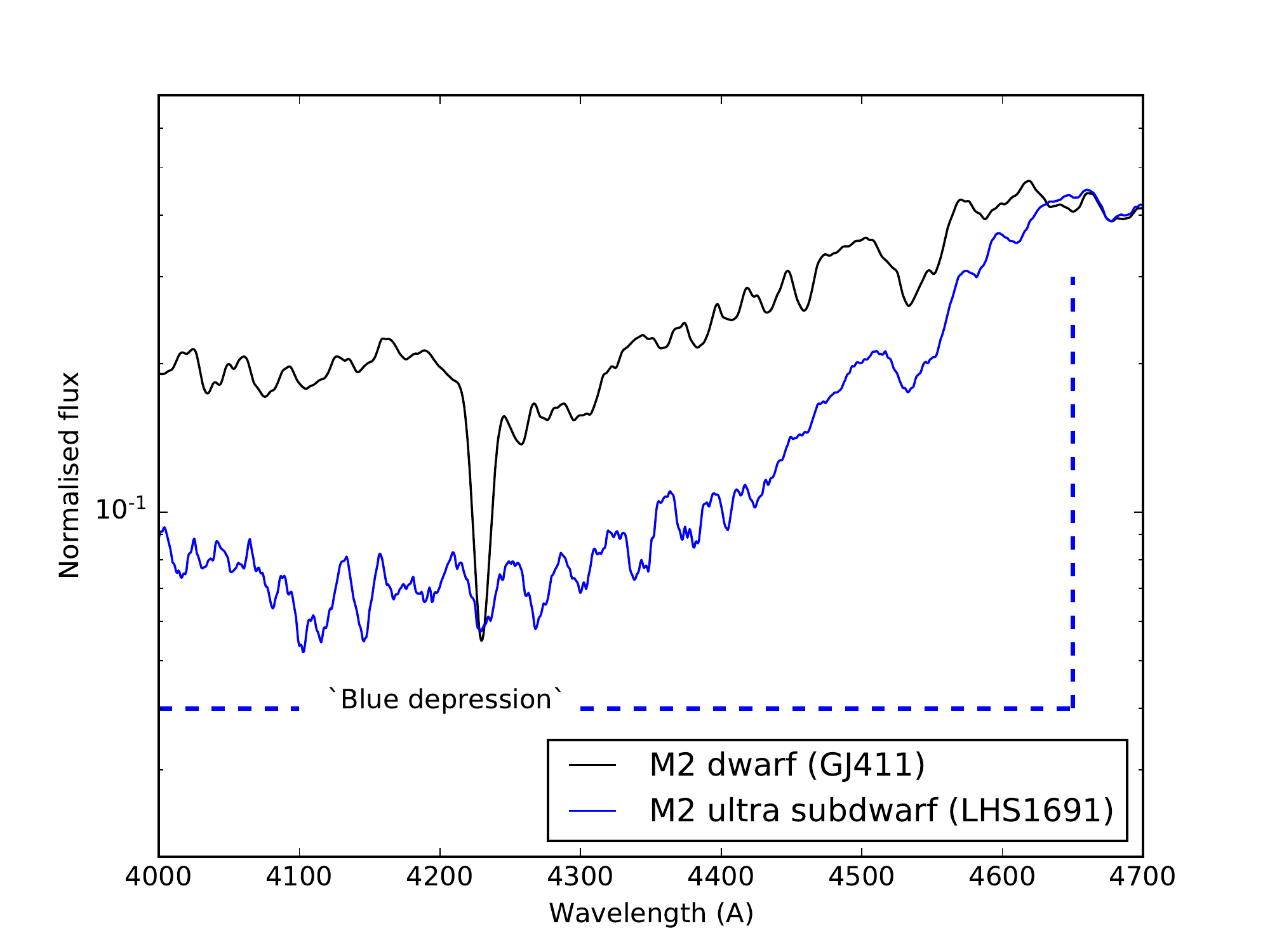}
\includegraphics[width=\columnwidth]{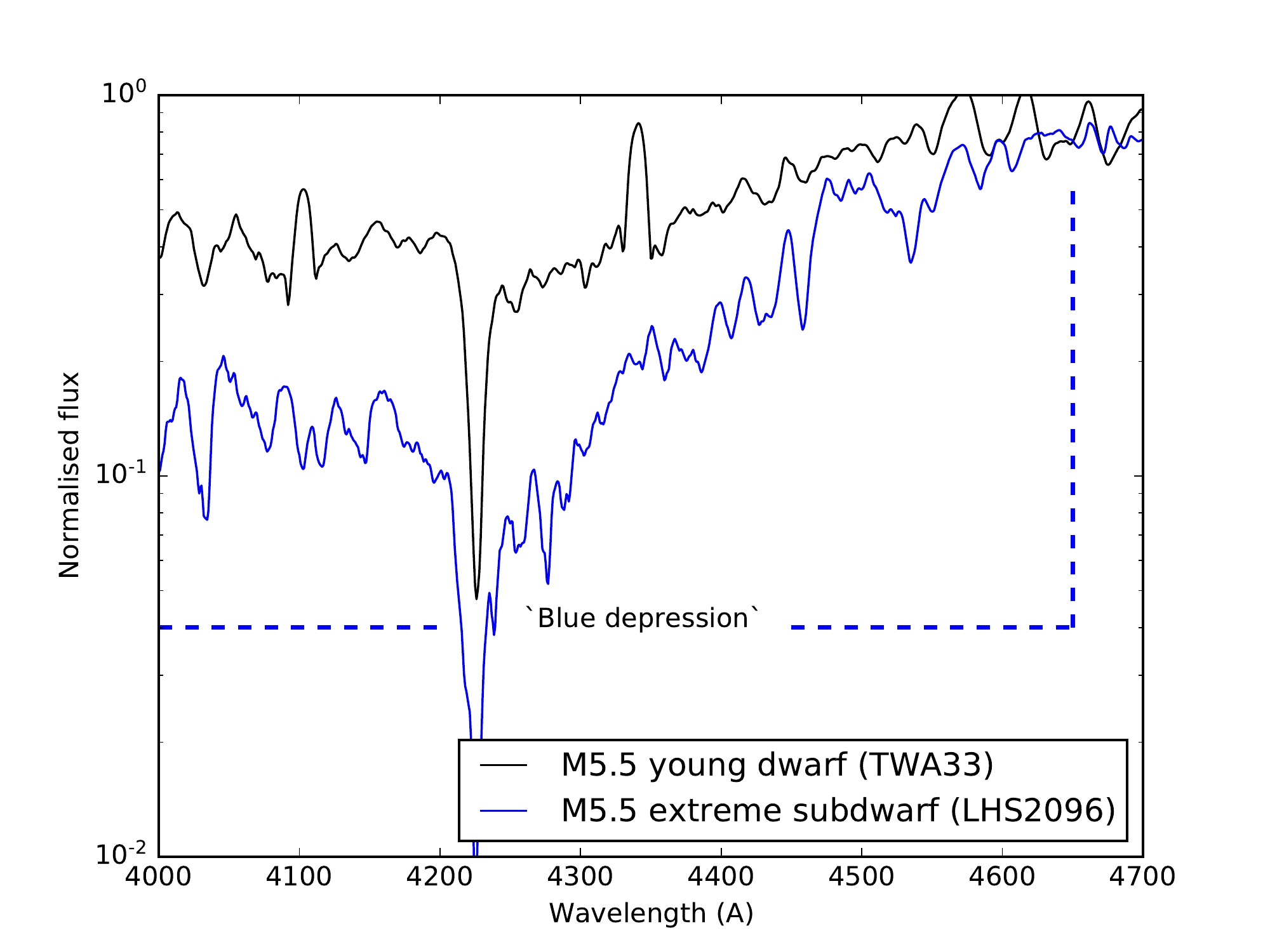}
\caption{The plots show the change in spectral feature in cooler subdwarfs, The upper plot shows the strong core of the calcium line in the M2 dwarf (Gl411). This is in contrast to the relatively weaker calcium core of the M2 ultra subdwarf (LHS1691) from  \citet{atlas}. In the lower plot, the core and the wings of the Ca feature in the M5 subdwarf (LHS2096) plotted in blue is significantly deeper than in the young M5.5 dwarf (TWA33) plotted in black.}
\label{met_grav}
\end{figure}

\begin{figure}
\centering
\includegraphics[width=\columnwidth]{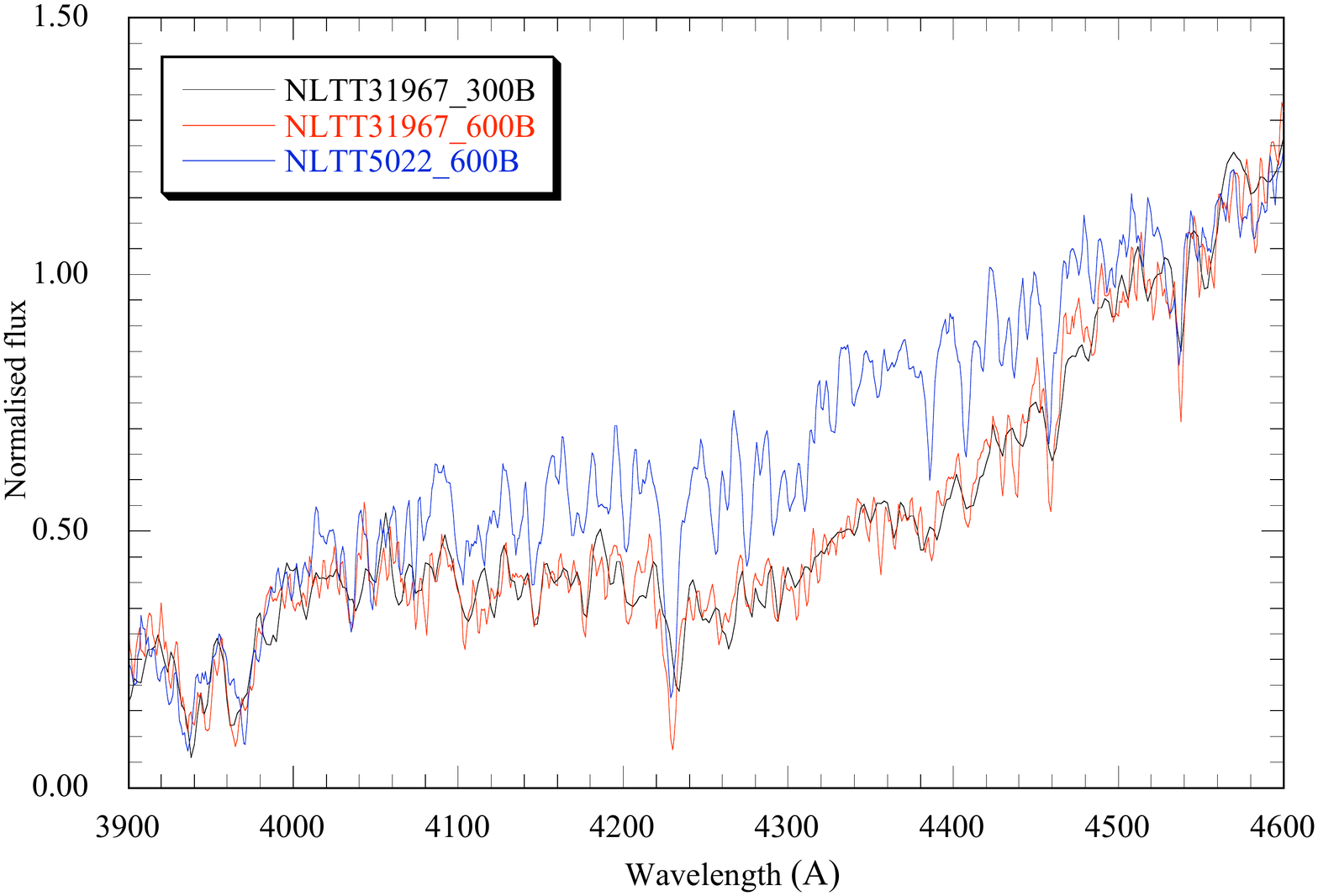}
\includegraphics[height=6cm,trim={1cm 8cm 1cm 8cm},clip,width=\columnwidth]{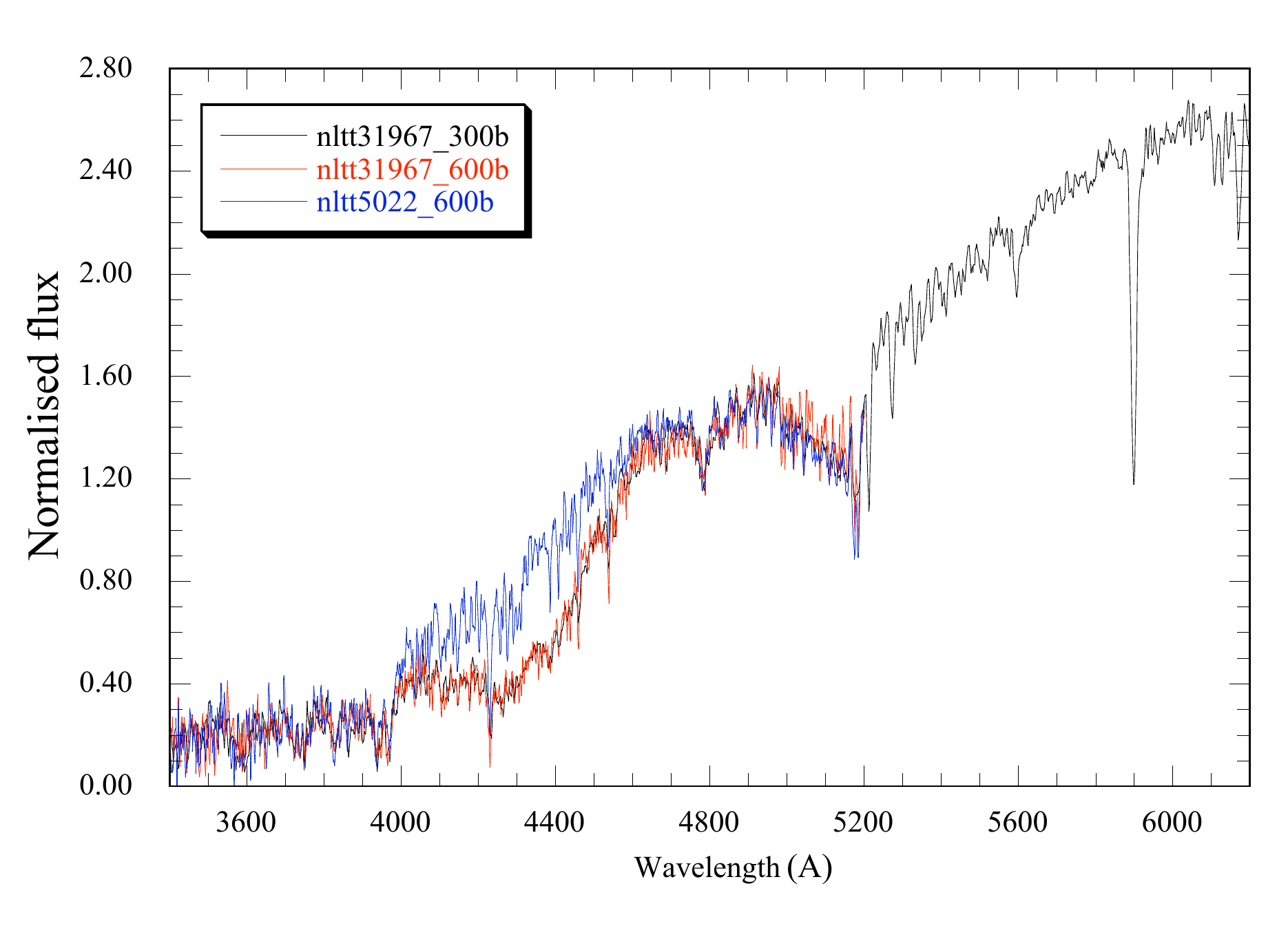}
\caption{The plots show M subdwarfs NLTT31967 and NLTT5022 with similar low metal abundance but with slightly different temperature. The upper plot shows the onset of the blue depression and the lower plot a zoom out where the spectra can be seen to closely match at wavelengths in the regions outside the 4000-4500\AA~region. NLTT31967 shows the blue depression, NLTT5022 does not. The observational spectra for NLTT31967 were taken with 300B and 600B settings and show that resolution is not the cause of the blue depression. }
\label{onset}
\end{figure}

\section{Observations}
We consider a range of observations of M dwarfs taken with different instruments and resolutions taken by some of the authors,  colleagues mentioned in the acknowledgements, from the literature, and also from telescope archives. The SDSS M dwarfs archive  \cite{sdss} \footnote{https://github.com/jbochanski/SDSS-templates/tree/master/data}
 and that of \cite{atlas} are particularly useful as they provide a wide range of spectral types with relatively uniform signal-to-noise with careful consideration for spectral type. However, we note that some of these spectra have rather low signal-to-noise in the spectral region of interest and that in the \cite{atlas} spectra the telluric bands have not been removed and that we found it difficult to remove these without leaving residuals. Thus, we use ESO archive processed data and supplement these datasets particularly with data of our own from the WiFeS instrument on the Australian National University (ANU) 2.3m telescope at Siding Springs \citep{dopita2010} and data from the ISIS instrument at the William Herschel Telescope on La Palma taken as part of the calibration dataset for \citet{pinfield2003}. Since a number of the spectra used for our study have not been published before we provide further details of their acquisition and reduction.
 
The flux calibration of WiFeS spectra were made with the 1\% Bohlin spectrophotometric standards (e.g., \citet{bohlin}).  The importance of observing a smooth spectrum star is set out in section 6 of \citet{b1999} with a preference for EG131, L745-46a, LTT4364 and VMa2. While these all work well at red optical wavelengths, EG131 and L745-46a are best in the blue. EG131 and L745-46a have an extremely shallow and broad Halpha line and a couple of very weak and shallow HeI lines (in particular 4471\AA), in addition to the deep Ca H and K lines in L745-46a. For data recovered from the ESO UVES archive where it is unlikely to have been used in the data reduction, the optimum calibration procedure has been to obtain a WiFEeS or other well flux calibrated intermediate or low resolution spectra of the UVES object. The UVES or other echelle spectrum is then smoothed to the same resolution as the fluxed WiFeS spectrum which is then divided by the WiFeS spectrum into the smoother echelle spectrum to obtain the flux corrections for the unsmoothed echelle spectrum. In particular this procedure is followed in the case of the UVES spectra of GJ109 and GJ551, the HARPS spectrum of GJ551, and the Magellan/MIKE spectrum (from David Yong) of NLTT5022. 

Here we present M1 to M9 spectra  though we can see evidence in literature spectra that a strong 4000--4500\AA~absorption feature probably does persist to later spectral types. Although many spectra of cooler objects do exist their signal-to-noise tends to be poor blueward of 5000\AA. An example of this is the combined SDSS spectrum of all observed M9 dwarfs labelled as m9.all.na.k \citep{sdss}, it is clear that the flux rapidly drops below 4500\AA~but the signal-to-noise is too poor to consider further. As can be seen from the higher resolution ISIS/WHT spectra shown in the lower plot in Fig. \ref{bluedepression}, later type spectra are increasingly impacted by chromospheric emission lines. Based on our comparisons of similar spectral types we do not find any evidence that chromospheric emission is related to the blue depression (e.g., fig. 7 in \citet{bessell2011}). Nonetheless, chromospheric emission from H$\alpha$ in M dwarfs is a well quantified indicator of age (e.g., \cite{kiman21} and young M dwarf stars may be pre-main sequence stars and have higher luminosity and lower gravity. For the later spectral type M dwarfs many do have emission lines and it is instructive if the spectra have sufficient resolution and signal-to-noise to resolve these lines.

\subsection{A depression in blue optical spectra of M dwarfs}
In Fig. \ref{bluedepression} it can be seen how an absorption feature from 4000-4500\AA~increases as a function of spectral type. The upper plot shows a sequence of averaged spectra from the SDSS \citep{sdss} and in the lower plot for the bright nearby M dwarfs GJ411 (M2), GJ699 (M4) and GJ406 (M6). In the lower plot it can be seen that apart from a broad absorption feature other spectral features in this region for GJ699 and GJ406  appear to get relatively weaker in comparison to GJ411. This is also the case for the SDSS spectra though  there is a significant decrease in signal-to-noise toward later spectral types. Given that we are investigating a number of different targets with different characteristics we include Table \ref{table} summarising some major observable parameters for the objects.
%For the particular prominent Ca{~\sc{i}} feature at 4500\AA~ case its strength appears broadly similar in all the spectra.

 In Fig. \ref{met_grav} we can get a sense of how the 4000-4500\AA~region is modified as a function of metallicity and gravity. In the upper plot, we consider the M2 old-disk main sequence star GJ411 with a gravity about log $g\sim$4.9 and the ultra subdwarf LHS1691, log $g\sim$5.2.
The Ca{~\sc{i}} 4227\AA~feature is very weak in the subdwarf as might be expected. However, the absorption feature from 4000-4500\AA~appears to have broadened in width and shows a steep decline blueward of 4600\AA. Indeed when we examine our carefully calibrated WiIFeS spectrum of LHS1691 we can not see any evidence of features blueward of 5600\AA.
%~is reminiscent of a much cooler spectral type than M2. 

We can also see  an apparent broadening in the behaviour of the blue depression in the lower plot of Fig. \ref{met_grav} with cooler dwarfs. In this case we consider the object TWA33 which is definitively in a young star forming region ($\sim$10Myr, e.g.,  \citep{Schneider}) and thus has a low gravity, log $g\sim$4.0. We compare TWA33 to an M5.5 extreme subdwarf LHS2096 (log g$\sim5.5)$ to ensure comparison with an object which is definitively older and of higher gravity. Despite the much higher abundance of TWA33, its lower gravity (log g$\sim4)$ appears to weaken the Ca {\textsc{i}} feature indicating that pressure is relatively more important that abundance. Notably the absorption feature in the M5 subdwarf from 4000-4500\AA~is no more visible than at M2 in the upper plot of Fig. \ref{met_grav}. Overall then the blue depression absorption feature is stronger at lower metallicity and higher gravity. The Ca{~\sc{i}} feature appears to behave broadly as expected. Lowering the metallicity increases the gravity and the pressure, and lowering the temperature also increases the gravity and the pressure. So the common feature is increased pressure causes the blue depression.

%In this case TWA33 has a gravity of log $g\sim$4.0 while for LHS2096 log $g\sim5.5$}. In general, the behaviour of metal poor stars to present the blue depression 4000-4500\AA~absorption feature as relatively broader and deeper continues to later M spectral types. The lower plot of Fig. \ref{met_grav} is intended to show the impact of gravity. {\bf 

%Although it does not have the prominent chromospheric emission features presented by GJ406 in Fig. \ref{bluedepression}, 

It is instructive to investigate the first appearance of the blue depression in the hotter M dwarfs. Fig. \ref{onset} shows a comparison of two M subdwarfs NLTT31967 and NLTT5022 with similar low abundance and temperature of around 3900K. This temperature was derived by a fit of the 4000-9500\AA~spectra using BT-Settl synthetic spectra. Most of the diagnostic ability of the models is found to reside in the continuum slope. NLTT31967 with $M_{\rm G}=6.58$ shows the blue depression whereas NLTT5022  with $M_{\rm G}=6.19$ does not. They both show a similar strength atomic Ca{~\textsc{i}} at 4227\AA~absorption feature. The spectra of the two objects appear to be well matched at wavelengths outside the 4000-4500\AA~region.

\begin{figure}
\centering
\includegraphics[width=\columnwidth]{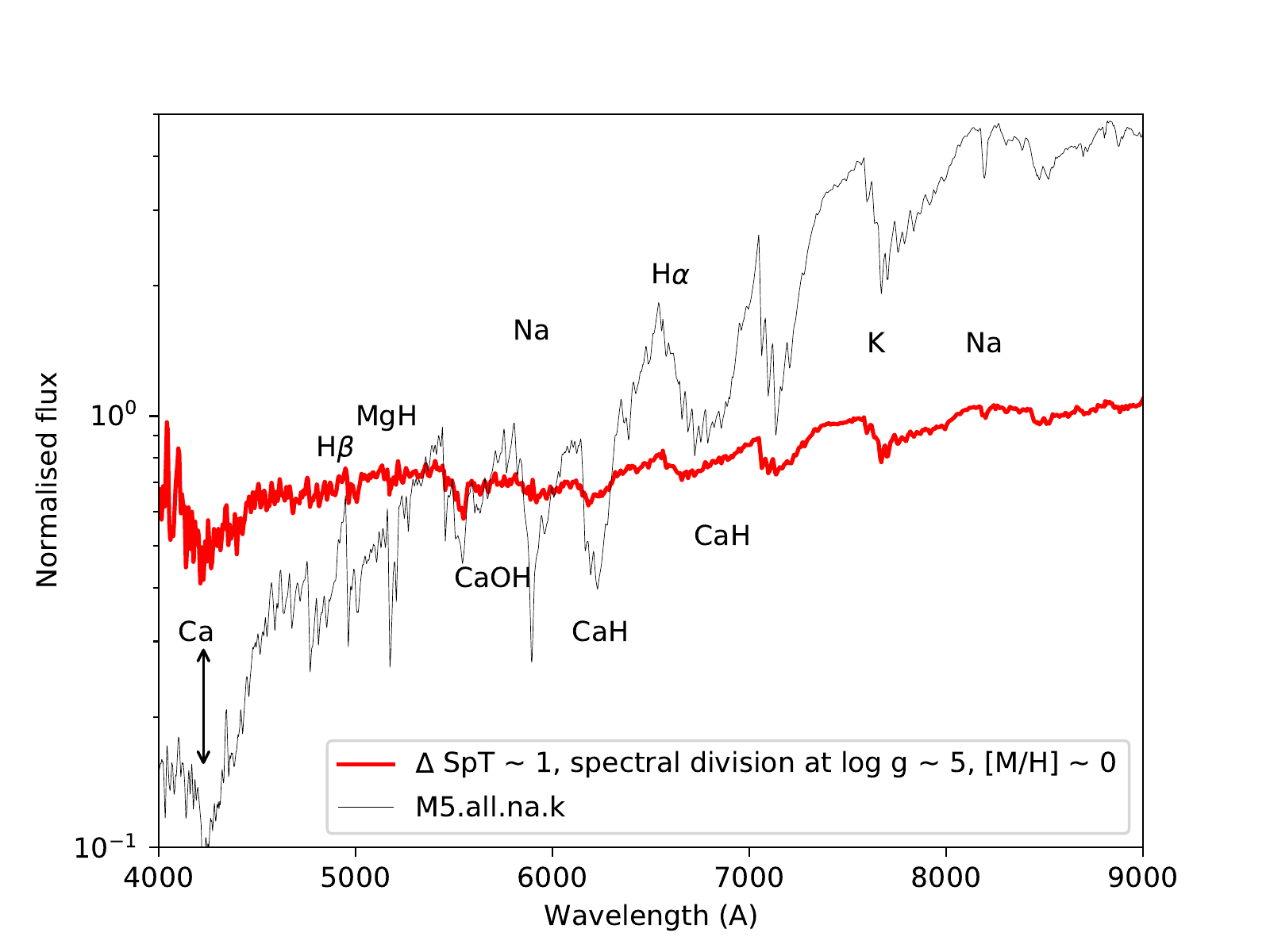}
\includegraphics[width=\columnwidth]{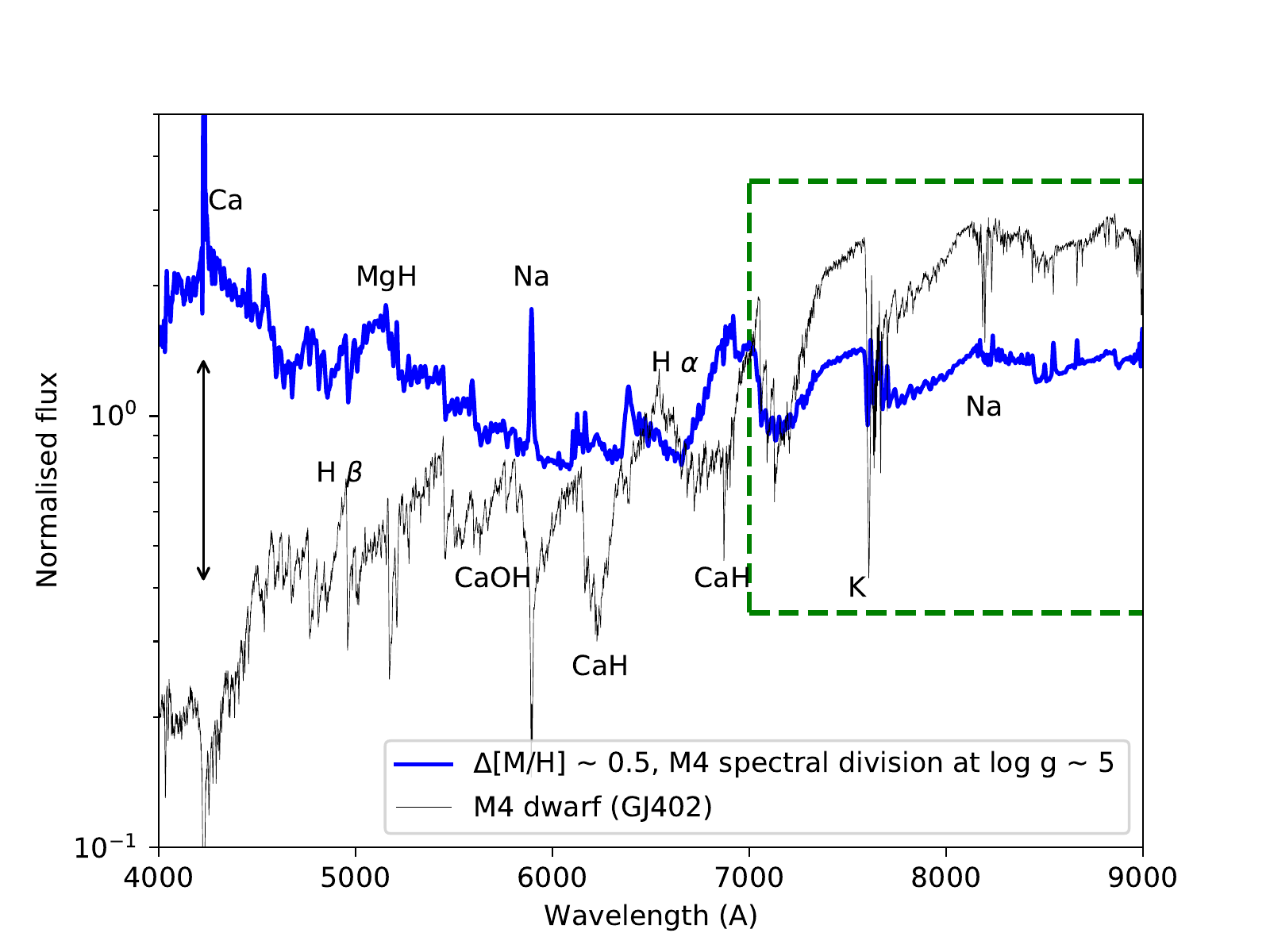}
\includegraphics[width=\columnwidth]{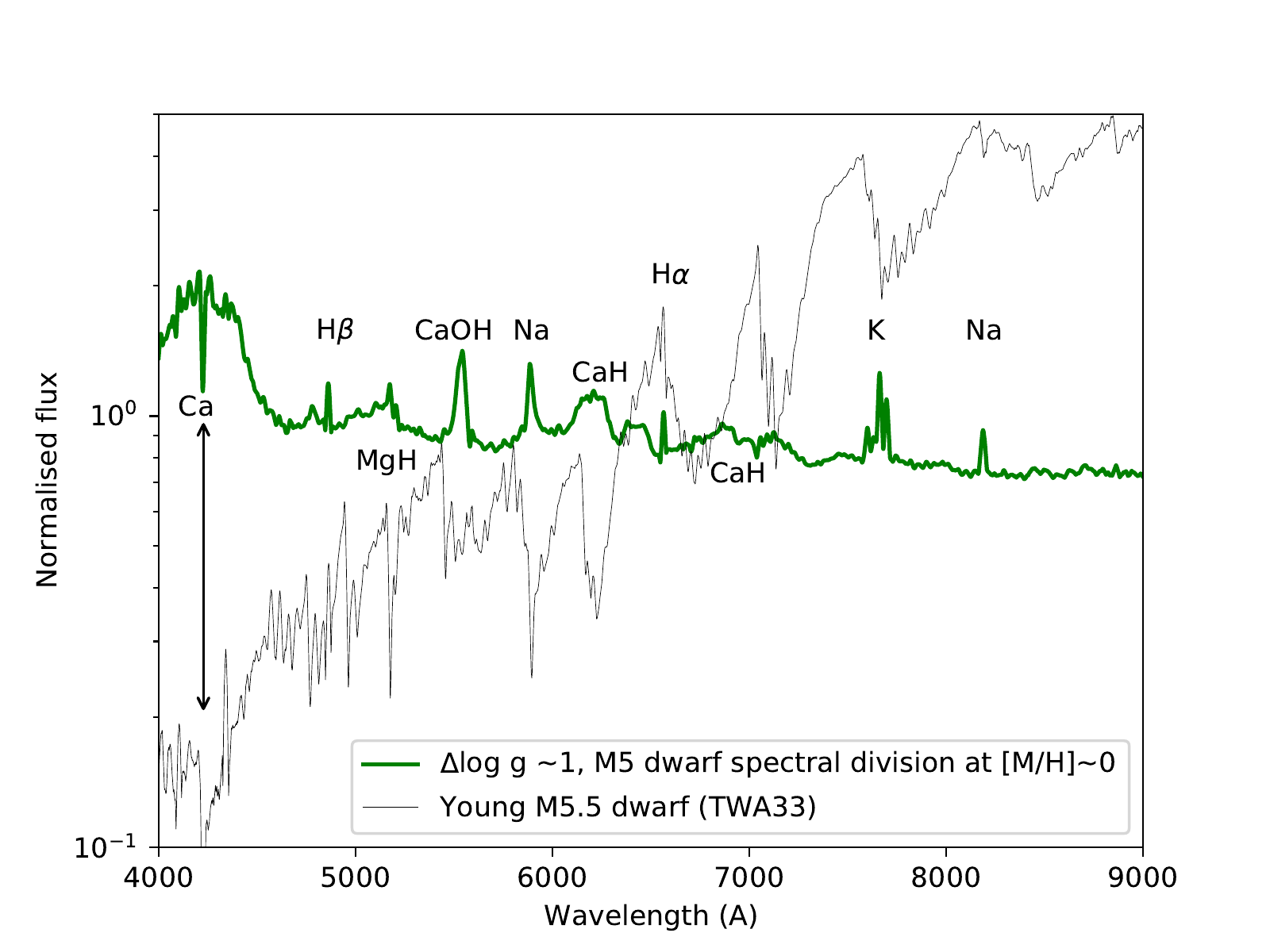}
\caption{The plots indicate the sensitivity of the blue depression to temperature (top), metallicity (middle) and gravity (bottom). In the top plot the red line represents a change in spectral type of 1 (approximately 150~K) and is made from the division of an SDSS average M5 dwarf spectrum (m5.all.na.K) by an average M4 dwarf spectrum (m4.all.na.k) with the M5 spectrum shown in grey. The middle plot, shows a blue line made from dividing the M4 dwarf (GJ402)  by an M4 subdwarf (LHS2674) to represent a metallicity change of around 0.5~dex. The green line in the lower plot is intended to represent a gravity change of approximately 1~dex. It is made from the division of a low gravity M5.5 (TWA33) by an average of the GJ551 (M5.5) and GJ299 (M4.5) so as to produce a spectrum approximately matched in colour. All plots identify the major spectral features in M dwarfs and show a double headed arrow to point to draw attention to the blue depression feature.}
\label{metal_gravity}
\end{figure}

%Beyond 6500\AA~the divided green spectrum is reminiscent of the original M5.5 but with features diminished in strength. Only the Na feature at 8200\AA~prominently gives away the low gravity nature of the M5.5 young dwarf TWA33. From 4500 to 6500 the divided green spectrum is relatively flat with gravity sensitive feature such as MgH, Na and CaH giving away the low gravity nature of TWA33. 
%Beyond  6500\AA~the red spectrum indicates similar features but reduced in strength due to lower metallicity. At wavelengths bluer than around 6000\AA, there is a significant underlying slope and indication that lower metallicity can lead to a stronger blue depression across a region extending as far as 6000\AA. 
%The red line shows a gentle upward slope and reduced feature strength. 
%how lower metallicity can lead to a stronger blue depression across a region extending as far as 6000\AA. 
%\begin{figure}
%\centering
%\includegraphics[width=\columnwidth]{index}
%\caption{The plot shows an index based on the spectral sensitivity of the blue depression. The black points and the joining black line are based on the SDSS averaged spectra shown in Fig. 1.  The other objects are ones used to illustrate the sensitivity of the blue depression to metallicity (LHS1691, LHS2674, LHS2096) and gravity (TWA33).}
%\label{index}
%\end{figure}

We can also examine spectra by dividing them. This helps to show the relative importance of different spectral features. The three plots in Fig. \ref{metal_gravity} are shown to indicate how different spectral features react to temperature (upper), metallicity (middle) and gravity (lower) across the optical regime for relatively small spectral changes around spectral types from M4 to M6. The red line in the top plot shows the impact of a spectral class change of 1 from the spectral division of averaged M5 by M4 spectra from the SDSS archive.  The red divided spectra resembles the M5 spectra plotted in grey with weaker features as would be expected for the spectral division of two objects with similar spectral types. As anticipated in Fig. 1, the divided spectrum indicates the blue depression has some significant temperature sensitivity.

The middle plot shows a blue line to represent a change in metallicity of approximately 0.5dex based on the division of an M4 dwarf (GJ402) by an M4 subdwarf (LHS2674) from \citet{atlas}. In this case we chose GJ402 because it does not have  particularly prominent spectral emission features associated with youth and a lower gravity. The M4 dwarf is also shown in grey as a comparison. The middle region of the plot from 6000-7000\AA~is strongly influenced by CaH and H$\alpha$. Redward of 7000\AA, marked with green dashed area, the divided blue spectrum is much flatter. In this green dashed area the divided spectrum resembles the M4 spectra with weaker features as would be expected for the spectral division of two objects with similar spectral types but one with significantly lower abundance. From 4000-6000\AA~the blue divided spectrum slopes gently upward with a bump around 4000-4500\AA~anticipated by Figs \ref{met_grav} and \ref{onset}. The upward slope might partly be caused by GJ402 being somewhat redder and brighter than LHS2674 ($\Delta$$(G-K)=0.5$, $\Delta$$M_{\rm K} =1.44$). It is also plausible that the influence of the blue depression feature in the 4000-4500\AA~region might extend much further for lower metallicity M dwarfs. In particular, the very varied behaviour of spectral features with wavelength emphasises the difficulties of spectral typing for low metallicity M dwarfs.

In the lower plot of Fig. \ref{metal_gravity} we again divide spectra of a similar spectral types but here we also seek to examine the impact of gravity. In this case we make comparisons between WiFeS/ANU spectra of the low gravity (log $g \sim$4.0 M5.5 dwarf TWA33 whose spectrum is shown in grey with GJ299 (M4.5, log $g \sim$5.0) and GJ551 (M5.5, log $g \sim$5.0). In this case, the green spectrum is TWA33 ($G-K=4.25$) divided by averaged spectrum of GJ299 ($G-K=3.74$) and GJ551 ($G-K=4.60$).  Relative to the upper plots, the green divided spectrum in the lower plot is rather flat without any particular slope. At wavelengths blueward of 4500\AA, the green spectrum shows the distinctive  4000-4500\AA~feature. As with the other plots, well known spectral features are identified. 
%The upper and lower red divided spectrum indicates that the blue depression presents very differently for a change in metallicity of approximately 0.5 dex and temperature change of one spectral class.
%also that the slope on the left-hand side of the upper plot from 6000\AA  consequence of the 4000-4500\AA~feature extending redward to 6000\AA~due to decreased metallicity. 

Overall, Fig. \ref{metal_gravity} illustrates that the 4000-4500\AA~feature is the single most sensitive feature in these M dwarf spectra. It is found to increase in strength towards lower temperatures and higher gravities in a relatively similar manner. As also indicated in  Fig. \ref{met_grav} a decrease in metallicity can lead to an increase in relative strength and a broader shape for the blue depression. Presumably, towards lower metallicities, the blue depression increasingly dominates over other opacities in the blue optical region.

\begin{figure}
\centering
\includegraphics[width=\columnwidth]{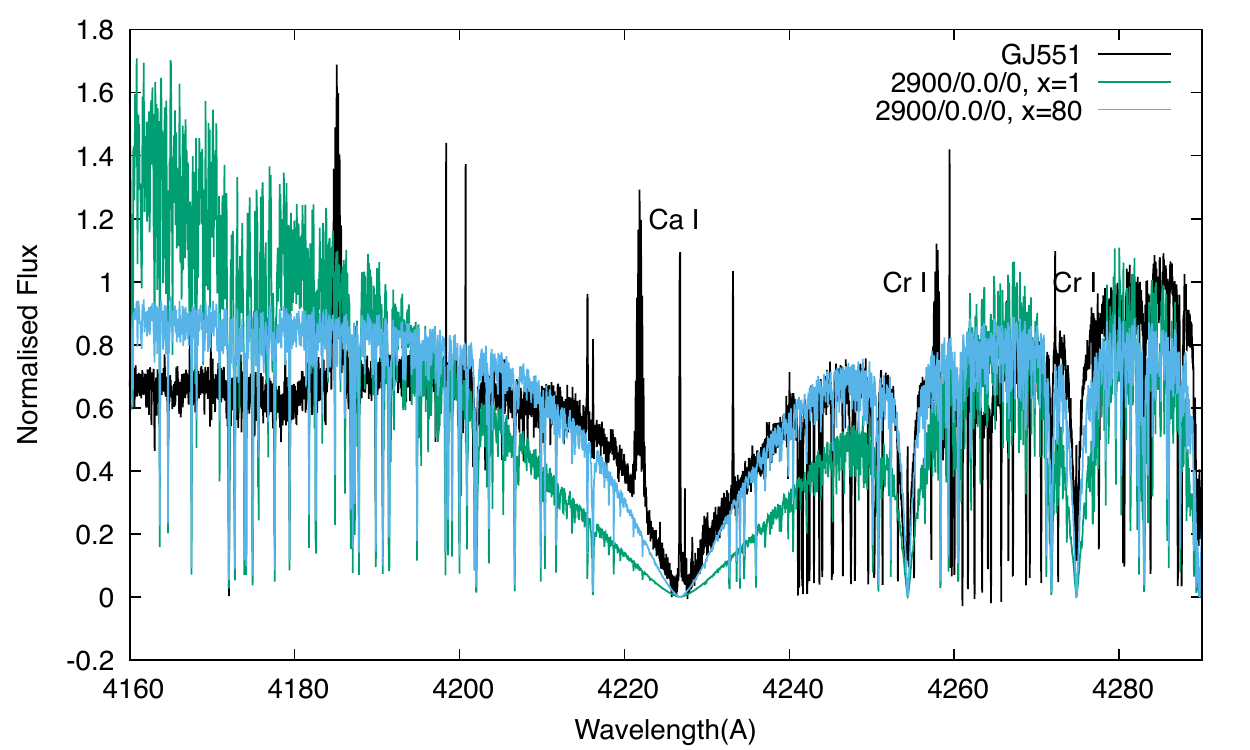}
\includegraphics[width=\columnwidth]{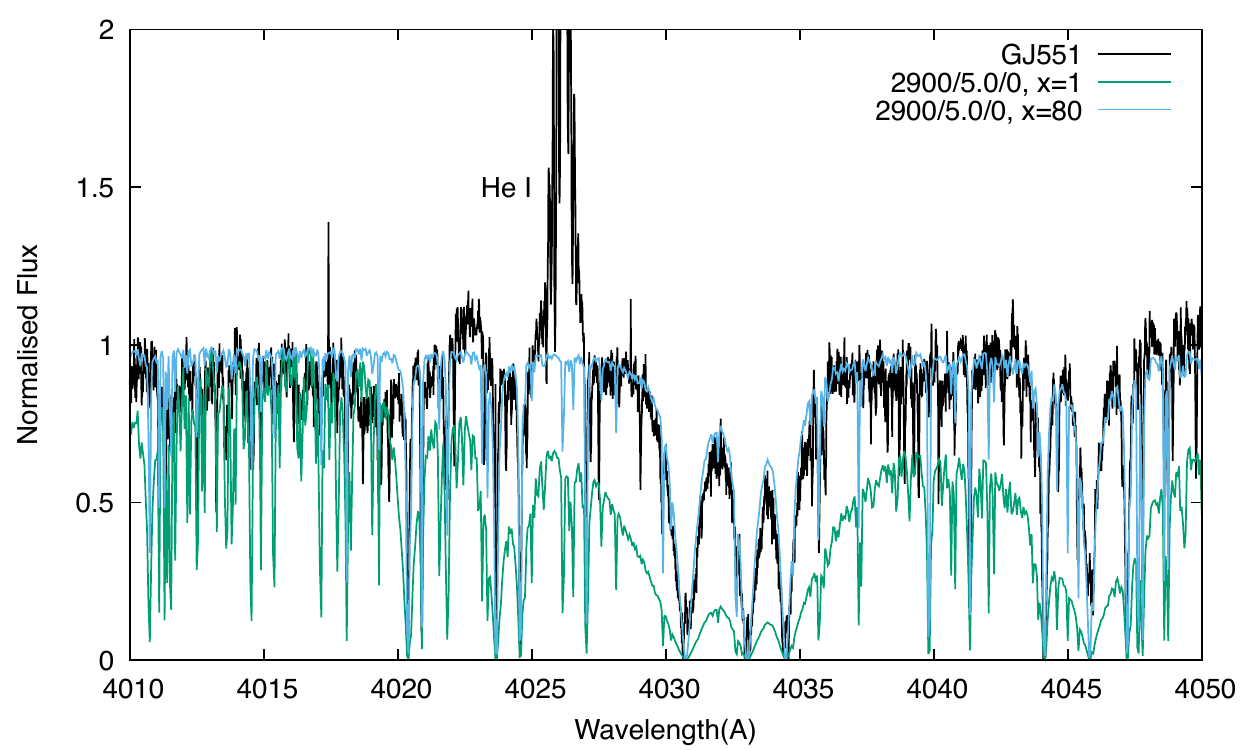}
\includegraphics[width=\columnwidth]{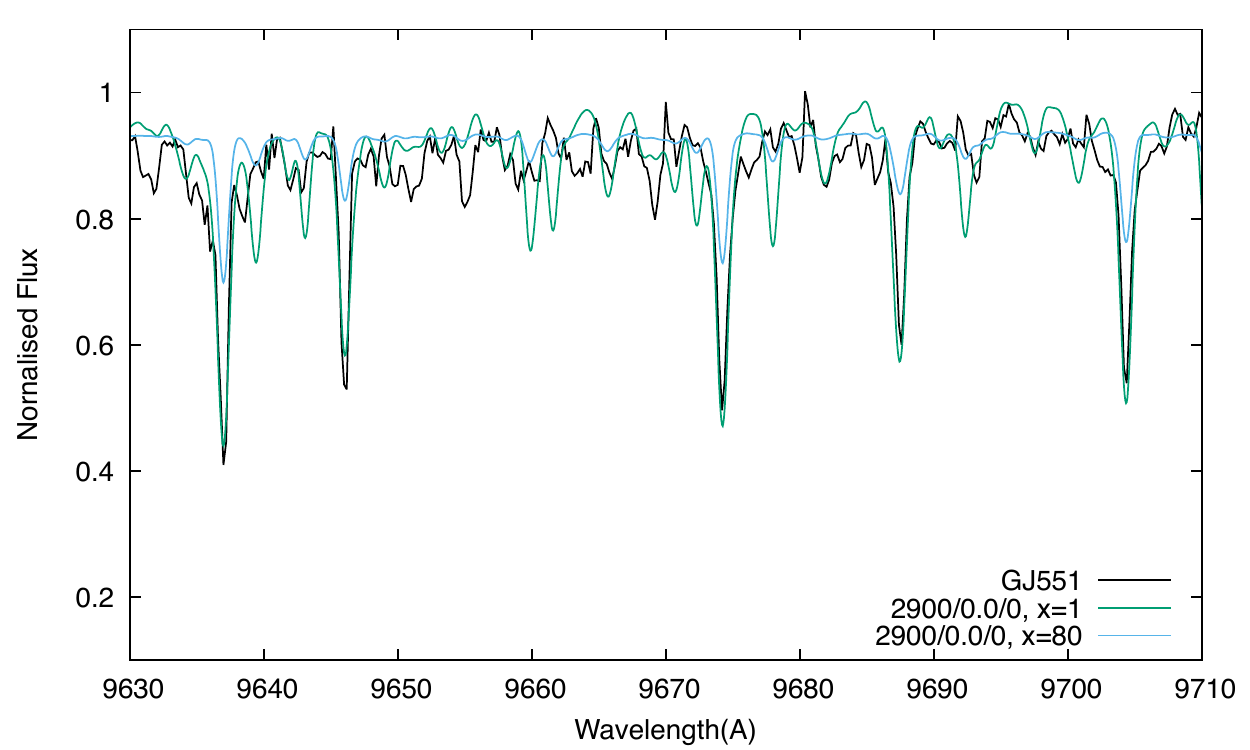}
\caption{Spectra of GJ551 in black compared with a `standard' model of solar metallicity with a temperature of 2900K and gravity of log g = 5 in green (labelled as x=1) which would be expected to produce a good spectral fit. The model in blue has had the continuum opacity enhanced by 80 (labelled as x=80). The upper plot shows that the immediate region around the 4227\AA~range. The middle plot shows a spectral region close to 4227\AA~with a variety of absorption features also better fit by the model with enhanced opacity. The lower plot shows a spectral region far away where the enhanced opacity provides a poor fit and would not be appropriate.}
\label{highres}
\end{figure}

In the past this broad depression in the blue optical spectra of M dwarfs has been noted by a number of authors (e.g., \citet{lindblad1935,lindblad1935nat,1943,vardya,warner,akeg1980,bessell2011,pavl17,pav22}) and modelled by an increase in the continuum opacity of this region (e.g., \citet{vardya}) and for Proxima Cen (GJ551) by a factor 40-80 relative to expected in order to match the observed line strengths of spectral features \citep{pavl17,pav22}. The upper plot of Fig. \ref{highres} illustrates how a fit is obtained for the Ca{~\textsc{i}} 4227\AA~line through an increase of the continuum opacity by 80. The middle plot shows a representative region 200\AA~to the blue which also requires the opacity increase for a reasonable fit. However, such an arbitrary enhancement is not appropriate to fit features much further from the blue depression, e.g., the lower plot of  Fig. \ref{highres} shows a redder wavelength where a poor fit is obtained for the enhanced continuum opacity. 

The necessity for additional opacity is appreciated elsewhere; for example, figs 7 and 25 of \cite{Herczeg} and fig. 6 of \citet{cap2023} that shows the M3.5V star GJ555 and a 3200K synthetic spectrum indicating a much stronger Ca{~\textsc{i}} 4227\AA~line than observed as well as a poor fit in the surrounding spectral continuum. The explanation for this 4000-4500\AA~spectral featfBure reported by \citet{lindblad1935,lindblad1935nat} and independently as the `Lindblad depression' by \citet{1943} has been attributed as the Ca$_2$ quasi-molecule (e.g., \cite{lindblad1935}), to CaH absorption (e.g., see discussion \citet{weniger}), as a missing opacity (e.g., \citet{vardya}), and perhaps as a lack of 3D NLTE models (e.g., \cite{pavl17}). 

\citet{akeg1980} found that the strengths of the Ca{~\textsc{i}} 4227\AA~line and the `Lindblad depression' increase with decreasing metallicity and if interpreted correctly can be used to identify metal-poor stars. However, it appears that appropriate modelling of this spectral region has not been adequately captured by synthetic spectra computed for the grids of modern model atmospheres. Although some modern studies have used this absorption feature as a spectral diagnostic the relatively lower signal-to-noise obtained below 4500\AA~ for M dwarfs means this spectral region has been relatively neglected. 

\begin{figure}
\centering

\includegraphics[height=7cm,trim={1.8cm 1.8cm 5cm 17cm},clip]{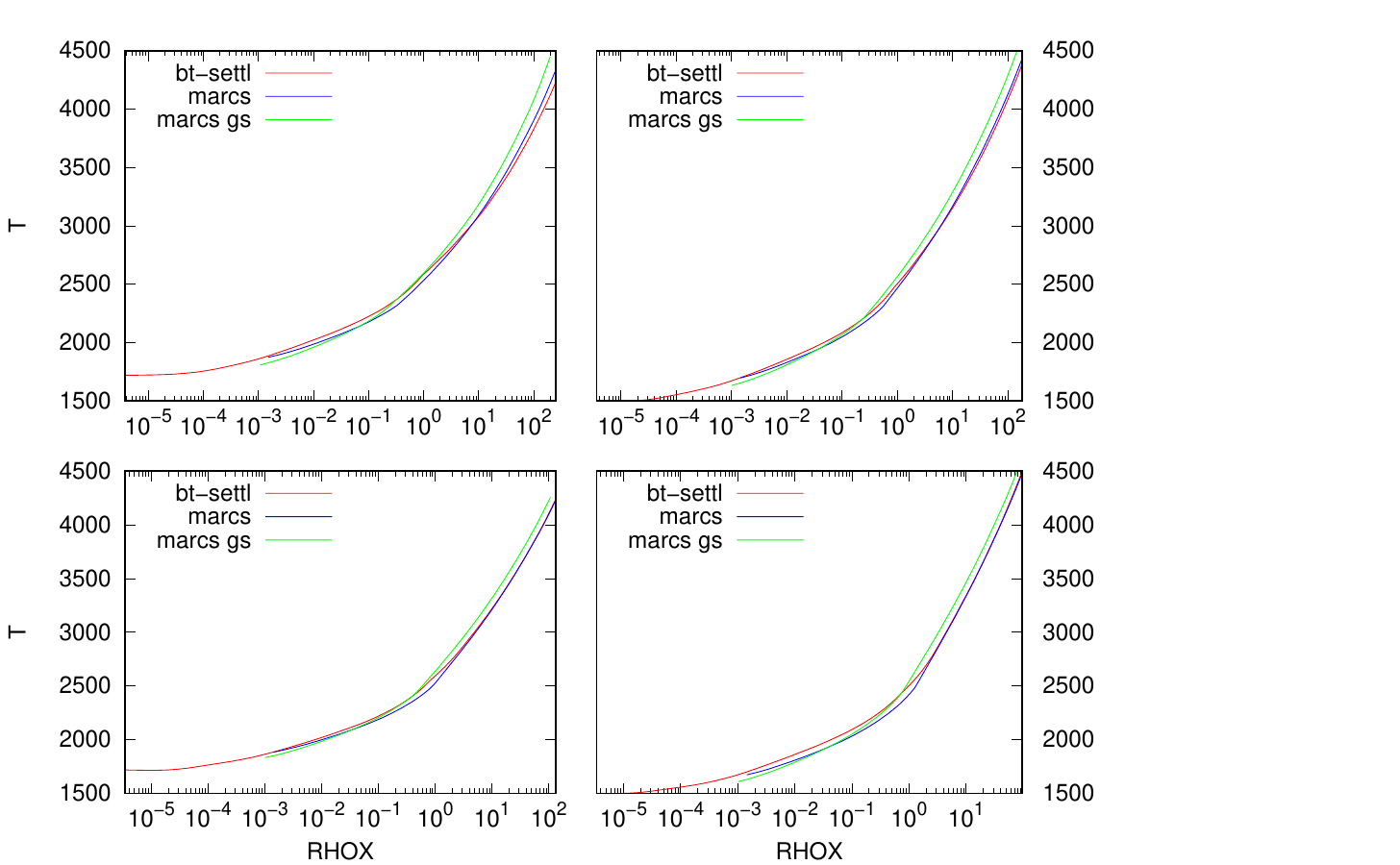}
\caption{The plot shows BT-Settl (in red) and MARCS (in blue) model structures for \Tef=3000K, [Fe/H]=0.0, v$_{\rm tan}$=2.0km/s, and log$g$=4.0 and 4.5 (upper plots) and 5.0 and 5.5 (lower plots). The axes are RHOX representing the total pressure divided by the models' nominal log $g$ and $T$ for temperature in K. The MARCS-GS model structure using \citet{gs98} for the abundance pattern is shown in green. For MARCS-GS the abundances are [Fe/H]=+0.05, [$\alpha$/Fe]=+0.11,  [C/Fe]=+0.08, [N/Fe]=+0.09 and  [O/Fe]=+0.12 whereas for the standard model (in blue) these parameters are 0.0. } 
\label{MS}
\end{figure}

\begin{figure}
\centering
\includegraphics[width=1.05\columnwidth]{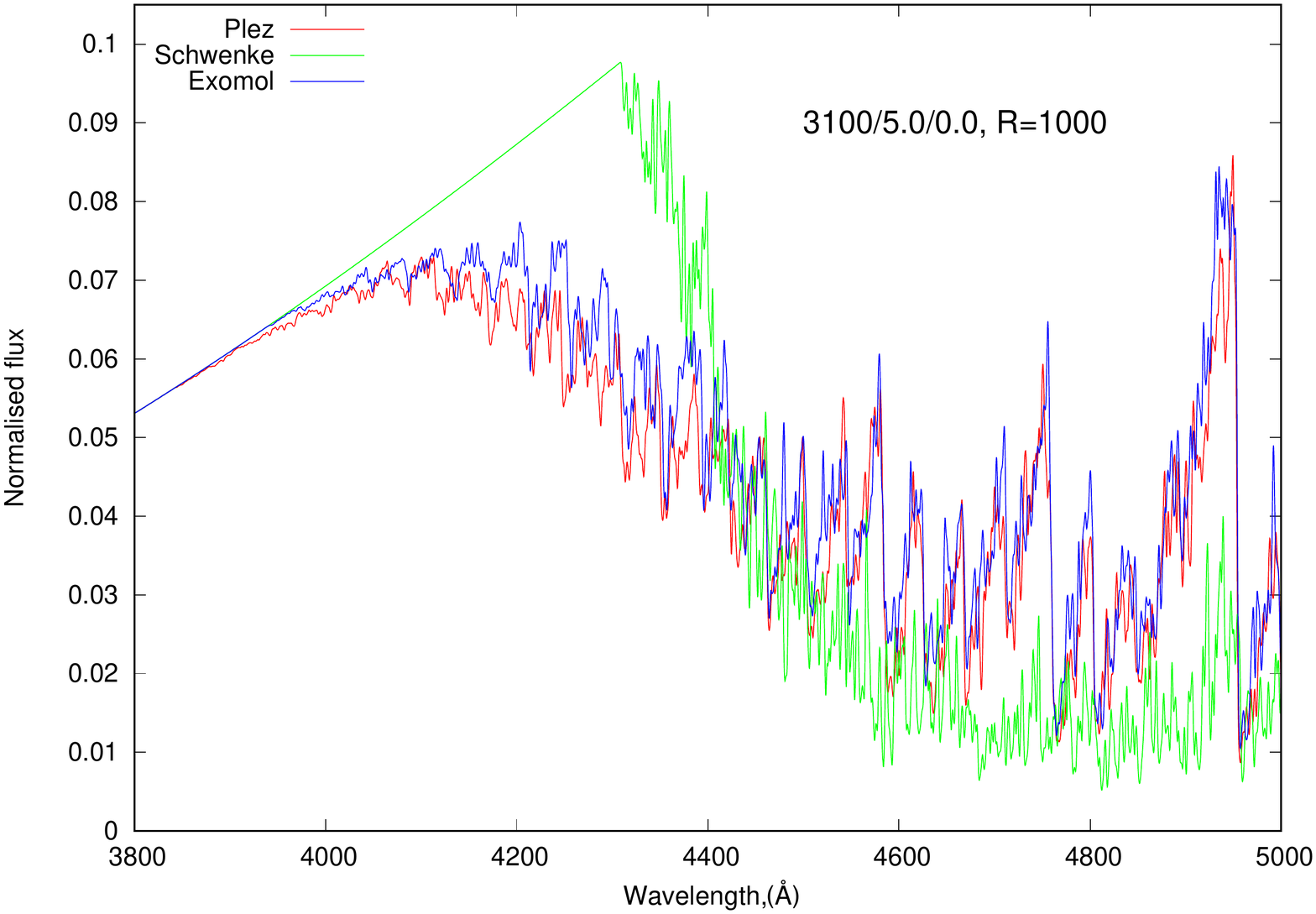}
\caption{The plot compares the evolution in available synthetic linelists for TiO for \citet{plez98},  \citet{Schwenke98} and EXOMOL \citet{McKemmish19} computations.} 
\label{tio}
\end{figure}

\section{Model atmospheres}
To  better understand the modelling of the 4000-4500\AA~region we consider a number of grids of models producing synthetic spectra for M dwarfs including MARCS \citep{gustafsson}, ATMO2020 \citep{phillips}, NextGen \citep{nextgen}, BT-Settl \citep{btsettl}. We note the extensive work on the variations in derived spectral parameters by \citet{crist} from the use of different model atmospheres. For example, they find that the temperatures of MARCS models are on average about 30~K higher than from the PHOENIX models including BT-Settl, metallicities are offset by around 0.4~dex, and log $g$ values lower by about 0.30~dex. We also find some modest differences between the models and so examine their underlying model structures. In  Fig. \ref{MS}, we show model structures for 3000K for MARCS and BT-Settl and also consider the impact of using a different abundance pattern labelled as MARCS GS (based on the abundances from \cite{gs98}). The differences between models structures which might provide for line formation taking place in relatively different places within the atmosphere are seen to be relatively consistent. We also find a relative similarity between model structures at 2500 and 3500K and so do not envisage that the differences between model structures can be responsible for large scale discrepancies between observed and synthetic spectra seen in Fig. \ref{highres}.

In general the spectra of M dwarfs are envisaged as a background of molecular absorption with some atomic absorption features still strong enough to be visible. In the optical regime, the opacity caused by the TiO band system is particularly prevalent (see \citet{pav14}). As experiments and ab initio calculations have been able to more accurately assess higher energy transitions the modelling of TiO has improved. In Fig. \ref{tio} it can be seen how different line lists have improved in terms of identifying the primarily short-wavelength bands head as well as more subtle spectral features from \citet{plez98} to \citet{Schwenke98} and latterly \citet{McKemmish19} .

When calculating synthetic spectra the “line by line” approximation is used and in particular we use (1) the list of atomic lines from the VALD database \citep{ryab15}, (2) lists of CN and MgH lines \citep{kuru18}, (3) lists of CrH and FeH lines (\citet{burr02} and \citet{duli03}), (4) lists of \hho absorption lines calculated by \citet{barb06}. Absorption line profiles for atomic lines were taken from VALD3, or in the case of absence from VALD3, they were determined in terms of the Voigt 
function and damping constants were
determined using the Unsold approximation \citep{unso55}. The microturbulent 
velocity here was assigned to be constant over the depth of the atmosphere and equal to \Vt = 2 km/s.
Spectra were computed by Wita 6 program \citep{pav14} with step 0.02\AA, and then convolved with a Gaussian FWHM and rotation profile by \citet{grim} to match the spectral comparison. 
Here we consider a number of modified modelling solutions to explain the 4000--4500\AA~blue depression feature.
%Where appropriate rotational broadening is applied following \cite{gray76}. 
%As the red optical spectra of cool stars became increasingly well modelled it became evident that the resonance lines of K and Na played a strong role in governing the appearance of different spectral types. identification of the L dwarf spectral type. Short \& Burrows and Burrows et al. Big efforts to determine resonance lines of L dwarfs. \\

\begin{figure}
\centering
\includegraphics[width=\columnwidth]{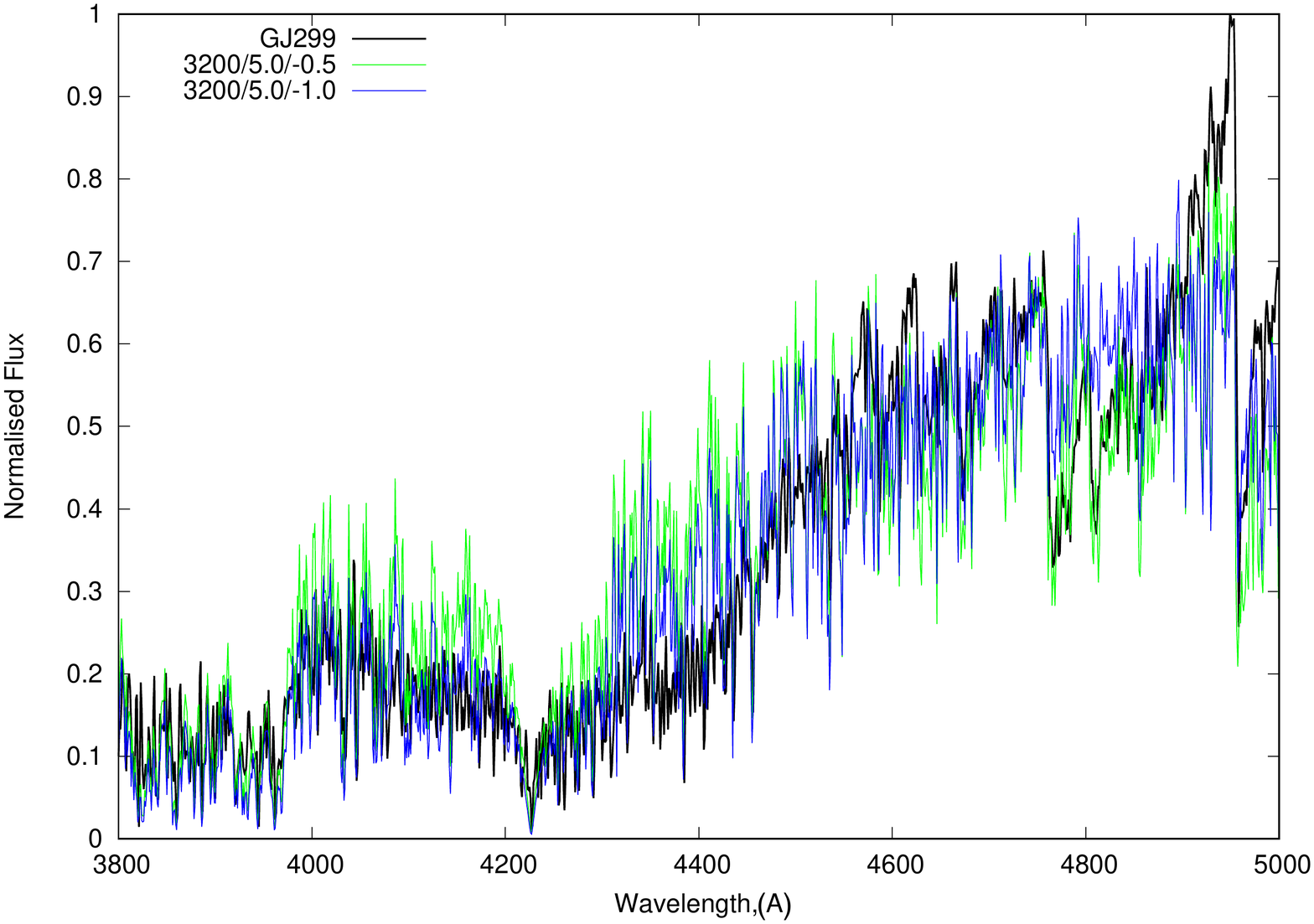}
\includegraphics[width=\columnwidth]{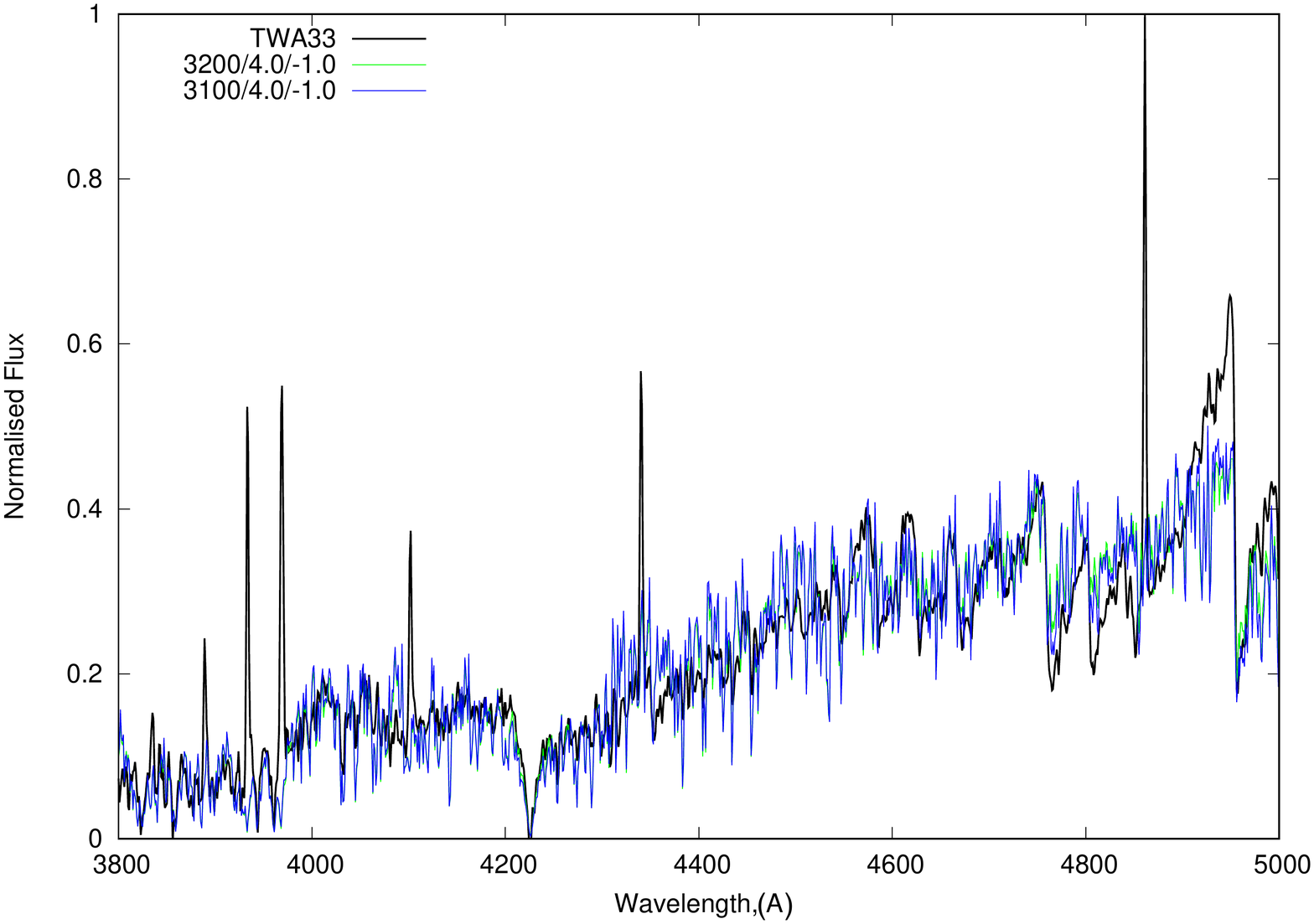}
\caption{Some M dwarf spectra can be fit rather well by reducing the metallicity of the synthetic spectra. The upper plot shows the M4.5 GJ299 and the lower plot shows the young M5.5 star TWA33. The observed spectra in black are relatively well fit by synthetic spectra with low metallicities.}
 \label{metal}
 \end{figure}
 
 \begin{figure}
\centering
\includegraphics[width=\columnwidth]{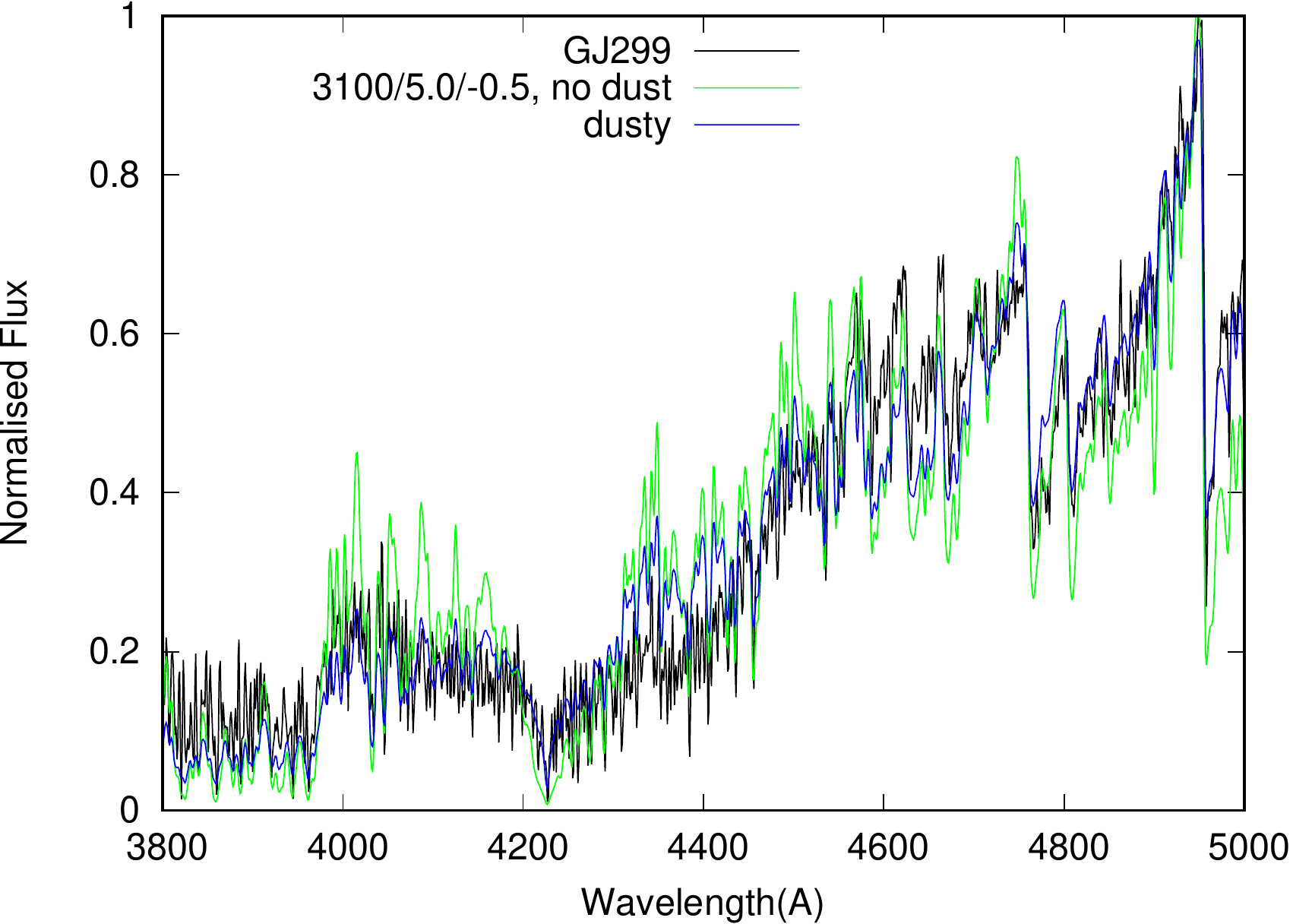}
\includegraphics[width=\columnwidth]{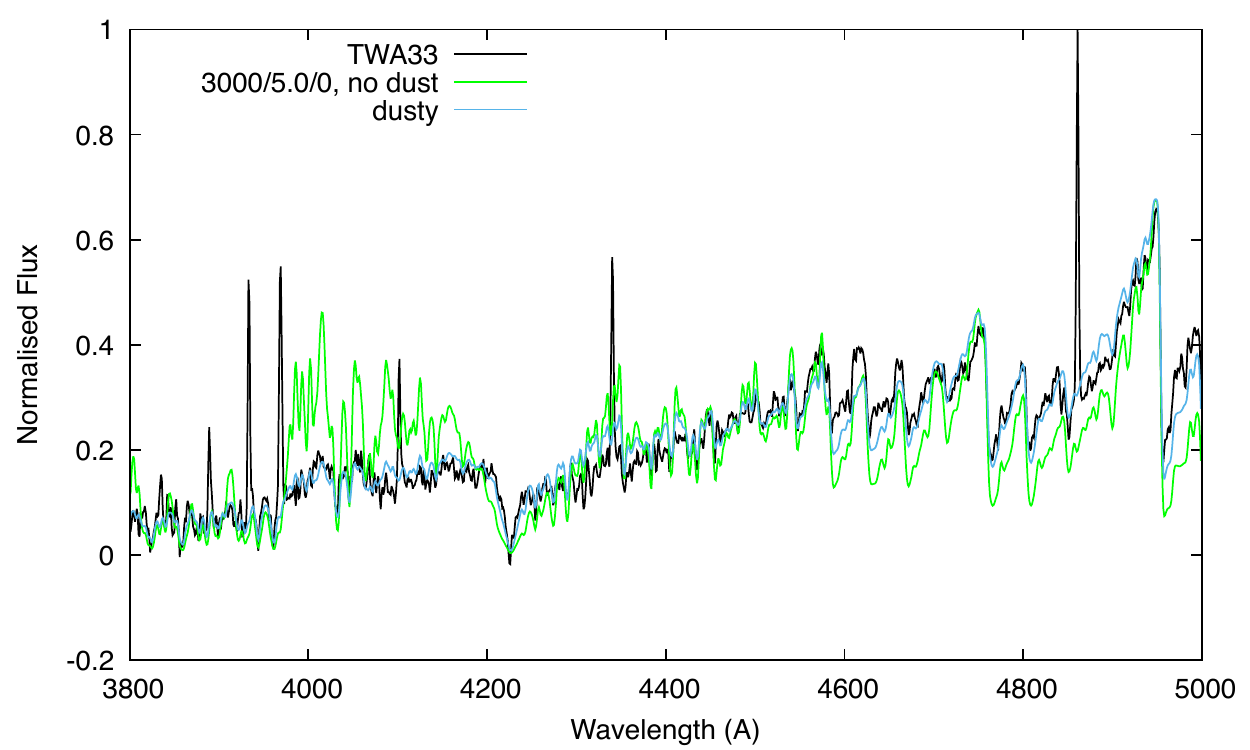}
\caption{Some M dwarf spectra can be fit rather well by introduction of dust into the synthetic spectra. The comparisons between observed and synthetic spectra with (blue) and without (green) dust suggest some improvement in the fit can be made with the addition of dust.}
 \label{dust}
 \end{figure}

\begin{figure}
\centering
\includegraphics[height=5.5cm,trim={2cm 1cm 1cm 1cm},clip,width=1.05\columnwidth]{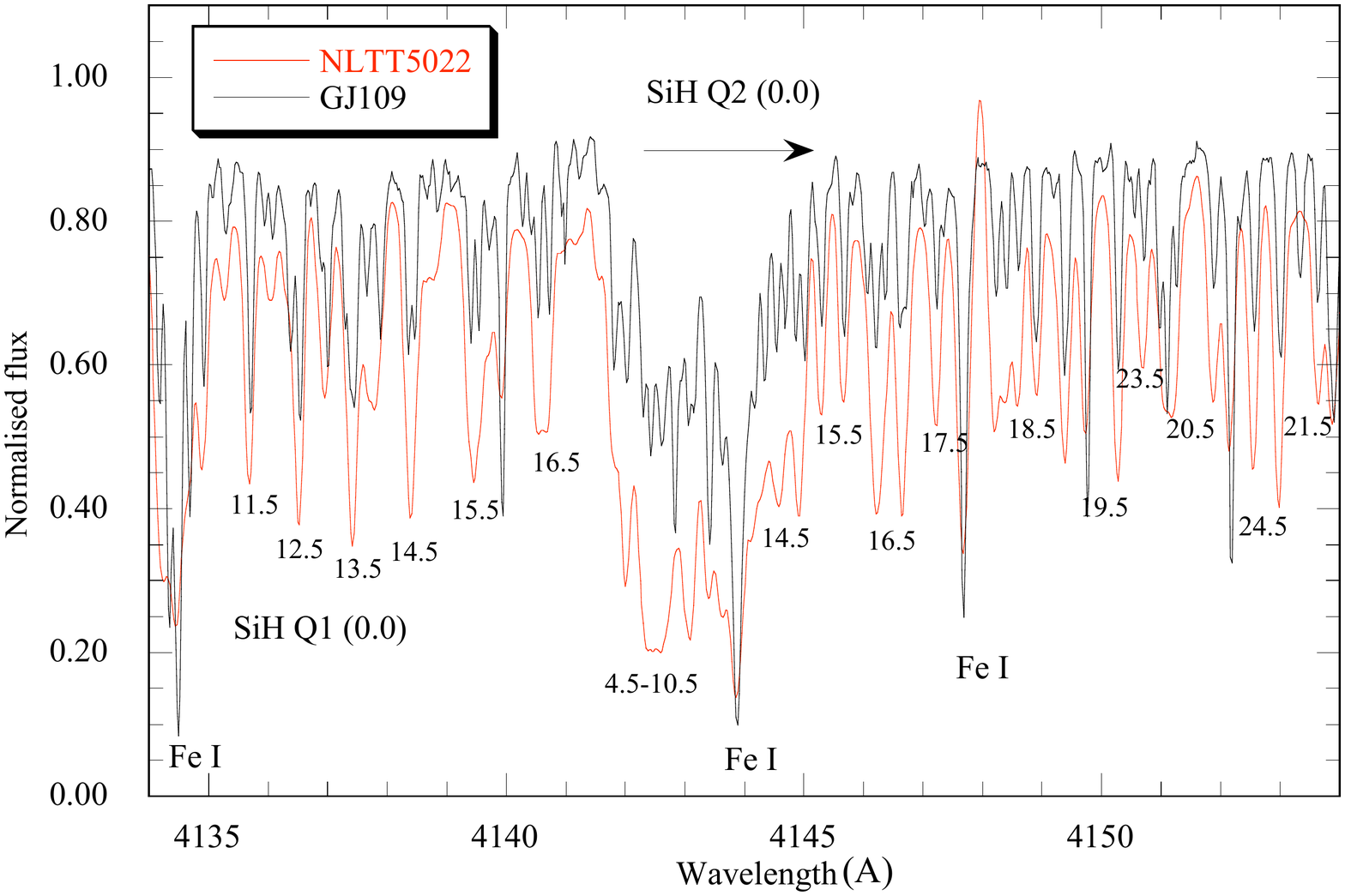}
\includegraphics[height=5.5cm,trim={2cm 1cm 0cm 1cm},clip,width=1.05\columnwidth]{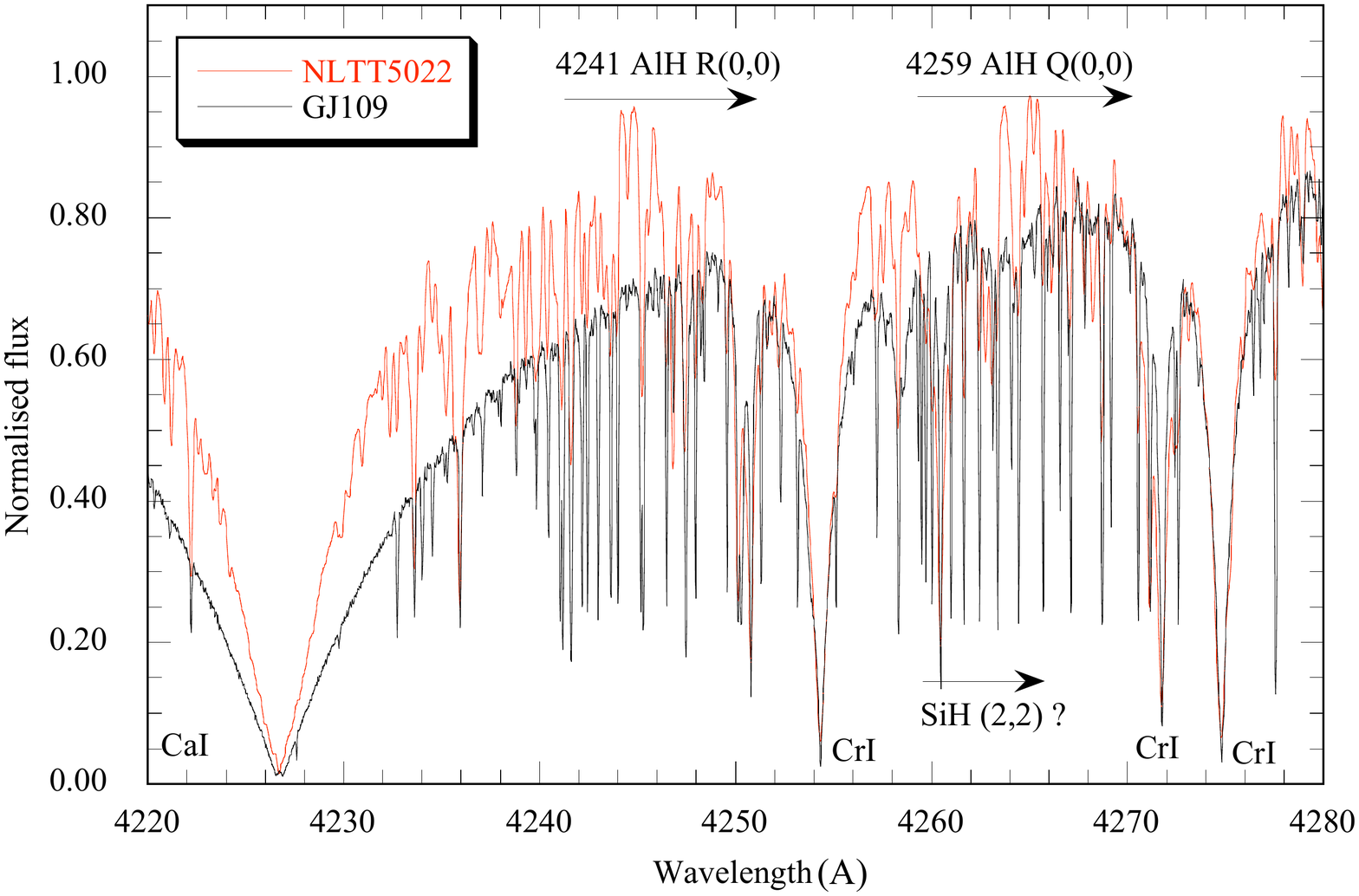}
\caption{The upper plot indicates SiH identifications for the Q1 and Q2 bands in the M3 dwarf GJ109 and an M0 subdwarf NLTT5022 with a metallicity of [Fe/H]$\sim$--2. The lower plot illustrates two bands of AlH and its impact showing its increasing relative importance toward lower temperatures.} 
\label{SiH}
\end{figure}

\begin{figure}
\centering
\includegraphics[height=6cm,trim={1.5cm 1cm 1cm 2cm},clip,angle=-90,width=1.05\columnwidth]{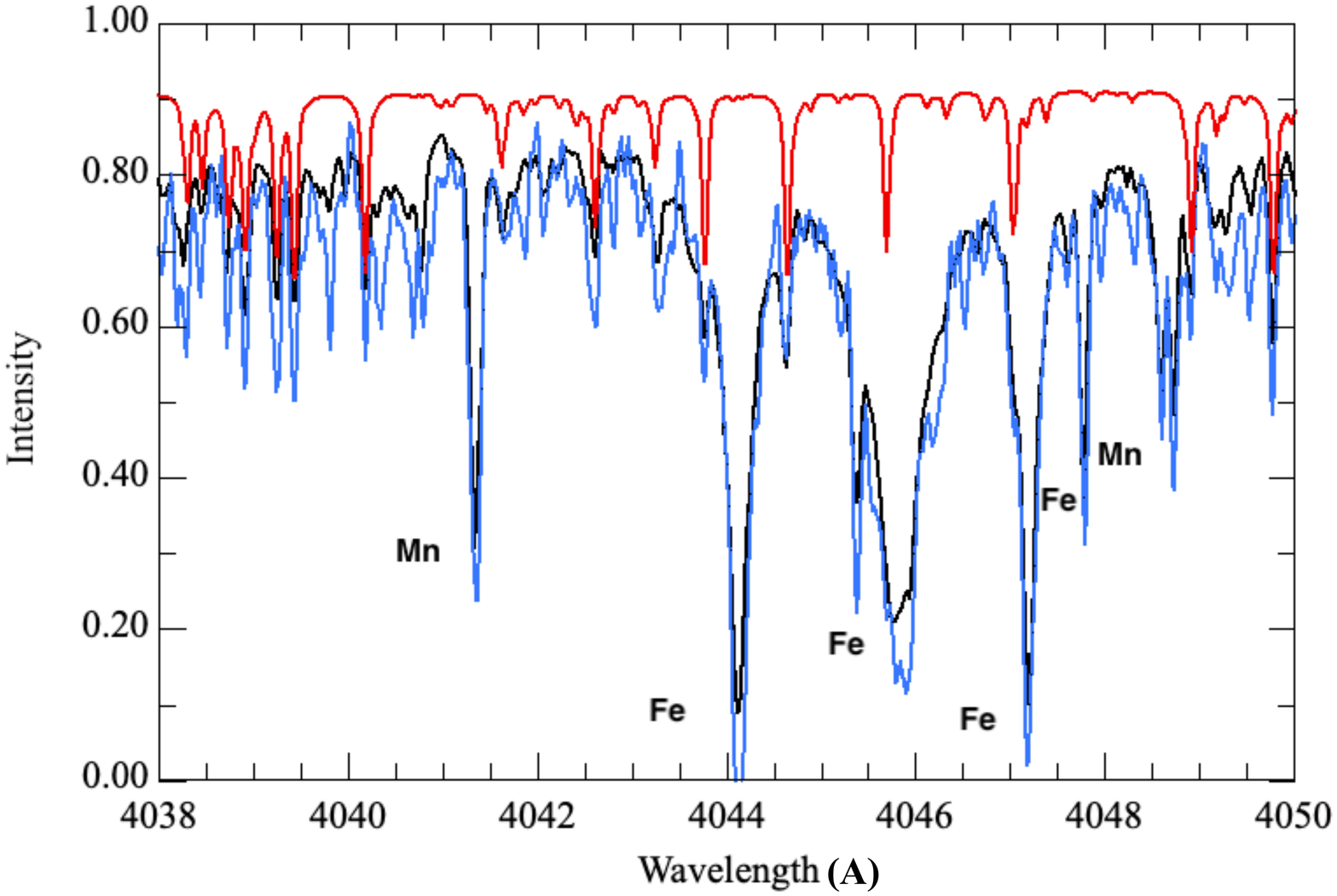}
\includegraphics[height=6cm,trim={1.5cm 1.5cm 7cm 17cm},clip,width=\columnwidth]{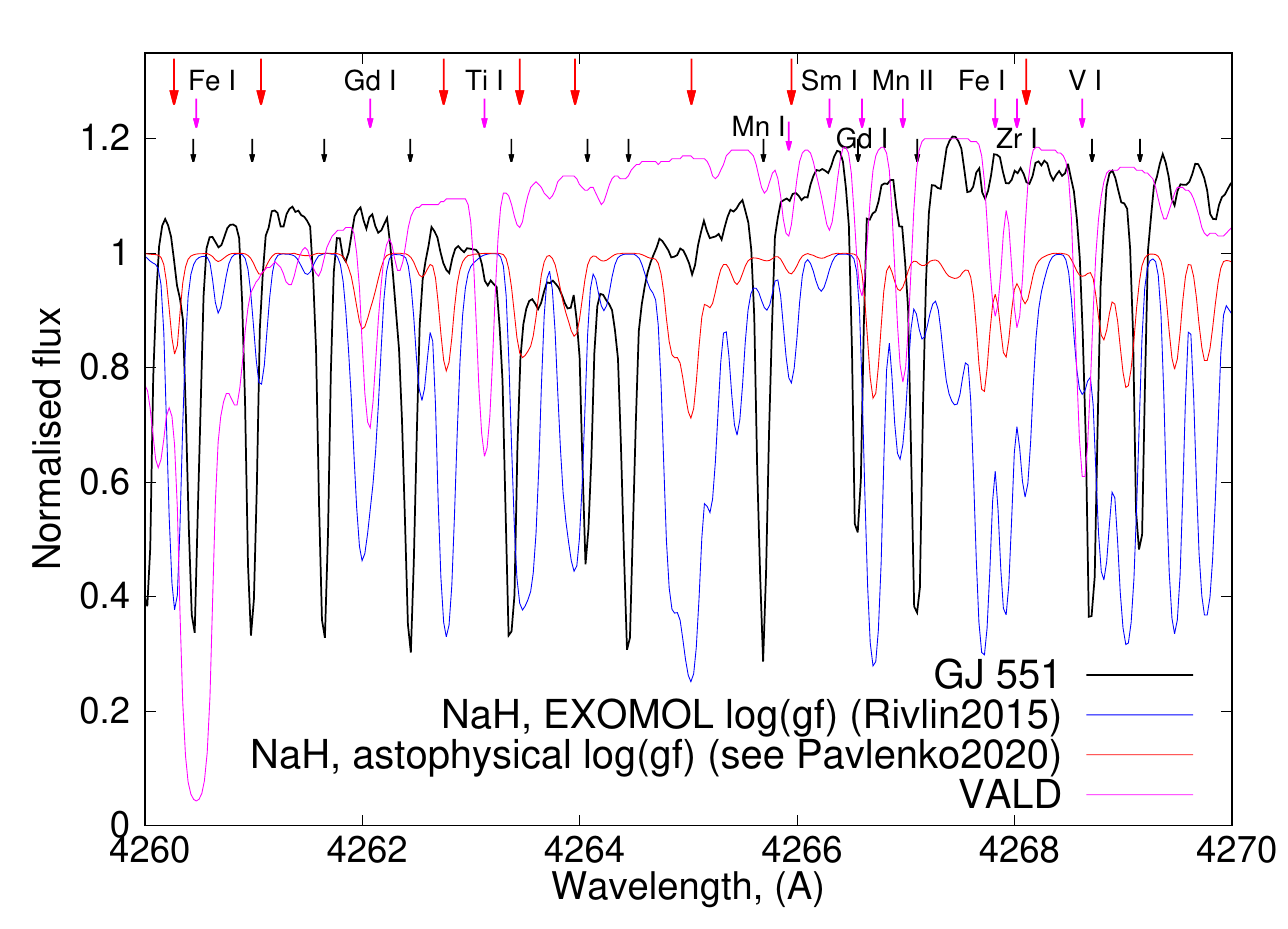}
\caption{The plot shows NaH and atomic transitions with observed spectra for GJ551 for different regions adjacent to the 4227\AA~Ca{~\sc{i}} line. The upper plot shows a bluer region where NaH can be identiifed, NaH is plotted in red above the observed HARPS (black) and UVES (blue) spectra. However, in the lower plot on the red side of Ca{~\sc{i}} line, we struggle to identify NaH and instead find AlH. The coloured lines in the plot are the result of the division of synthetic spectra with/without NaH for different oscillator strengths: in blue based on \citet{rivlin15}, in red based on  \citet{yp2020}. The magenta line represents atomic lines in this spectral region from VALD \citep{ryab15}. The red upper arrows in the plot denote the coincidence of synthetic and observed NaH lines based on oscillator strengths. The shorter magenta arrows denote atomic features and are labelled individually. Shorter black arrows in the plot identify AlH features based on Table A1 in \citet{pav22}.} 
\label{NaH}
\end{figure}

\subsection{Metallicity and dust}
There has been a long standing concern with models for M dwarfs that metallicities are not well constrained and that there might not be adequate treatment of dust opacity (e.g.,  \citet{Tsuji,jone97}). We examine spectral fits to two nearby M dwarfs, GJ299 known to be slightly metal poor (e.g., \citet{Jones1996}) and TWA33 known to be young \citep{Schneider}. 

In Fig. \ref{metal} and \ref{dust} we show two modelling solutions of metallicity and dust that can be used to improve the overall fit across the 4000--4500\AA~ region. We find the best fits of the observed spectra of TWA33 and GJ299 to synthetic spectra using a least-squared  minimisation procedure \citep{jones2002}. The upper plot of Fig. \ref{metal} indicates that lowering the metallicity to --1 provides a substantial improvement relative to a --0.5 model. However, the absorption in the 4300--4500\AA~region is not quite sufficient to reproduce the observations for GJ299 and the need to select an even lower metallicity for a slightly metal poor M dwarf in the solar neighbourhood seems inappropriate. In the lower plot of Fig. \ref{metal} the lower gravity of the baseline model preferred for TWA33 at first sight produces a rather good fit. This situation might be anticipated because in Fig.  \ref{met_grav} and \ref{metal_gravity} it could be seen that the blue absorption feature largely disappeared in the relatively low gravity TWA33. In this case the choice of such a low metallicity model seems particularly inappropriate when modelling a star from a nearby star forming region with no evidence for any peculiar metallicity. 
%We choose this low metallicity model as a means to try to fit the detailed spectral features. 
Scrutiny of the figure also suggests that the largely overlapping model lines in green and blue are stronger than the observed features. As with the upper plot further iterations of the chosen synthetic model can be made. 

In Fig. \ref{dust} we consider the same objects investigated in Fig. \ref{metal} using dust as a possible solution. \citet{pavl07} considered the lack of synthetic spectral fits as perhaps arising from the heating produced by atmospheric dust.  We use a layer of dust following the prescription of \citet{pavl07}. Fig. \ref{dust} indicates how dust  opacity can be adjusted to provide a superficially improved fit. However, the addition of dust opacity has an impact in other regions beyond our 4000--4500\AA~region of the interest. For example, the lower plot of Fig. \ref{dust} also shows how the addition of dust causes the observed TiO band at 4950\AA~to become too weak in the synthetic spectra. This is also the case for modelled features to the blue of 3850\AA~in TWA33. We note that \citet{Herczeg} use the strength of the Ca{~\sc{i}} feature as a function of spectral type to measure the optical veiling in young M dwarfs. However in the absence of identified optical veiling in our targets, our dust solutions are somewhat artificial particularly above 3000K where dust rapidly disappears and so provides a relatively improbable explanation at higher temperatures.

As with our experiments with metallicity in Fig. \ref{metal}, dust provides a ready means to get reasonable fits at lower temperatures across this spectral region. However, in both cases such fine tuning does not provide a satisfactory solution. In particular, it leads to the situation where spectral regions that were hitherto well reproduced are simultaneously worst fit. Moreover such opacities do not naturally produce a particular concentration in the 4000--4500\AA~region in the manner suggested by Figs \ref{bluedepression} to \ref{metal_gravity}.

\subsection{Other molecules}

M dwarfs contain a rich plethora of molecules some of which might have been relatively neglected in model atmosphere calculations. Of particular interest are molecules with low dissociation potentials such as SiH (3.34eV), AlH (2.99eV) and NaH (1.96eV) which might have transitions in the 4000--4500\AA~spectral range. 

SiH is considered by \citet{yurchenko18} and is seen in the Sun and in K giants. Here we identify SiH in M dwarfs though we find that it is most strongly present in higher temperature M subdwarfs, e.g., upper plot of Fig. \ref{SiH}. Although SiH transitions are relatively spread out, they do not extend across the 4000--4500\AA~region and the strongest lines are shortward of 4250\AA. 

Similar to SiH, AlH transitions are relatively localised and mostly occur between 4065--4090 and 4240--4280\AA~\citet{bessell2011,pav22}. In the lower plot of Fig. \ref{SiH} two bands of absorption lines can be seen that become considerably stronger towards lower temperature and higher metallicity. Although AlH is an important spectral diagnostic in this spectral region, its transitions do not extend blueward of 4065\AA~where the dissociation limit can be robustly identified and it has too few transitions to provide a continuum opacity.

NaH is an abundant species at cool temperatures and the A-X band mainly covers the region between 3500 and 4600\AA. Unlike the relatively restricted distribution of AlH and SiH transitions, the NaH molecule presents many closely spaced lines spread out across a wider wavelength region.  We can identify a number of NaH transitions in the spectral region particularly between 3500 and 4600\AA. Fig. \ref{NaH} gives an example of the character of NaH absorption lines in red based on \citet{rivlin15}. Using HARPS and UVES spectra for GJ551 in the lower plot, we can identify the strongest NaH features as appearing in the observed spectra and thus NaH is a relevant opacity source.  The theoretically strongest NaH transitions can be found in the spectra though these are predominantly around 4000\AA~where we only identify the strongest ones; others are predicted to be several orders of magnitude weaker. In the lower plot of Fig. \ref{NaH} we plot a somewhat redder region as an example that AlH acan more easily be identified. Since we struggle to reliably identify NaH transitions beyond 4350\AA~ they do not provide a simple solution to explain the shape of the 4000--4500\AA~region. 

%No trace in the spectrum of Prox Cen of any NaH lines, although the synthetic spectrum generated from my Stancil and Yanbo NaH linelist indicates hundreds of lines should be evident. (b) Identify AlH lines in M dwarfs at M3 and in Prox Cen M5.5.Normalised spectrum as on ther panels, the blue shows normalised to the local continuum spectrum. Picture around CaHK attached for Proxima. Section 3.1.5 of https://arxiv.org/pdf/1706.04678.pdf about absorption in the 3800–4200Å region Exomol wavelengths look promisingly for spectroscopic comparisons, e.g., Fig. 5 of https://arxiv.org/pdf/1506.00174.pdf

\begin{figure}
\centering
\includegraphics[width=1.05\columnwidth]{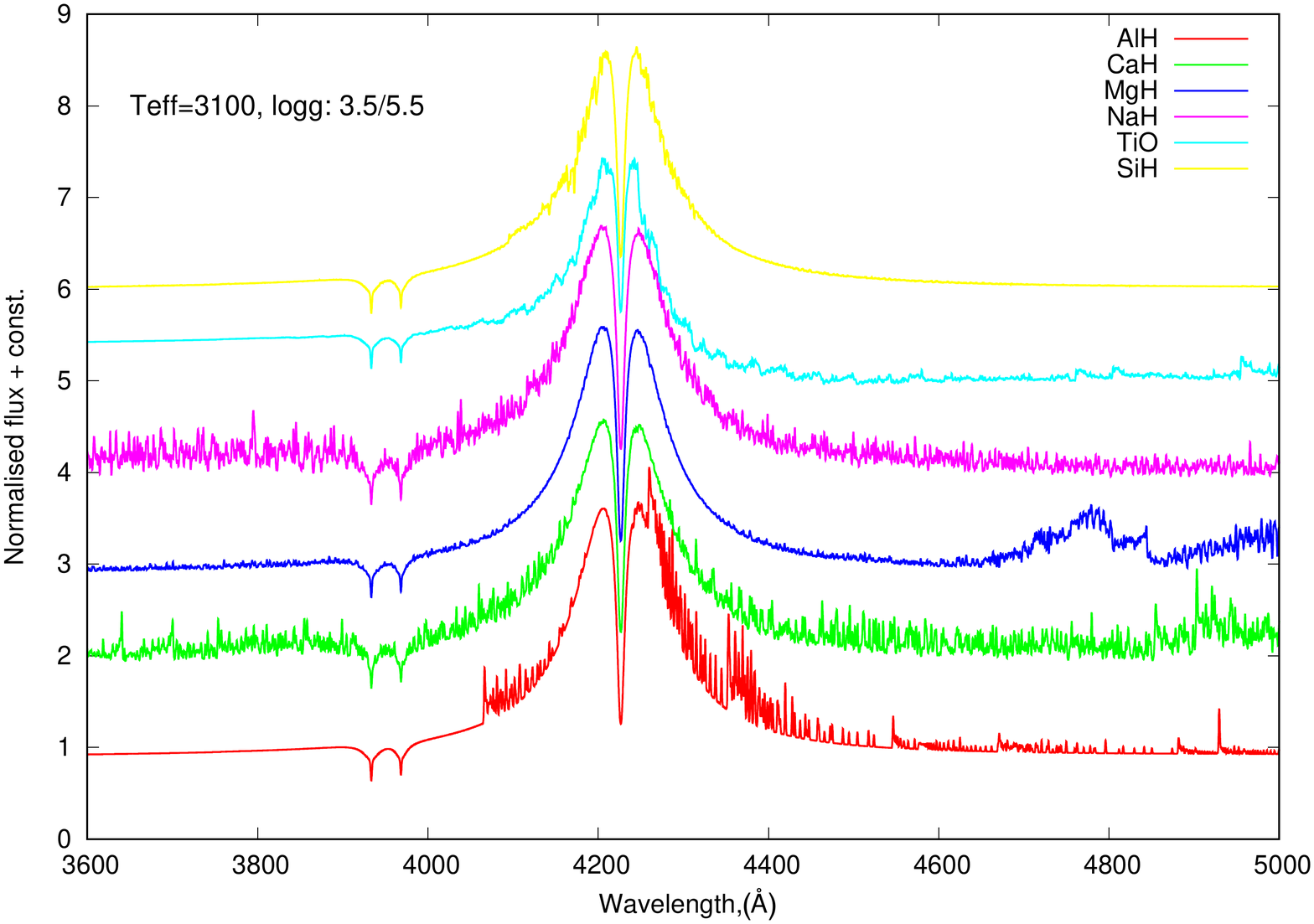}
\caption{The ratio of two synthetic spectra with 3100K at two different gravities, log $g$ = 3.5 and 5.5 is shown. This illustrates the gravity dependence of the shape of Ca{~\textsc{i}} line along with other molecules in this spectral region.}
\label{vdw}
\end{figure}

In Fig.\ref{vdw} we investigate the gravity dependence of different molecular features across the 4000--4500\AA~region. The relatively modest changes in Fig.\ref{vdw} suggest that the gravity sensitivity of CaH, MgH, NaH, TiO SiH and AlH is small with only AlH offering significant identifiable transitions increasing to lower gravities. For the case of AlH, the lower plot of Fig. \ref{SiH} and \citet{pav22} indicate that it is well enough understood and is not a continuum opacity across the 4000--4500\AA~region. 

\begin{figure}
\centering
\includegraphics[width=\columnwidth]{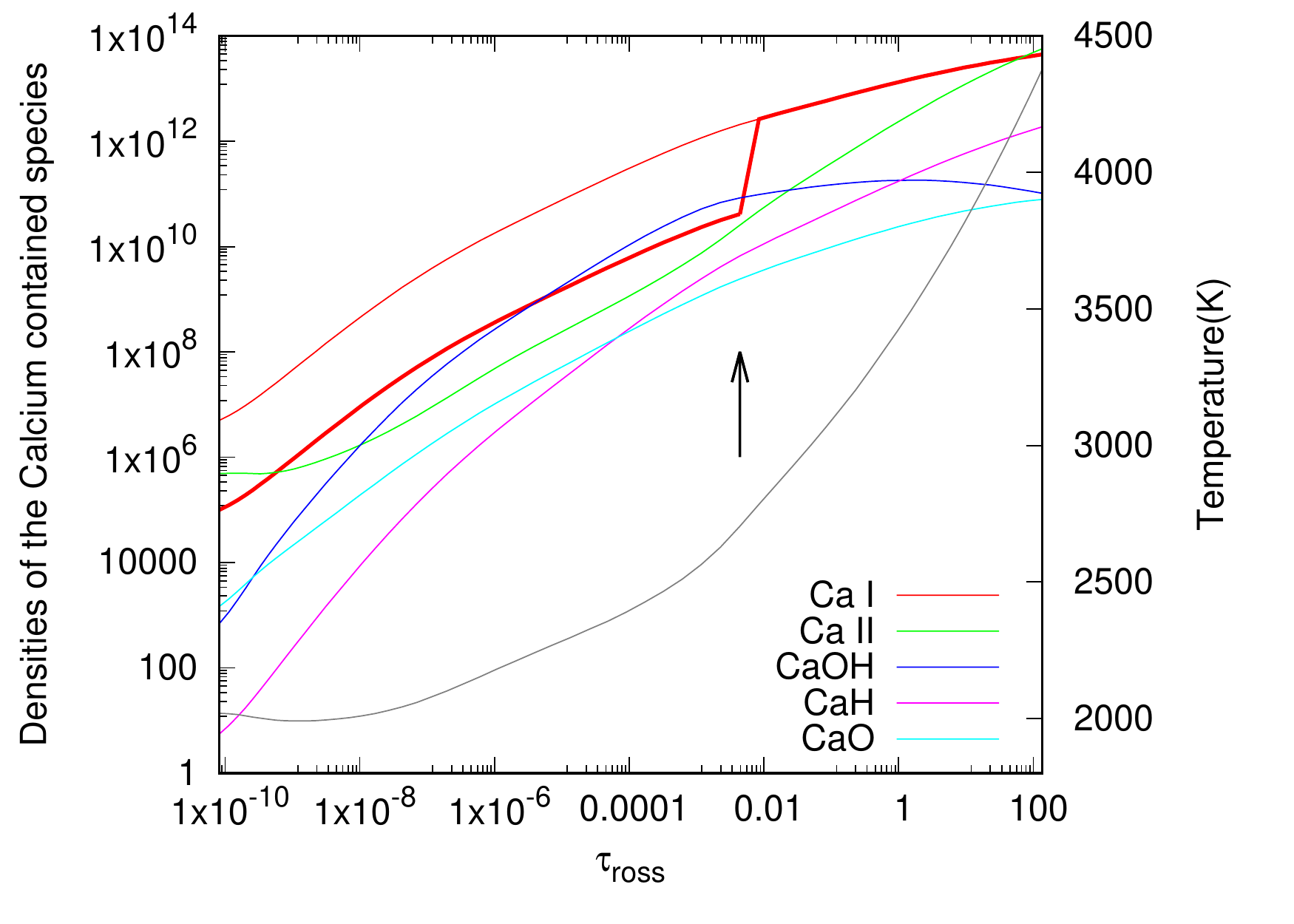}
\includegraphics[height=4cm,trim={1.7cm 1cm 2.15cm 12cm},clip,width=0.49\columnwidth]{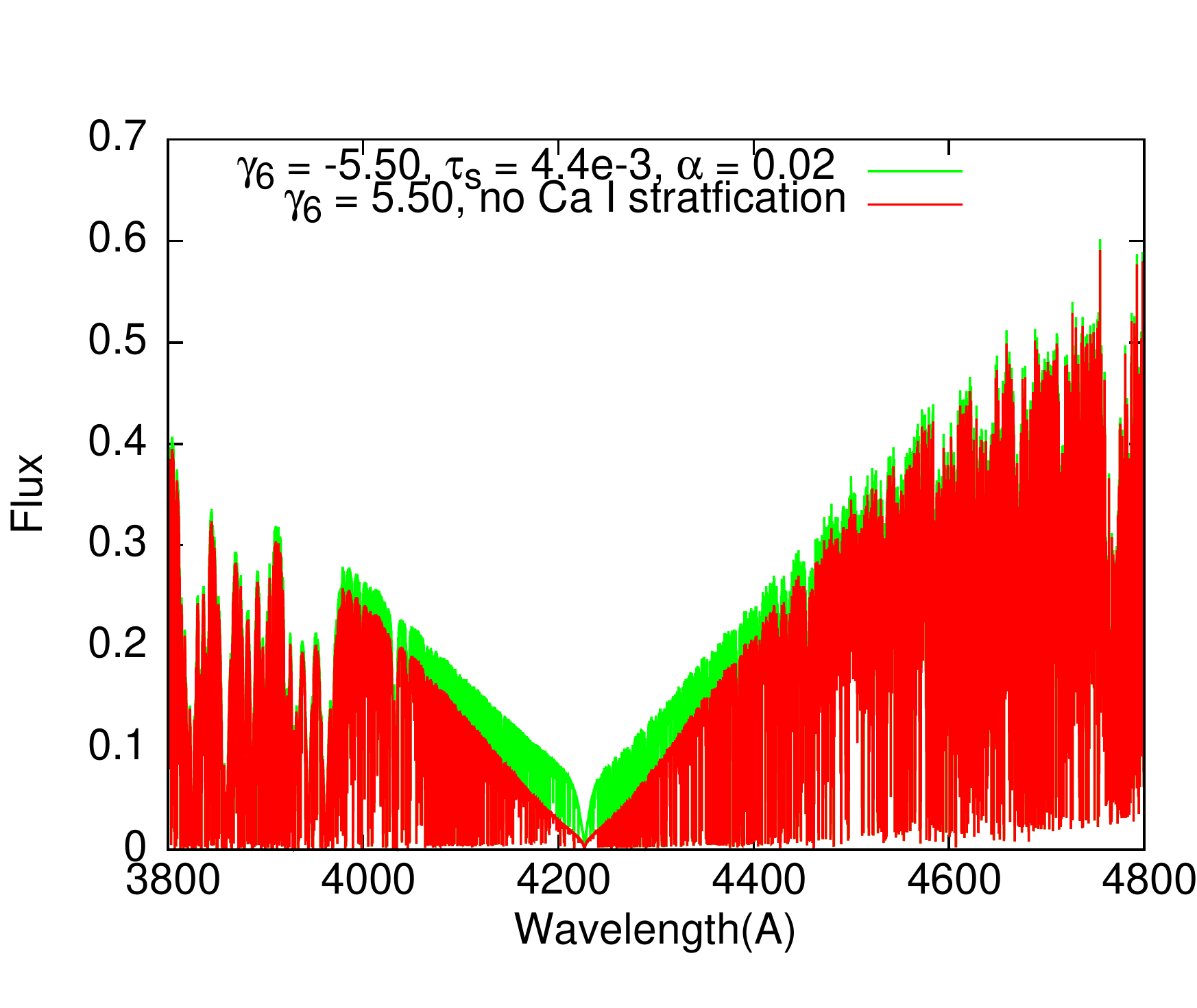}
\includegraphics[height=4cm,trim={1.7cm 1cm 2.15cm 12cm},clip,width=0.49\columnwidth]{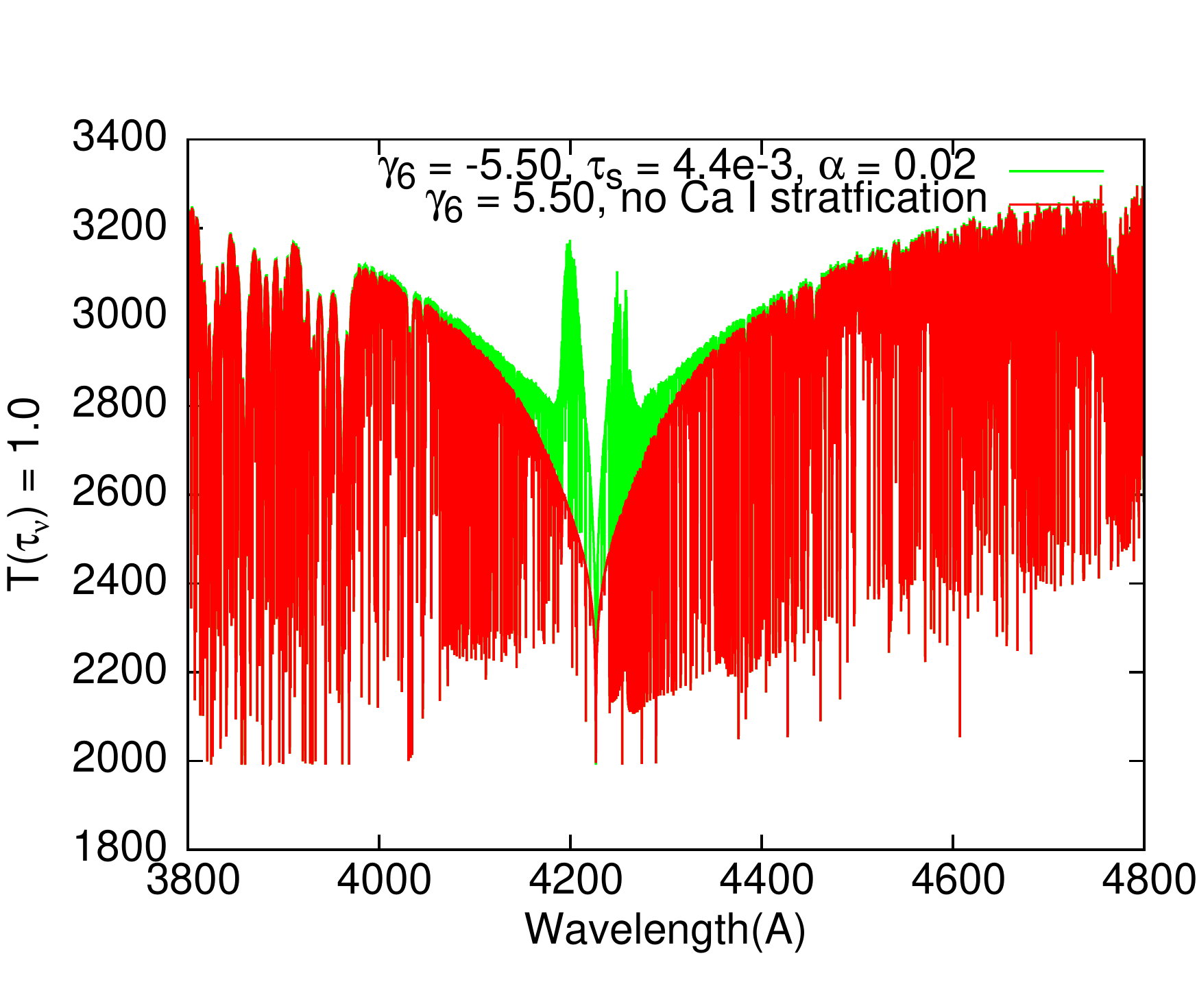}
\caption{The upper plot shows a model atmosphere structure taken from BT-Settl grid. Red thin and thick 
lines show the molecular densities of Ca{~\textsc{i}} in the ``normal and stratified'' model atmospheres on the left-hand y axis. Other colours show the densities of other Ca species. 
An arrow shows the depth location  ($\tau_{\rm{ross}}$ = 3.4e-4) where stratiftication is on, i.e.
above this point molecular densities of Ca{~\textsc{i}} are reduced by factor $\alpha$ = 0.02. The grey line uses the right-hand y axis and shows temperature. The lower left-hand plot shows the
impact of different stratification parameters on the spectrum around the 4227\AA~Ca{~\sc{i}} line. The lower right-hand plot depicts the temperature at optical depth $\tau_\nu=1$.}
\label{strat}
\end{figure}

In principle, incorrect dissociation constants would affect the computed number densities of molecular lines. We ran a few experiments looking at the impacts on lines from varying D$_{\rm0}$ for NaH using the substantially different available values in the literature:  experimental 1.876eV \citep{gurvits1982thermodynamical} and ab initio 2.036eV \citep{tsuji73}  and 1.958eV \citep{le2013dpf}. No interesting differences in our results were obvious and it should be noted that here is no particular evidence that such a spread in literature values is appropriate based on \citep{barklem} who find the experimental value should be adjusted to 1.886eV using updated data for the ground state and an improved partition function. The most significant D$_{\rm0}$ change in the dissociation constants of interest indicated by \citep{barklem} is for CaH changing from 1.7  to 2.28 eV. This serves to make CaH features stronger and otherwise the purple line labelling CaH in the upper plot of Fig. \ref{strat} would be further to the bottom right and CaH less important. Since we use the \citep{barklem} values this is already accounted for in our models though even substantial further changes in the D$_{\rm0}$ for CaH would not seem enough to alter its relative lack of importance (e.g., Fig. \ref{vdw}).

\begin{figure}
\centering
\includegraphics[height=7cm,trim={2cm 3cm 5cm 18cm},clip,width=1.2\columnwidth]{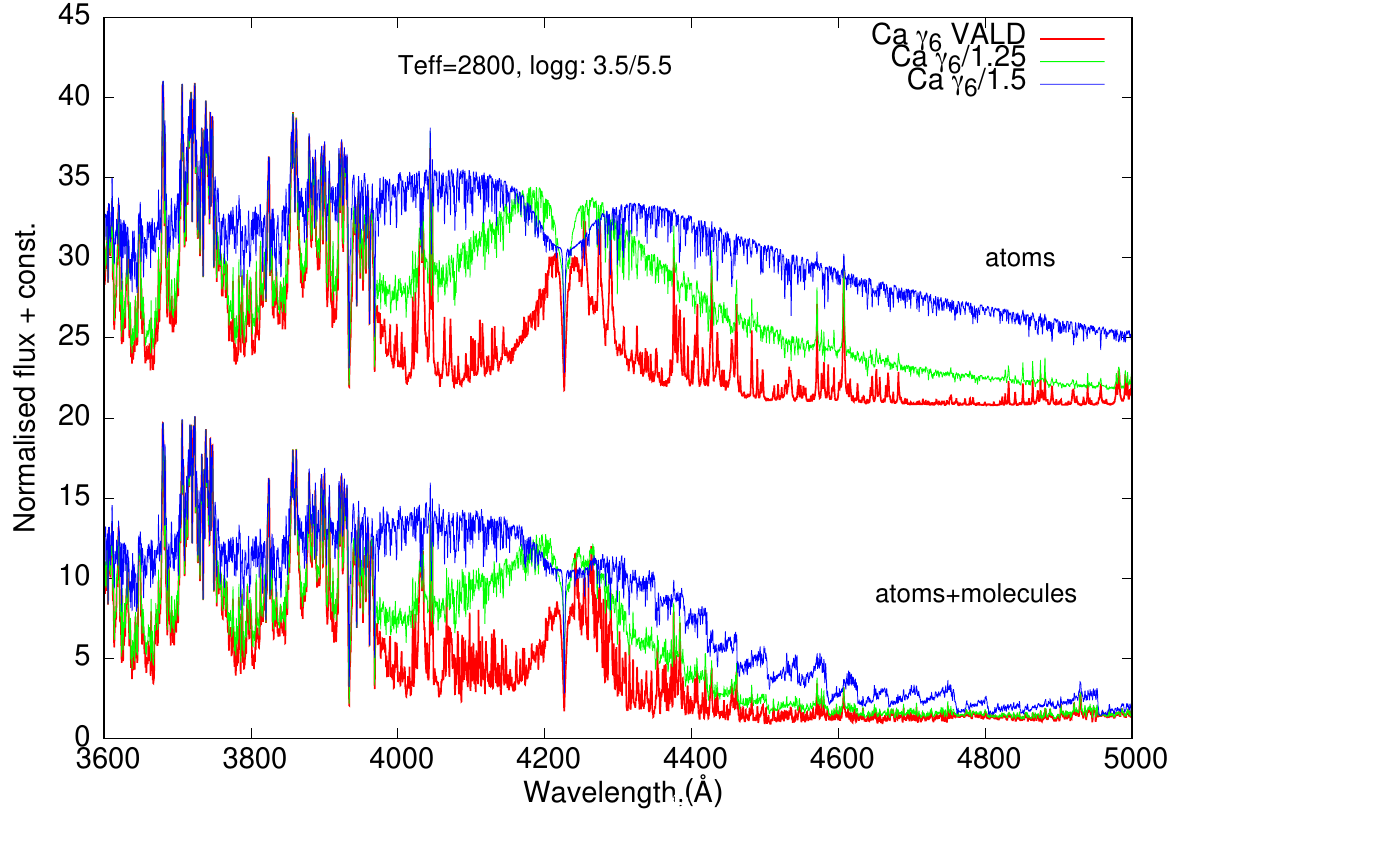}
\caption{The upper plot shows the ratio of spectral energy distributions for log $g$ = 3.5 to 5.5. It also shows the impact of different damping constants of Ca{~\textsc{i}} line for model atmospheres 2800/3.5/0.0 and 
2800/5.5/0.0.}
\label{strat0}
\end{figure}

\begin{figure}
\centering
\includegraphics[width=\columnwidth]{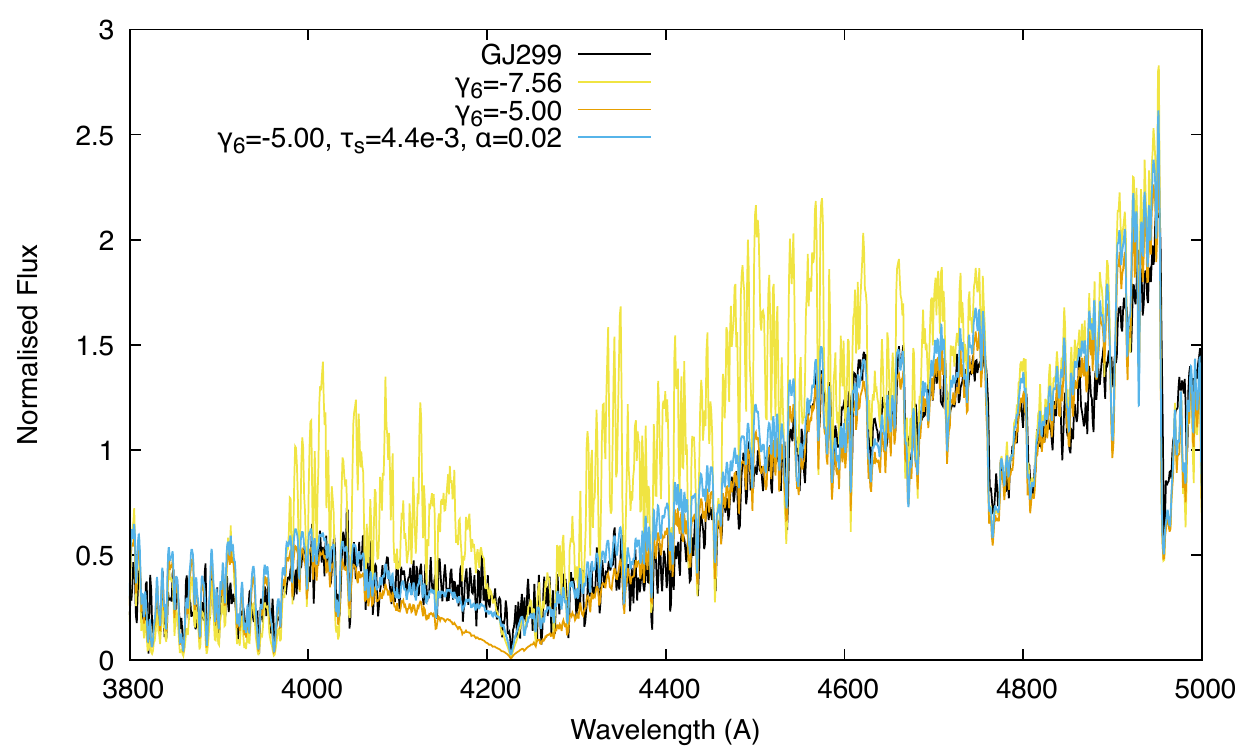}
\caption{Fit of our synthetic spectra to the observed SED of GJ299
across spectral range of 3800-5000 \AA. A good solution is found for
\ts = 3.4e-3, $\alpha$ = 0.02, \gg = -5.50.}
\label{vdw_strat}
\end{figure}

\subsection{Ca{~\sc{i}} resonance line parameters and atmospheric stratification}

The Ca{~\sc{i}} line at 4227\AA~along with the Ca{~\sc{ii}}  lines at 3934 and 3968\AA~are well known and the later are utilised as the primary spectral diagnostic of activity in the Sun and other stars since \citet{preston}. The behaviour of the ionised Ca{~\sc{ii}} lines appears to be in line with expectations and is not examined further here. However, the neutral line does not appear to have been so widely used particularly in such cool stars and is susceptible to possible issues with its line strength and broadening not considered by standard model atmospheres.

Given the interest to produce the broad absorption feature suggested by Fig. \ref{metal_gravity} we make changes to the damping constant for the Ca{~\sc{i}} resonance line with the observational constraint that the absorption feature only appears in 4000--4500\AA~region at higher gravity, and is pronounced at lower metallicity. In Fig. \ref{strat0} we consider solar metallicity models of 2800 K with log $g$=3.5 and 5.5 and make theoretical plots analogous to the empirical red lines in the Fig. \ref{metal_gravity} plots. We investigate the effect of modifying the VALD \citep{ryab15} damping constant of log$_{10}$ $\gamma_6$= --7.57 for the Ca{~\sc{i}} line. We show variations of log$_{10}$ $\gamma_6$=VALD, VALD/1.25 and VALD/1.5 in Fig. \ref{strat0}, where we adopt a  shorthand syntax $\gamma_{\rm 6}$=VALD/1.25 to represent $\gamma_{\rm 6}$ = log$_{\rm 10}$ ($\gamma_{\rm 6}$ from VALD) / 1.25. It can be seen that these modifications from the standard VALD value do have a dramatic impact on the Ca{~\sc{i}} line at 4227\AA~but that these changes are most dramatic in the core region. So while such changes do produce the required significant line wing changes that are observed blueward of 4100\AA~or redward of 4400\AA~the side effect is a very large change in the shape of line core region (green and blue synthetic spectra) which is not observed.

In general terms the formation of the strong wings of the resonance lines of metals in cool dense atmospheres is well-studied in the framework of the semi-stationary theory of line broadening (see \citet{alla82, alla03},\citet{burr03}, \citet{pavl07}). In order to provide a theoretical investigation  of the collisional effects in the blue wing of the Ca{~\sc{i}}  line perturbed by H, He and H$_{\rm 2}$, unified line profile  calculations and  molecular data are both required. The first step will be the determination of 
accurate  potential energies and electronic transition dipole moments of Ca{~\sc{i}} perturbed by H$_2$ (Thierry Leininger, private communication). Lacking a full treatment of broadening such as done for potassium by \citet{burr03}, we investigate an alternative solution by modifying the position of calcium in the photosphere. This approach is justified in the sense that there is considerable model uncertainty in the temperature versus pressure profile in cool stellar atmospheres and that the empirical constraints on the structure largely come from a good match between synthetic and observed spectra. 

Stratification of atmospheres is another approach that has been employed elsewhere for example in the case of the Sun \citep{solanki02}. Here we have the situation where the Ca{~\sc{i}} resonance line occurs in a region of relatively low opacity and so we are interested in the extent to which its line shape is dependent on exactly where its line formation happens within the local radiation field. Namely, we consider that in the atmosphere of M dwarfs above the depth \ts molecular densities $N(i)$ are reduced in comparison with the equilibrium values ($N_{\rm equil}$) according to $N(i) = \alpha N_{\rm equil}$. The upward pointing arrow in the upper plot of Fig. \ref{strat} locates our arbitrary modification to the density of calcium containing species in an M dwarf model atmosphere.

The impact of the stratification on the flux and temperature of the photosphere with wavelength across the Ca{~\sc{i}} line can be seen in lower plots of Fig. \ref{strat}. Stratification can have the desired effect of producing an enhancement in the wings of Ca{~\sc{i}} over a considerable range but without unduly impacting the line core. From the perspective of the model atmospheres we are left with the parameters \ts and $\alpha$ which can be determined from comparison with observations. 
%While these parameters enable stratification to provide a perfectly reasonable fit this provides a rather contrived solution.

In Fig. \ref{vdw_strat} we compare different values of $\gamma_{\rm 6}$ and use \ts and $\alpha$ to modify the line broadening of the wings for the Ca{~\sc{i}} line to fit the M4.5 dwarf GJ299. The synthetic spectral lines with  $\gamma_6$ = --5.00 and --7.56 show the problem of simply modifying $\gamma_6$ where it is difficult to find a value which leaves the  Ca{~\sc{i}} line core intact and simultaneously provides appropriately broadened wings. It can also be seen how stratification modifies the extreme line wing broadening which is introduced by modification of the Van der Waals damping constant. Stratification can suggest a fit to the overall shape of the blue depression as well modifying the modelled strength of atomic features within this region but at the same time not impacting Ca{~\sc{ii}}  lines at 3934 and 3968\AA~and the developing system of TiO bands towards redder wavelengths. Further fine tuning of this stratification could improve the match between empirical and synthetic spectra and enable stratification to provide a perfectly reasonable albeit contrived solution. 

%We also check the impact of other opacities, particularly the hydrides discussed above on the Ca{~\sc{i}} line in order to check the extent to which these other opacities might combine together to play a role in shaping the observed blue depression. Following Fig. \ref{vdw}, we divide different gravity models by one another in order to maximise our sensitivity to see these different effects in Fig. \ref{vdw_other}. It can also be seen that NaH and OH do have some impact on the Ca{~\sc{i}} lines.

\begin{figure}
\centering
\includegraphics[height=6cm,trim={2.5cm 1cm 1cm 1cm},clip,width=1.1\columnwidth]{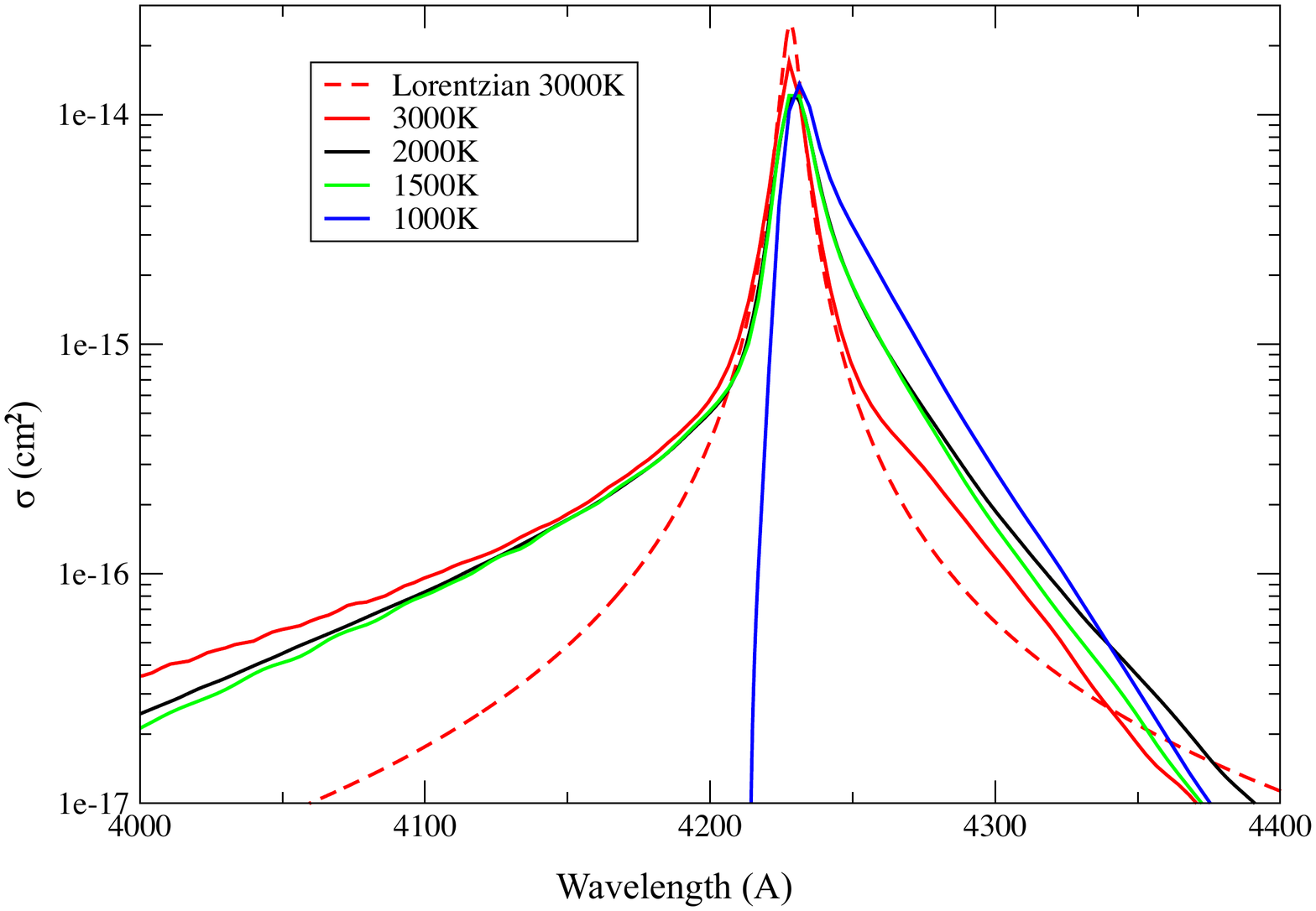}
\includegraphics[width=\columnwidth]{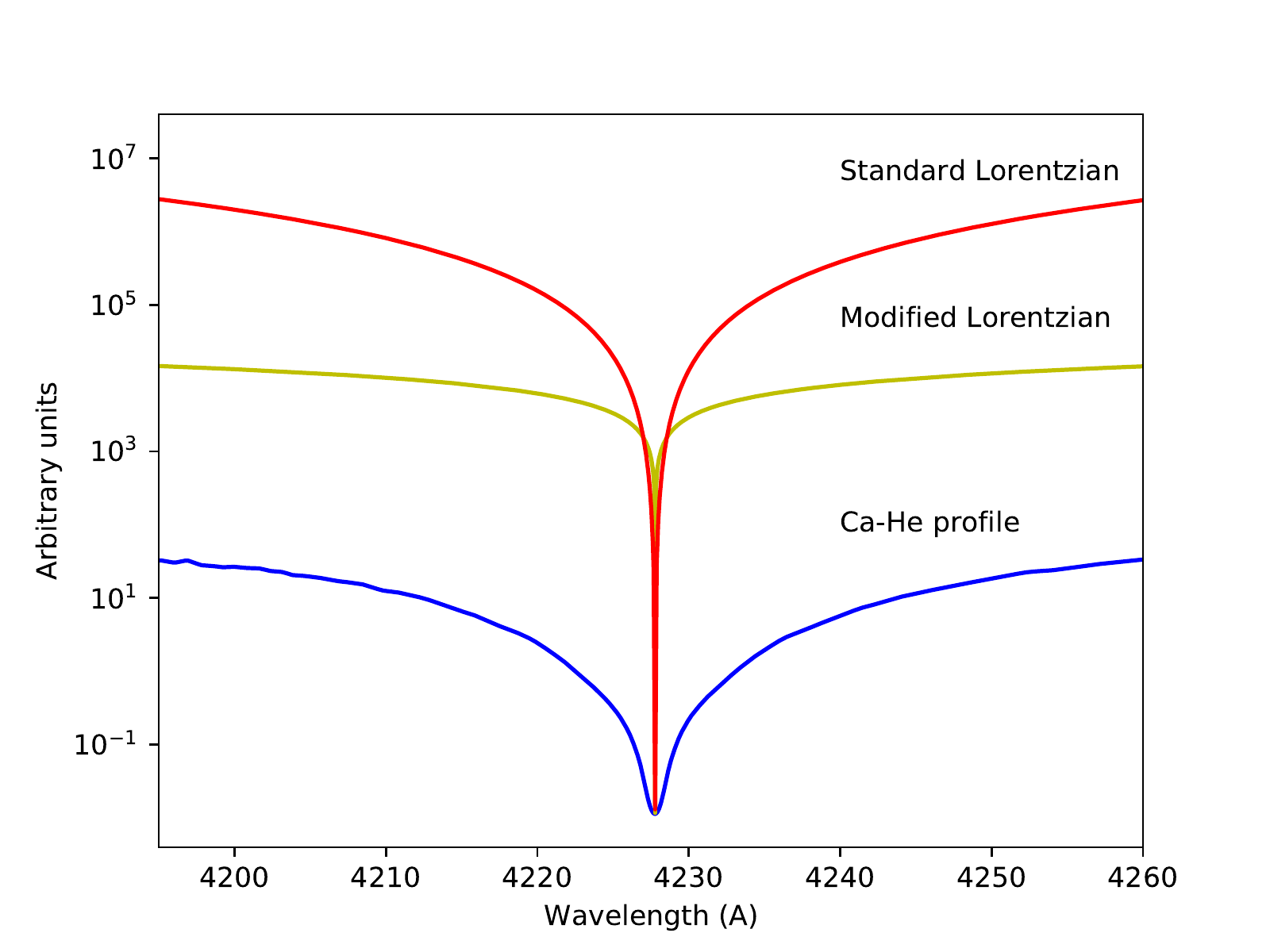}
\caption{The upper plot shows comparison between unified line profile calculations for $T$ = 3000, 2000, 1500 and 1000K based on broadening by Helium ($n_{\rm He}$ = 10$^{21}$cm$^{-3}$) extending the calculations of \citet{blouin} and based on \citet{nallard1999} in comparison with a 3000K Lorentzian line profile (dashed red). The lower plot shows a zoom in of the line core region. It is inverted to illustrate the wide range of possible line core and wing shapes introduced by different model assumptions. The Ca-He 4000K unified profile from \citet{blouin} based on broadening by Helium ($n_{\rm He}$ = 10$^{20}$cm$^{-3}$) is shown in blue along with the standard Lorentzian in red and the `modified' Lorentzian `fit' used for Fig. \ref{lineprofile} in yellow. }
\label{lineprofiles}
\end{figure}

\begin{figure}
\centering
\includegraphics[width=\columnwidth]{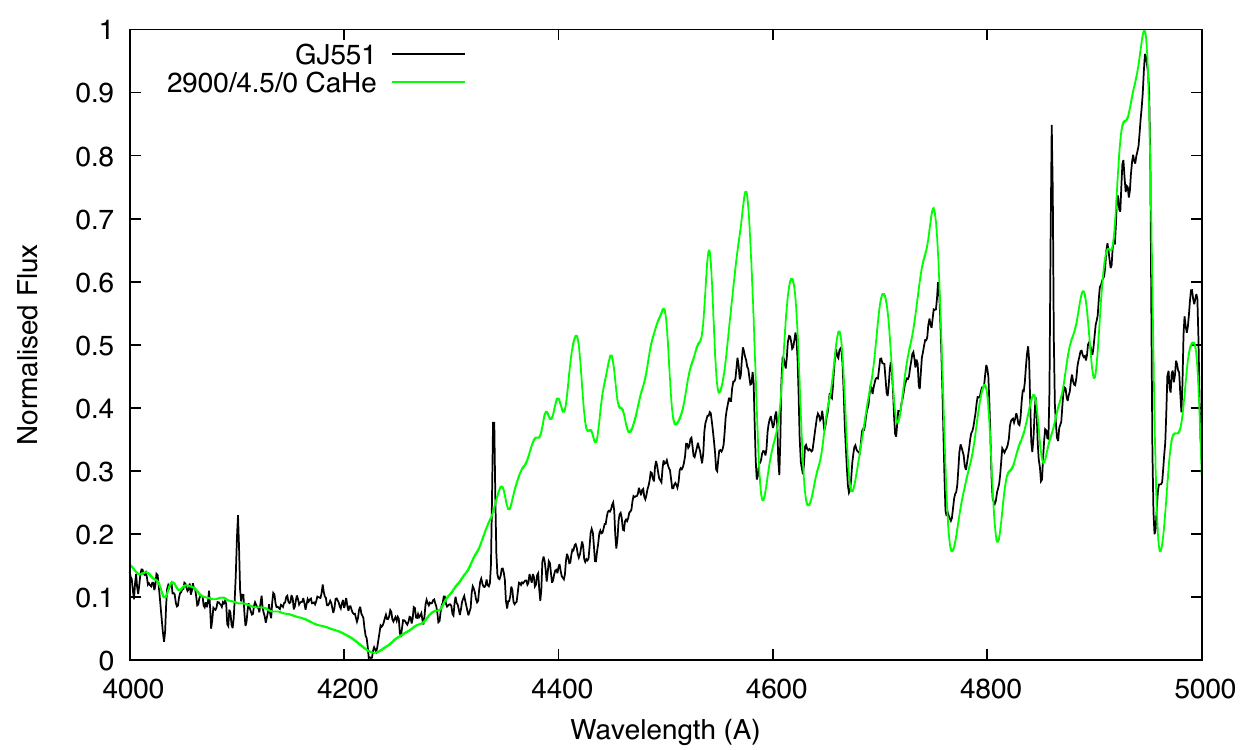}
\includegraphics[width=\columnwidth]{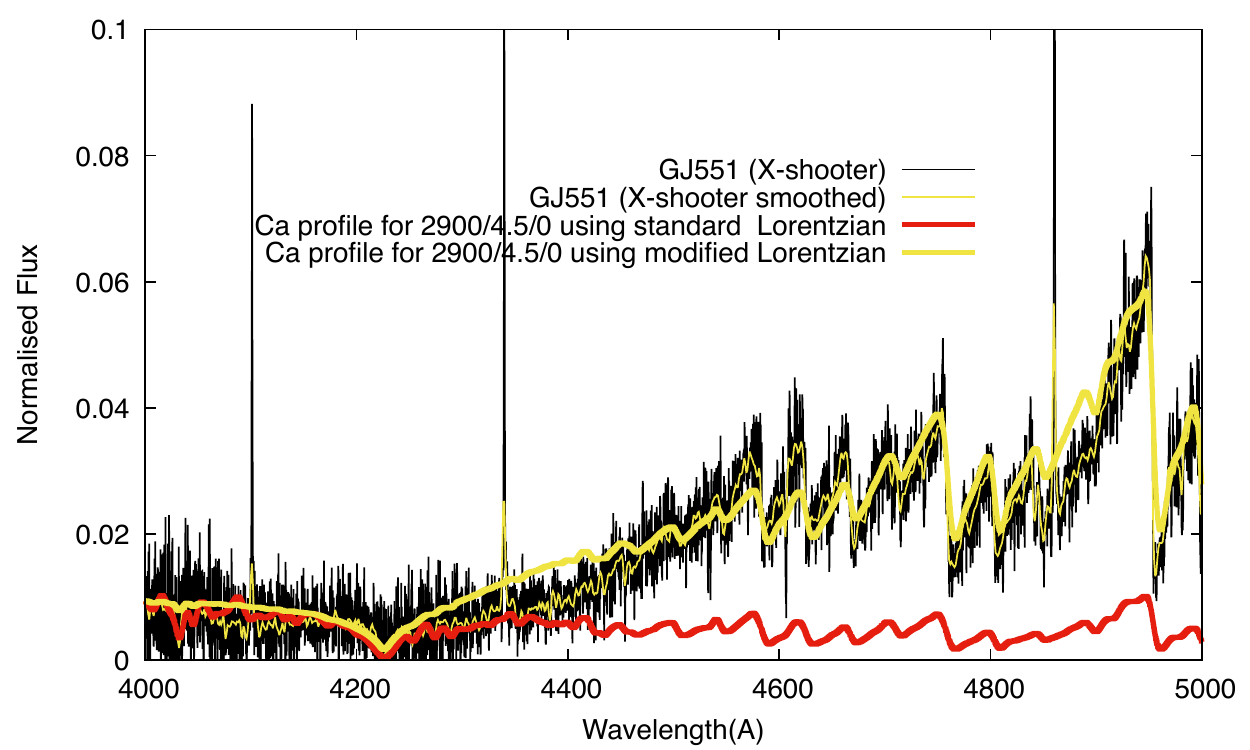}
\includegraphics[width=\columnwidth]{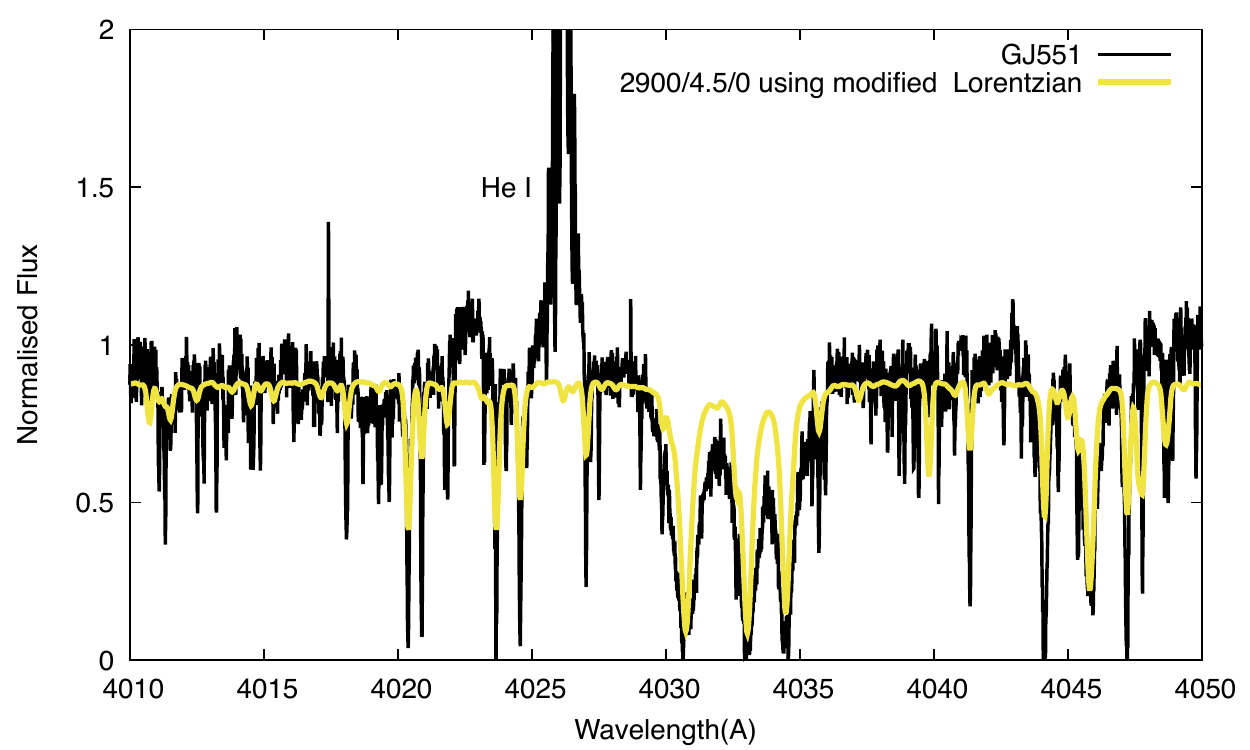}
\caption{The upper plot shows comparison between standard (green) Lorentzian line profiles along with the unified line profile calculation for T=4000K based on broadening by Helium ($n_{\rm He}$ = 10$^{21}$cm$^{-3}$) extending the calculations of \citet{blouin} and based on \citet{nallard1999}. The middle plot shows the `modified' Lorentzian `fit' as a thick yellow line to an X-shooter spectrum of GJ551 in black and binned thin yellow. The red line in the middle plot shows the standard Lorentzian leaves too little opacity to explain strength of atomic features and overall spectral energy distribution across 4000-4600\AA~region.The lower plot is made using the `modified' Lorentzian `fit' and corresponds to the same bluer spectral region as the lower plot of Fig. \ref{highres}.}
\label{lineprofile}
\end{figure}

\subsection{A `modified' Lorentzian solution}

The interactions between pairs of neutral atoms and the corresponding perturbations of atomic levels known as resonance broadening is classically a concern for Hydrogen and Helium atoms in hotter stars. In cooler stars reliable synthetic line profiles calculations sometimes need to consider line broadening from neutral hydrogen collisions. This is has been borne out in spectacular fashion particularly in the L and T dwarfs with the sodium and potassium resonance line at 5900\AA~and 7700\AA~which increase in strength from the M through the L dwarfs into the T dwarf regime. 

The work of \cite{blouin} indicates that line broadening can provide a suitable solution for the line shape of the Ca{~\sc{i}} in 4000-5000K DZ white dwarfs and the resonance profile line shape resembles the sought after line shape to explain the appearance of the 4000-4500\AA~region. The upper plot of Fig. \ref{lineprofiles} extends the calculations of \citet{blouin} to lower temperatures and illustrates the importance of temperature and the line shape of a broadened Ca{~\sc{i}} line in a pressure broadened environment with He. The Ca-He profile is not symmetric and this lack of symmetry is something which is found in observations of white dwarfs and well matched when using high-quality Ca-He potentials \citep{blouin}. Similar calculations are not yet available for Ca-H and Ca-H$_2$ broadening which would be appropriate for M dwarfs. However, the situation for Rb/Cs-He/H$_2$ broadening may be seen as promising \citep{na06} and might be analogous to the study of Na and K in \citet{nallard2019}. Initial calculations by \citet{kielkopf} suggest that the unified profiles for Na-He and Na-H${_2}$ are not significantly different and the quasi-molecular line satellites become closer to the main line when using more accurate potential data. As an interim alternative to an appropriately broadened profile we investigate a modified Van der Waals broadening as a plausible solution. 

In a crude attempt to adequately model the calcium line broadening we empirically modify the Lorentzian used to represent the calcium line in order to fit the measured continuum. The profile of our modified Lorentz is determined by the formula
\begin{equation}
\label{eq:kasttype} 
(-p_p*x^2)+p_s*a_{\rm damping}/{x^{p_x}}
\end{equation}
where ${a_{\rm damping}}$ = ($\gamma_{\rm R} + \gamma_{\rm vdW}+\gamma_{\rm S})/\delta\nu_{\rm D}$, and
$\gamma_{\rm R}, \gamma_{\rm vdW}, \gamma_{\rm S}$ are damping parameters due to the natural, van-der Waals and Stark broadening and 
$x= \delta\lambda/\delta\lambda_{\rm{Doppler}}$. In the lower plot of Fig. \ref{lineprofiles} we invert the plot and zoom in on the line core to illustrate schematically how the different the alternative line profiles are to the standard Lorentzian plotted in red. 
%For comparison, the blue line in the lower plot of Fig. \ref{lineprofiles} shows the Ca-He unified profile. 
%As a crude way of investigating the importance of line broadening, we adjust the standard Voigt profile of the Ca{~\sc{i}} line at 4227\AA~using the damping constant.

From the conventional approximation for the Voigt profile, $p_p, p_s, p_x$= 1, 1, 2, (e.g., \citet{Gibson}), we adjusted the parameters of our `modified' Lorentzian by eye in order to fit the observed spectral energy distribution. Although in principle, a Lorentzian allows for control of line wings and line core in practise this required some iterations to obtain the reasonable `fit' shown in in the middle plot of Fig. \ref{lineprofile}. For this we use $p_p, p_s, p_x$ = 0.0001, 0.001, 0.5 to give an approximate fit to the smoothed spectrum taken with the ESO X-shooter instrument (programme 092.D-0300 employed by \citet{pavl17}). This choice also provides a similarly improved fit to other regions, e.g., in the lower plot of Fig. \ref{lineprofile}. This choice of this spectrum is driven by the desire to ensure that the flux calibration of the empirical spectrum is robust. X-shooter is a well established instrument for determining the spectral energy distributions of objects (e.g., \citep{xshoot}).

Although our `modified' Lorentzian solution provides only an approximate solution it gives an illustration of how a strong line may be modelled with the Lorentzian dominated wings of a standard Voigt profile.  Our modified synthetic spectrum plotted in yellow in Fig. \ref{lineprofile} enables a much more realistic fit. It is notable that the red line in the middle plot of Fig. \ref{lineprofile} represents the situation without modification to the models. Without the very significant increase in the opacity in the 4000-4500\AA~region, models predict that the M dwarf flux does not increase significantly across this region and remains relatively flat even by the start of the TiO bands. A similar result is apparent for GJ551 in fig. 2 of \citet{bessell2011} for synthetic spectra from MARCS. In reality all M dwarfs do exhibit a rise in the flux across this region to the TiO bands. 

\section{Discussion}
While this paper is concerned with extensive spectral synthesis, we do not update existing literature values for effective temperature, metallicity and gravity. Gravity remains a difficult parameter to fit precisely from the spectra. For the moment, we consider that it is as reliable to use the isochrone gravities for M stars and iterate with the temperature and metallicity. Though the problem of the paper is precisely that we do not have a good spectral model. A key purpose of this paper is to highlight the importance of line broadening calculations specifically for Calcium. Such calculations have the potential to reduce the number of free parameters needed to provide reliable parameters from spectral synthesis.

We have presented a range of hypotheses to understand the potential reasons for the blue depression in the energy distribution of M dwarfs across the 4000-4500\AA~region though we note that all plausible modelling issues have not been exhausted. For example, we do not consider non-thermal equilibrium effects which can alter the overall ionisation balance in late-type stars particularly toward those with lower metallicities, e.g., \citet{nlte}. Though as with most of the presented alternative explanations, non-thermal equilibrium is likely to only provide a partial solution to the blue depression and at least for solar type metallicities would primarily impact the line core, e.g., \citet{nlte1}. We find that the simplest solution is a proper treatment for the pressure broadening of Ca{~\sc{i}} 4227\AA.

In recent years there have been a wide variety of systematic efforts to determine opacities appropriate for cool objects. These opacities have provided a ready explanation for most of the unknown features in cool stars. Here we focussed on a blue depression in M dwarfs in the blue optical. We concur with \citet{lindblad1935,lindblad1935nat} that the best explanation for the blue depression is the broadening of the calcium resonance line. This feature appears as the strongest and most sensitive feature in the optical spectral of M dwarfs. While the importance of careful line treatment for strong lines has been implemented for other resonance lines our work emphasises the need for a proper treatment of strong atomic line parameters particularly for Calcium as well as the identification of other relevant molecular species.

It is already clear from studies such as \citet{lodieu} that a number of well-studied atomic lines do not behave quite as expected. Such identifications can easily be attributed to non-solar abundance patterns or more exotic explanations but as the quality and quantity of empirical spectra improve it is vital to resolve issues associated with line parameters.  Multi-object spectrographs provide the tools to be able to take large numbers of M dwarf spectra and to derive their detailed characteristics (e.g. \citet{ding}). These can be important for a range of different reasons. For example, the M dwarfs may be intrinsically interesting themselves, due to their companions, or as a statistical chronometer to map out the history and evolution of our Galaxy. It is likely that existing analyses will have biases introduced by using grids of synthetic spectra that lack a systematic  consideration of strong line broadening. This is particular important where astrophysical parameters are increasingly determined by algorithms blind to underlying opacity issues.

The identified optical blue depression shows considerable sensitivity within the observed spectra of similar spectral types and so provides the potential as useful diagnostic spectral region for M dwarfs. Since calcium broadening is the single most sensitive feature in M dwarfs it might be used to help resolve the degeneracies found in the analysis of subdwarfs. \citet{jao} find there is a complex phase space in the appearance of very explicit metallicity indicators. They note that CaH is impacted in complicated ways by combinations of temperature, metallicity and gravity. A solution to this is to introduce a further step in the analysis of subdwarfs focussed on gravity determination \citet{zhang}, the sensitivity of calcium broadening to gravity and metallicity would seem to have promise for this step. More simply, a filter centred on the 4227\AA~line of similar width to a Str{\"o}mgren filter would be sensitive to metallicity and gravity. For example, low gravity objects would present as relatively brighter than field objects of the same colour due to their reduced opacity in the 4100-4400\AA~region and low metallicity objects would present as fainter.
%St$\rm\ddot{o}$mgren
%Str{\"o}mgren

Although the general causes of line broadening are relatively well understood, the details of the calculations are complex and require considerable effort. Calculations need to consider the broadening of different alkali lines with a range of different molecules across a large range of temperatures, pressures and abundances. Such calculations are essential for synthetic spectra to match the spectra of M dwarfs. In the meantime, it may sometimes be practical to use a `modified Lorentzian' whose shape is empirically determined. This can be thought of as analogous to the use of astrophysical oscillator strengths. While it will always be desirable to use the proper line parameters for strong lines it is notable that even after two decades the proper treatment of collisional broadening for the well studied potassium resonance doublet is not complete e.g., \citet{phillips}.

\section*{Acknowledgements}

We are grateful to the referee for very thoughtful feedback on the original version of the manuscript which greatly improved the clarity. We are very grateful to David Yong for a Magellan/MIKE spectrum of NLTT5022 taken in 2008, to Anna Frebel for the first high resolution blue spectrum of Proxima Centauri with the MIKE spectrograph in July 2009, and helpful questions and comments from Thomas Nordlander and Mike Ireland to MB. Data from the WiFeS instrument on the Australian National University 2.3m telescope at Siding Springs were vital for spectrophotometry. The ESO Science Archive Facility helped enabled MB to definitively identify NaH in 2018 and AlH, SiH in 2019 and provided a number of useful spectra. The ISIS instrument on the William Herschel Telescope, installed in the Spanish Observatorio del Roque de los
Muchachos of the Instituto de Astrofísica de Canarias, in the island of La Palma provided some of the spectra for this study. HJ was supported by STFC grant ST/R000905/1, YP and YL were funded as part of the routine financing
programme for institutes of the National Academy of Sciences
of Ukraine. YP acknowledges financial support from Jesús Serra Foundation through its “Visiting Researchers Programme” and from the visitor programme of the Centre of Excellence “Severo Ochoa” award to the Instituto de Astrofísica de Canarias (CEX2019-000920-S).
%Funded by Chinese Academy of Sciences President’s International Fellowship Initiative. Grant No. 2020VMA0033. 
This research has made use of the
VALD database operated at Uppsala University, and NASA’s Astrophysics Data System
Bibliographic Services (ADS).

%%%%%%%%%%%%%%%%%%%%%%%%%%%%%%%%%%%%%%%%%%%%%%%%%%
\section*{Data Availability}
The observational spectra used for this study are available at CDS via anonymous ftp to cdsarc.u-strasbg.fr (130.79.128.5) 
or via https://cdsarc.unistra.fr/viz-bin/cat/J/MNRAS.



\bibliographystyle{mnras}
\bibliography{mnemonic,b1,yp} % if your bibtex file is called example.bib

\begin{thebibliography}{}
\makeatletter
\relax
\def\mn@urlcharsother{\let\do\@makeother \do\$\do\&\do\#\do\^\do\_\do\%\do\~}
\def\mn@doi{\begingroup\mn@urlcharsother \@ifnextchar [ {\mn@doi@}
  {\mn@doi@[]}}
\def\mn@doi@[#1]#2{\def\@tempa{#1}\ifx\@tempa\@empty \href
  {http://dx.doi.org/#2} {doi:#2}\else \href {http://dx.doi.org/#2} {#1}\fi
  \endgroup}
\def\mn@eprint#1#2{\mn@eprint@#1:#2::\@nil}
\def\mn@eprint@arXiv#1{\href {http://arxiv.org/abs/#1} {{\tt arXiv:#1}}}
\def\mn@eprint@dblp#1{\href {http://dblp.uni-trier.de/rec/bibtex/#1.xml}
  {dblp:#1}}
\def\mn@eprint@#1:#2:#3:#4\@nil{\def\@tempa {#1}\def\@tempb {#2}\def\@tempc
  {#3}\ifx \@tempc \@empty \let \@tempc \@tempb \let \@tempb \@tempa \fi \ifx
  \@tempb \@empty \def\@tempb {arXiv}\fi \@ifundefined
  {mn@eprint@\@tempb}{\@tempb:\@tempc}{\expandafter \expandafter \csname
  mn@eprint@\@tempb\endcsname \expandafter{\@tempc}}}

\bibitem[\protect\citeauthoryear{{Ake} \& {Greenstein}}{{Ake} \&
  {Greenstein}}{1980}]{akeg1980}
{Ake} T.~B.,  {Greenstein} J.~L.,  1980, \mn@doi [\apj] {10.1086/158299}, \href
  {https://ui.adsabs.harvard.edu/abs/1980ApJ...240..859A} {240, 859}

\bibitem[\protect\citeauthoryear{Allard \& Kielkopf}{Allard \&
  Kielkopf}{1982}]{alla82}
Allard N.,  Kielkopf J.,  1982, \mn@doi [Rev. Mod. Phys.]
  {10.1103/RevModPhys.54.1103}, 54, 1103

\bibitem[\protect\citeauthoryear{{Allard} \& {Spiegelman}}{{Allard} \&
  {Spiegelman}}{2006}]{na06}
{Allard} N.~F.,  {Spiegelman} F.,  2006, \mn@doi [\aap]
  {10.1051/0004-6361:20054485}, \href
  {https://ui.adsabs.harvard.edu/abs/2006A&A...452..351A} {452, 351}

\bibitem[\protect\citeauthoryear{{Allard}, {Royer}, {Kielkopf}  \&
  {Feautrier}}{{Allard} et~al.}{1999}]{nallard1999}
{Allard} N.~F.,  {Royer} A.,  {Kielkopf} J.~F.,   {Feautrier} N.,  1999,
  \mn@doi [\pra] {10.1103/PhysRevA.60.1021}, \href
  {https://ui.adsabs.harvard.edu/abs/1999PhRvA..60.1021A} {60, 1021}

\bibitem[\protect\citeauthoryear{{Allard}, {Allard}, {Hauschildt}, {Kielkopf}
  \& {Machin}}{{Allard} et~al.}{2003}]{alla03}
{Allard} N.~F.,  {Allard} F.,  {Hauschildt} P.~H.,  {Kielkopf} J.~F.,
  {Machin} L.,  2003, \mn@doi [\aap] {10.1051/0004-6361:20031299}, \href
  {https://ui.adsabs.harvard.edu/abs/2003A&A...411L.473A} {411, L473}

\bibitem[\protect\citeauthoryear{{Allard}, {Homeier}  \& {Freytag}}{{Allard}
  et~al.}{2011}]{btsettl}
{Allard} F.,  {Homeier} D.,   {Freytag} B.,  2011, in {Johns-Krull} C.,
  {Browning} M.~K.,   {West} A.~A.,  eds,  Astronomical Society of the Pacific
  Conference Series Vol. 448, 16th Cambridge Workshop on Cool Stars, Stellar
  Systems, and the Sun. p.~91 (\mn@eprint {arXiv} {1011.5405})

\bibitem[\protect\citeauthoryear{{Allard}, {Spiegelman}, {Leininger}  \&
  {Molliere}}{{Allard} et~al.}{2019}]{nallard2019}
{Allard} N.~F.,  {Spiegelman} F.,  {Leininger} T.,   {Molliere} P.,  2019,
  \mn@doi [\aap] {10.1051/0004-6361/201935593}, \href
  {https://ui.adsabs.harvard.edu/abs/2019A&A...628A.120A} {628, A120}

\bibitem[\protect\citeauthoryear{{Allende Prieto}}{{Allende
  Prieto}}{2023}]{cap2023}
{Allende Prieto} C.,  2023, \mn@doi [arXiv e-prints]
  {10.48550/arXiv.2303.14340}, \href
  {https://ui.adsabs.harvard.edu/abs/2023arXiv230314340A} {p. arXiv:2303.14340}

\bibitem[\protect\citeauthoryear{{Allers} \& {Liu}}{{Allers} \&
  {Liu}}{2013}]{allers}
{Allers} K.~N.,  {Liu} M.~C.,  2013, \mn@doi [\apj]
  {10.1088/0004-637X/772/2/79}, \href
  {https://ui.adsabs.harvard.edu/abs/2013ApJ...772...79A} {772, 79}

\bibitem[\protect\citeauthoryear{{Almendros-Abad}, {Mu{\v{z}}i{\'c}},
  {Moitinho}, {Krone-Martins}  \& {Kubiak}}{{Almendros-Abad}
  et~al.}{2022}]{almendrosabad}
{Almendros-Abad} V.,  {Mu{\v{z}}i{\'c}} K.,  {Moitinho} A.,  {Krone-Martins}
  A.,   {Kubiak} K.,  2022, \mn@doi [\aap] {10.1051/0004-6361/202142050}, \href
  {https://ui.adsabs.harvard.edu/abs/2022A&A...657A.129A} {657, A129}

\bibitem[\protect\citeauthoryear{{Barber}, {Tennyson}, {Harris}  \&
  {Tolchenov}}{{Barber} et~al.}{2006}]{barb06}
{Barber} R.~J.,  {Tennyson} J.,  {Harris} G.~J.,   {Tolchenov} R.~N.,  2006,
  \mn@doi [\mnras] {10.1111/j.1365-2966.2006.10184.x}, \href
  {https://ui.adsabs.harvard.edu/abs/2006MNRAS.368.1087B} {368, 1087}

\bibitem[\protect\citeauthoryear{{Barklem} \& {Collet}}{{Barklem} \&
  {Collet}}{2016}]{barklem}
{Barklem} P.~S.,  {Collet} R.,  2016, \mn@doi [\aap]
  {10.1051/0004-6361/201526961}, \href
  {https://ui.adsabs.harvard.edu/abs/2016A&A...588A..96B} {588, A96}

\bibitem[\protect\citeauthoryear{{Bessell}}{{Bessell}}{1999}]{b1999}
{Bessell} M.~S.,  1999, \mn@doi [\pasp] {10.1086/316454}, \href
  {https://ui.adsabs.harvard.edu/abs/1999PASP..111.1426B} {111, 1426}

\bibitem[\protect\citeauthoryear{{Bessell}}{{Bessell}}{2011}]{bessell2011}
{Bessell} M.~S.,  2011, in {Johns-Krull} C.,  {Browning} M.~K.,   {West} A.~A.,
   eds,  Astronomical Society of the Pacific Conference Series Vol. 448, 16th
  Cambridge Workshop on Cool Stars, Stellar Systems, and the Sun. p.~131

\bibitem[\protect\citeauthoryear{{Blouin}, {Allard}, {Leininger}, {Gad{\'e}a}
  \& {Dufour}}{{Blouin} et~al.}{2019}]{blouin}
{Blouin} S.,  {Allard} N.~F.,  {Leininger} T.,  {Gad{\'e}a} F.~X.,   {Dufour}
  P.,  2019, \mn@doi [\apj] {10.3847/1538-4357/ab1266}, \href
  {https://ui.adsabs.harvard.edu/abs/2019ApJ...875..137B} {875, 137}

\bibitem[\protect\citeauthoryear{{Bochanski}, {West}, {Hawley}  \&
  {Covey}}{{Bochanski} et~al.}{2007}]{sdss}
{Bochanski} J.~J.,  {West} A.~A.,  {Hawley} S.~L.,   {Covey} K.~R.,  2007,
  \mn@doi [\aj] {10.1086/510240}, \href
  {https://ui.adsabs.harvard.edu/abs/2007AJ....133..531B} {133, 531}

\bibitem[\protect\citeauthoryear{{Bohlin}}{{Bohlin}}{2007}]{bohlin}
{Bohlin} R.~C.,  2007, in {Sterken} C.,  ed.,  Astronomical Society of the
  Pacific Conference Series Vol. 364, The Future of Photometric,
  Spectrophotometric and Polarimetric Standardization. p.~315 (\mn@eprint
  {arXiv} {astro-ph/0608715})

\bibitem[\protect\citeauthoryear{{Buder} et~al.,}{{Buder} et~al.}{2018}]{buder}
{Buder} S.,  et~al., 2018, \mn@doi [\mnras] {10.1093/mnras/sty1281}, 478, 4513

\bibitem[\protect\citeauthoryear{Burrows \& Volobuyev}{Burrows \&
  Volobuyev}{2003}]{burr03}
Burrows A.,  Volobuyev M.,  2003, \mn@doi [The Astrophysical Journal]
  {10.1086/345412}, 583, 985

\bibitem[\protect\citeauthoryear{{Burrows}, {Ram}, {Bernath}, {Sharp}  \&
  {Milsom}}{{Burrows} et~al.}{2002}]{burr02}
{Burrows} A.,  {Ram} R.~S.,  {Bernath} P.,  {Sharp} C.~M.,   {Milsom} J.~A.,
  2002, \mn@doi [\apj] {10.1086/342242}, \href
  {https://ui.adsabs.harvard.edu/abs/2002ApJ...577..986B} {577, 986}

\bibitem[\protect\citeauthoryear{{Cristofari} et~al.,}{{Cristofari}
  et~al.}{2022}]{crist}
{Cristofari} P.~I.,  et~al., 2022, \mn@doi [\mnras] {10.1093/mnras/stab3679},
  \href {https://ui.adsabs.harvard.edu/abs/2022MNRAS.511.1893C} {511, 1893}

\bibitem[\protect\citeauthoryear{{Dalton} et~al.,}{{Dalton}
  et~al.}{2014}]{dalton}
{Dalton} G.,  et~al., 2014, in {Ramsay} S.~K.,  {McLean} I.~S.,   {Takami} H.,
  eds,  Society of Photo-Optical Instrumentation Engineers (SPIE) Conference
  Series Vol. 9147, Ground-based and Airborne Instrumentation for Astronomy V.
  p. 91470L, \mn@doi{10.1117/12.2055132}

\bibitem[\protect\citeauthoryear{{Ding} et~al.,}{{Ding} et~al.}{2022}]{ding}
{Ding} M.-Y.,  et~al., 2022, \mn@doi [\apjs] {10.3847/1538-4365/ac6754}, \href
  {https://ui.adsabs.harvard.edu/abs/2022ApJS..260...45D} {260, 45}

\bibitem[\protect\citeauthoryear{Dopita et~al.,}{Dopita
  et~al.}{2010}]{dopita2010}
Dopita M.,  et~al., 2010, Astrophysics and Space Science, 327, 245

\bibitem[\protect\citeauthoryear{{Dulick}, {Bauschlicher}, {Burrows}, {Sharp},
  {Ram}  \& {Bernath}}{{Dulick} et~al.}{2003}]{duli03}
{Dulick} M.,  {Bauschlicher} C.~W. J.,  {Burrows} A.,  {Sharp} C.~M.,  {Ram}
  R.~S.,   {Bernath} P.,  2003, \mn@doi [\apj] {10.1086/376791}, \href
  {https://ui.adsabs.harvard.edu/abs/2003ApJ...594..651D} {594, 651}

\bibitem[\protect\citeauthoryear{{Feng} \& {Jones}}{{Feng} \&
  {Jones}}{2018}]{FJ18}
{Feng} F.,  {Jones} H.~R.~A.,  2018, \mn@doi [\mnras] {10.1093/mnras/stx2576},
  \href {https://ui.adsabs.harvard.edu/abs/2018MNRAS.473.3185F} {473, 3185}

\bibitem[\protect\citeauthoryear{{Gagn{\'e}} et~al.,}{{Gagn{\'e}}
  et~al.}{2017}]{G17}
{Gagn{\'e}} J.,  et~al., 2017, \mn@doi [\apjs] {10.3847/1538-4365/228/2/18},
  \href {https://ui.adsabs.harvard.edu/abs/2017ApJS..228...18G} {228, 18}

\bibitem[\protect\citeauthoryear{{Gibson}}{{Gibson}}{1973}]{Gibson}
{Gibson} E.~G.,  1973, {The quiet sun}.
 Vol. 303

\bibitem[\protect\citeauthoryear{{Gizis}}{{Gizis}}{1997}]{g97}
{Gizis} J.~E.,  1997, \mn@doi [\aj] {10.1086/118302}, \href
  {https://ui.adsabs.harvard.edu/abs/1997AJ....113..806G} {113, 806}

\bibitem[\protect\citeauthoryear{{Grevesse} \& {Sauval}}{{Grevesse} \&
  {Sauval}}{1998}]{gs98}
{Grevesse} N.,  {Sauval} A.~J.,  1998, \mn@doi [\ssr]
  {10.1023/A:1005161325181}, \href
  {https://ui.adsabs.harvard.edu/abs/1998SSRv...85..161G} {85, 161}

\bibitem[\protect\citeauthoryear{{Grim} \& {Staubach}}{{Grim} \&
  {Staubach}}{1996}]{grim}
{Grim} E.,  {Staubach} P.,  1996, in {Gustafson} B. A.~S.,  {Hanner} M.~S.,
  eds,  Astronomical Society of the Pacific Conference Series Vol. 104, IAU
  Colloq. 150: Physics, Chemistry, and Dynamics of Interplanetary Dust. p.~3

\bibitem[\protect\citeauthoryear{Gurvits, Veits  \& Medvedev}{Gurvits
  et~al.}{1982}]{gurvits1982thermodynamical}
Gurvits L.,  Veits I.,   Medvedev V.,  1982, Thermodynamical Properties of
  Specified Substances

\bibitem[\protect\citeauthoryear{{Gustafsson}, {Edvardsson}, {Eriksson},
  {J{\o}rgensen}, {Nordlund}  \& {Plez}}{{Gustafsson}
  et~al.}{2008}]{gustafsson}
{Gustafsson} B.,  {Edvardsson} B.,  {Eriksson} K.,  {J{\o}rgensen} U.~G.,
  {Nordlund} {\r{A}}.,   {Plez} B.,  2008, \mn@doi [\aap]
  {10.1051/0004-6361:200809724}, \href
  {https://ui.adsabs.harvard.edu/abs/2008A&A...486..951G} {486, 951}

\bibitem[\protect\citeauthoryear{{Hauschildt}, {Allard}, {Ferguson}, {Baron}
  \& {Alexander}}{{Hauschildt} et~al.}{1999}]{nextgen}
{Hauschildt} P.~H.,  {Allard} F.,  {Ferguson} J.,  {Baron} E.,   {Alexander}
  D.~R.,  1999, \mn@doi [\apj] {10.1086/307954}, \href
  {https://ui.adsabs.harvard.edu/abs/1999ApJ...525..871H} {525, 871}

\bibitem[\protect\citeauthoryear{{Herczeg} \& {Hillenbrand}}{{Herczeg} \&
  {Hillenbrand}}{2014}]{Herczeg}
{Herczeg} G.~J.,  {Hillenbrand} L.~A.,  2014, \mn@doi [\apj]
  {10.1088/0004-637X/786/2/97}, \href
  {https://ui.adsabs.harvard.edu/abs/2014ApJ...786...97H} {786, 97}

\bibitem[\protect\citeauthoryear{{Jao}, {Henry}, {Beaulieu}  \&
  {Subasavage}}{{Jao} et~al.}{2008}]{jao}
{Jao} W.-C.,  {Henry} T.~J.,  {Beaulieu} T.~D.,   {Subasavage} J.~P.,  2008,
  \mn@doi [\aj] {10.1088/0004-6256/136/2/840}, \href
  {https://ui.adsabs.harvard.edu/abs/2008AJ....136..840J} {136, 840}

\bibitem[\protect\citeauthoryear{{Jones} \& {Tsuji}}{{Jones} \&
  {Tsuji}}{1997}]{jone97}
{Jones} H. R.~A.,  {Tsuji} T.,  1997, \mn@doi [\apjl] {10.1086/310619}, \href
  {https://ui.adsabs.harvard.edu/abs/1997ApJ...480L..39J} {480, L39}

\bibitem[\protect\citeauthoryear{{Jones}, {Longmore}, {Allard}  \&
  {Hauschildt}}{{Jones} et~al.}{1996}]{Jones1996}
{Jones} H. R.~A.,  {Longmore} A.~J.,  {Allard} F.,   {Hauschildt} P.~H.,  1996,
  \mn@doi [\mnras] {10.1093/mnras/280.1.77}, \href
  {https://ui.adsabs.harvard.edu/abs/1996MNRAS.280...77J} {280, 77}

\bibitem[\protect\citeauthoryear{{Jones}, {Pavlenko}, {Viti}  \&
  {Tennyson}}{{Jones} et~al.}{2002}]{jones2002}
{Jones} H. R.~A.,  {Pavlenko} Y.,  {Viti} S.,   {Tennyson} J.,  2002, \mn@doi
  [\mnras] {10.1046/j.1365-8711.2002.05090.x}, \href
  {https://ui.adsabs.harvard.edu/abs/2002MNRAS.330..675J} {330, 675}

\bibitem[\protect\citeauthoryear{{Kesseli} et~al.,}{{Kesseli}
  et~al.}{2019}]{atlas}
{Kesseli} A.~Y.,  et~al., 2019, \mn@doi [\aj] {10.3847/1538-3881/aae982}, \href
  {https://ui.adsabs.harvard.edu/abs/2019AJ....157...63K} {157, 63}

\bibitem[\protect\citeauthoryear{{Kielkopf} \& {Allard}}{{Kielkopf} \&
  {Allard}}{2008}]{kielkopf}
{Kielkopf} J.~F.,  {Allard} N.~F.,  2008, in {Gigosos} M.~A.,  {Gonzalez}
  M.~A.,  eds,  American Institute of Physics Conference Series Vol. 1058,
  Spectral Line Shapes: Volume 15-19th International Conference on Spectral
  Line Shapes. pp 281--288, \mn@doi{10.1063/1.3026462}

\bibitem[\protect\citeauthoryear{{Kiman}, {Schmidt}, {Angus}, {Cruz}, {Faherty}
   \& {Rice}}{{Kiman} et~al.}{2019}]{kiman20}
{Kiman} R.,  {Schmidt} S.~J.,  {Angus} R.,  {Cruz} K.~L.,  {Faherty} J.~K.,
  {Rice} E.,  2019, \mn@doi [\aj] {10.3847/1538-3881/ab1753}, \href
  {https://ui.adsabs.harvard.edu/abs/2019AJ....157..231K} {157, 231}

\bibitem[\protect\citeauthoryear{{Kiman} et~al.,}{{Kiman}
  et~al.}{2021}]{kiman21}
{Kiman} R.,  et~al., 2021, \mn@doi [\aj] {10.3847/1538-3881/abf561}, \href
  {https://ui.adsabs.harvard.edu/abs/2021AJ....161..277K} {161, 277}

\bibitem[\protect\citeauthoryear{{Kirkpatrick}, {Henry}  \&
  {McCarthy}}{{Kirkpatrick} et~al.}{1991}]{kirkpatrick1991}
{Kirkpatrick} J.~D.,  {Henry} T.~J.,   {McCarthy} Donald~W. J.,  1991, \mn@doi
  [\apjs] {10.1086/191611}, \href
  {https://ui.adsabs.harvard.edu/abs/1991ApJS...77..417K} {77, 417}

\bibitem[\protect\citeauthoryear{{Kurucz}}{{Kurucz}}{2018}]{kuru18}
{Kurucz} R.~L.,  2018, {Including All the Lines: Data Releases for Spectra and
  Opacities through 2017}.
p.~47

\bibitem[\protect\citeauthoryear{{Le Roy}, {Walji}  \& {Sentjens}}{{Le Roy}
  et~al.}{2013}]{le2013dpf}
{Le Roy} R.~J.,  {Walji} S.,   {Sentjens} K.,  2013, \href
  {https://ui.adsabs.harvard.edu/abs/2013mss..confEMI01L} {p. EMI01}

\bibitem[\protect\citeauthoryear{Lindblad}{Lindblad}{1935a}]{lindblad1935}
Lindblad B.,  1935a, Stockholms Observatoriums Annaler, 12, 2

\bibitem[\protect\citeauthoryear{Lindblad}{Lindblad}{1935b}]{lindblad1935nat}
Lindblad B.,  1935b, Nature, 136, 67

\bibitem[\protect\citeauthoryear{{Lodieu}, {Allard}, {Rodrigo}, {Pavlenko},
  {Burgasser}, {Lyubchik}, {Kaminsky}  \& {Homeier}}{{Lodieu}
  et~al.}{2019}]{lodieu}
{Lodieu} N.,  {Allard} F.,  {Rodrigo} C.,  {Pavlenko} Y.,  {Burgasser} A.,
  {Lyubchik} Y.,  {Kaminsky} B.,   {Homeier} D.,  2019, \mn@doi [\aap]
  {10.1051/0004-6361/201935299}, \href
  {https://ui.adsabs.harvard.edu/abs/2019A&A...628A..61L} {628, A61}

\bibitem[\protect\citeauthoryear{{Mallinson}, {Lind}, {Amarsi}, {Barklem},
  {Grumer}, {Belyaev}  \& {Youakim}}{{Mallinson} et~al.}{2022}]{nlte}
{Mallinson} J.~W.~E.,  {Lind} K.,  {Amarsi} A.~M.,  {Barklem} P.~S.,  {Grumer}
  J.,  {Belyaev} A.~K.,   {Youakim} K.,  2022, arXiv e-prints, \href
  {https://ui.adsabs.harvard.edu/abs/2022arXiv221008880M} {p. arXiv:2210.08880}

\bibitem[\protect\citeauthoryear{{Mart{\'\i}n}, {Delfosse}, {Basri}, {Goldman},
  {Forveille}  \& {Zapatero Osorio}}{{Mart{\'\i}n} et~al.}{1999}]{martin1998}
{Mart{\'\i}n} E.~L.,  {Delfosse} X.,  {Basri} G.,  {Goldman} B.,  {Forveille}
  T.,   {Zapatero Osorio} M.~R.,  1999, \mn@doi [\aj] {10.1086/301107}, \href
  {https://ui.adsabs.harvard.edu/abs/1999AJ....118.2466M} {118, 2466}

\bibitem[\protect\citeauthoryear{{Mart{\'\i}n}, {Lodieu}  \& {del
  Burgo}}{{Mart{\'\i}n} et~al.}{2022}]{martin2022}
{Mart{\'\i}n} E.~L.,  {Lodieu} N.,   {del Burgo} C.,  2022, \mn@doi [\mnras]
  {10.1093/mnras/stab2969}, \href
  {https://ui.adsabs.harvard.edu/abs/2022MNRAS.510.2841M} {510, 2841}

\bibitem[\protect\citeauthoryear{{Mashonkina}, {Sitnova}  \&
  {Belyaev}}{{Mashonkina} et~al.}{2017}]{nlte1}
{Mashonkina} L.,  {Sitnova} T.,   {Belyaev} A.~K.,  2017, \mn@doi [\aap]
  {10.1051/0004-6361/201731236}, \href
  {https://ui.adsabs.harvard.edu/abs/2017A&A...605A..53M} {605, A53}

\bibitem[\protect\citeauthoryear{{McKemmish}, {Masseron}, {Hoeijmakers},
  {P{\'e}rez-Mesa}, {Grimm}, {Yurchenko}  \& {Tennyson}}{{McKemmish}
  et~al.}{2019}]{McKemmish19}
{McKemmish} L.~K.,  {Masseron} T.,  {Hoeijmakers} H.~J.,  {P{\'e}rez-Mesa} V.,
  {Grimm} S.~L.,  {Yurchenko} S.~N.,   {Tennyson} J.,  2019, \mn@doi [\mnras]
  {10.1093/mnras/stz1818}, \href
  {https://ui.adsabs.harvard.edu/abs/2019MNRAS.488.2836M} {488, 2836}

\bibitem[\protect\citeauthoryear{{Morel}}{{Morel}}{2018}]{M18}
{Morel} T.,  2018, \mn@doi [\aap] {10.1051/0004-6361/201833125}, \href
  {https://ui.adsabs.harvard.edu/abs/2018A&A...615A.172M} {615, A172}

\bibitem[\protect\citeauthoryear{{Pavlenko}}{{Pavlenko}}{2014}]{pav14}
{Pavlenko} Y.~V.,  2014, \mn@doi [Astronomy Reports]
  {10.1134/S1063772914110043}, \href
  {https://ui.adsabs.harvard.edu/abs/2014ARep...58..825P} {58, 825}

\bibitem[\protect\citeauthoryear{{Pavlenko}, {Jones}, {Mart{\'\i}n},
  {Guenther}, {Kenworthy}  \& {Zapatero Osorio}}{{Pavlenko}
  et~al.}{2007}]{pavl07}
{Pavlenko} Y.~V.,  {Jones} H.~R.~A.,  {Mart{\'\i}n} E.~L.,  {Guenther} E.,
  {Kenworthy} M.~A.,   {Zapatero Osorio} M.~R.,  2007, \mn@doi [\mnras]
  {10.1111/j.1365-2966.2007.12182.x}, \href
  {https://ui.adsabs.harvard.edu/abs/2007MNRAS.380.1285P} {380, 1285}

\bibitem[\protect\citeauthoryear{{Pavlenko}, {Su{\'a}rez Mascare{\~n}o},
  {Rebolo}, {Lodieu}, {B{\'e}jar}  \& {Gonz{\'a}lez Hern{\'a}ndez}}{{Pavlenko}
  et~al.}{2017}]{pavl17}
{Pavlenko} Y.,  {Su{\'a}rez Mascare{\~n}o} .~A.,  {Rebolo} R.,  {Lodieu} N.,
  {B{\'e}jar} V.~J.~S.,   {Gonz{\'a}lez Hern{\'a}ndez} J.~I.,  2017, \mn@doi
  [\aap] {10.1051/0004-6361/201730733}, \href
  {https://ui.adsabs.harvard.edu/abs/2017A&A...606A..49P} {606, A49}

\bibitem[\protect\citeauthoryear{{Pavlenko}, {Yurchenko}  \&
  {Tennyson}}{{Pavlenko} et~al.}{2020}]{yp2020}
{Pavlenko} Y.~V.,  {Yurchenko} S.~N.,   {Tennyson} J.,  2020, \mn@doi [\aap]
  {10.1051/0004-6361/201936811}, \href
  {https://ui.adsabs.harvard.edu/abs/2020A&A...633A..52P} {633, A52}

\bibitem[\protect\citeauthoryear{{Pavlenko}, {Tennyson}, {Yurchenko},
  {Schmidt}, {Jones}, {Lyubchik}  \& {Su{\'a}rez Mascare{\~n}o}}{{Pavlenko}
  et~al.}{2022}]{pav22}
{Pavlenko} Y.~V.,  {Tennyson} J.,  {Yurchenko} S.~N.,  {Schmidt} M.~R.,
  {Jones} H. R.~A.,  {Lyubchik} Y.,   {Su{\'a}rez Mascare{\~n}o} A.,  2022,
  \mn@doi [\mnras] {10.1093/mnras/stac2588}, \href
  {https://ui.adsabs.harvard.edu/abs/2022MNRAS.516.5655P} {516, 5655}

\bibitem[\protect\citeauthoryear{{Phillips} et~al.,}{{Phillips}
  et~al.}{2020}]{phillips}
{Phillips} M.~W.,  et~al., 2020, \mn@doi [\aap] {10.1051/0004-6361/201937381},
  \href {https://ui.adsabs.harvard.edu/abs/2020A&A...637A..38P} {637, A38}

\bibitem[\protect\citeauthoryear{{Pinfield}, {Dobbie}, {Jameson}, {Steele},
  {Jones}  \& {Katsiyannis}}{{Pinfield} et~al.}{2003}]{pinfield2003}
{Pinfield} D.~J.,  {Dobbie} P.~D.,  {Jameson} R.~F.,  {Steele} I.~A.,  {Jones}
  H.~R.~A.,   {Katsiyannis} A.~C.,  2003, \mn@doi [\mnras]
  {10.1046/j.1365-8711.2003.06630.x}, \href
  {https://ui.adsabs.harvard.edu/abs/2003MNRAS.342.1241P} {342, 1241}

\bibitem[\protect\citeauthoryear{{Plez}}{{Plez}}{1998}]{plez98}
{Plez} B.,  1998, \aap, \href
  {https://ui.adsabs.harvard.edu/abs/1998A&A...337..495P} {337, 495}

\bibitem[\protect\citeauthoryear{{Preston}}{{Preston}}{1959}]{preston}
{Preston} G.~W.,  1959, \mn@doi [\apj] {10.1086/146743}, \href
  {https://ui.adsabs.harvard.edu/abs/1959ApJ...130..507P} {130, 507}

\bibitem[\protect\citeauthoryear{{Reyl{\'e}}, {Jardine}, {Fouqu{\'e}},
  {Caballero}, {Smart}  \& {Sozzetti}}{{Reyl{\'e}} et~al.}{2021}]{reyle2021}
{Reyl{\'e}} C.,  {Jardine} K.,  {Fouqu{\'e}} P.,  {Caballero} J.~A.,  {Smart}
  R.~L.,   {Sozzetti} A.,  2021, \mn@doi [\aap] {10.1051/0004-6361/202140985},
  \href {https://ui.adsabs.harvard.edu/abs/2021A&A...650A.201R} {650, A201}

\bibitem[\protect\citeauthoryear{{Rivlin}, {Lodi}, {Yurchenko}, {Tennyson}  \&
  {Le Roy}}{{Rivlin} et~al.}{2015}]{rivlin15}
{Rivlin} T.,  {Lodi} L.,  {Yurchenko} S.~N.,  {Tennyson} J.,   {Le Roy} R.~J.,
  2015, \mn@doi [\mnras] {10.1093/mnras/stv979}, \href
  {https://ui.adsabs.harvard.edu/abs/2015MNRAS.451..634R} {451, 634}

\bibitem[\protect\citeauthoryear{Ryabchikova, Piskunov, Kurucz, Stempels,
  Heiter, Pakhomov  \& Barklem}{Ryabchikova et~al.}{2015}]{ryab15}
Ryabchikova T.,  Piskunov N.,  Kurucz R.~L.,  Stempels H.~C.,  Heiter U.,
  Pakhomov Y.,   Barklem P.~S.,  2015, \mn@doi [Physica Scripta]
  {10.1088/0031-8949/90/5/054005}, 90, 054005

\bibitem[\protect\citeauthoryear{{Schlieder}, {L{\'e}pine}, {Rice}, {Simon},
  {Fielding}  \& {Tomasino}}{{Schlieder} et~al.}{2012}]{schlieder2012}
{Schlieder} J.~E.,  {L{\'e}pine} S.,  {Rice} E.,  {Simon} M.,  {Fielding} D.,
  {Tomasino} R.,  2012, \mn@doi [\aj] {10.1088/0004-6256/143/5/114}, \href
  {https://ui.adsabs.harvard.edu/abs/2012AJ....143..114S} {143, 114}

\bibitem[\protect\citeauthoryear{{Schneider}, {Song}, {Melis}, {Zuckerman}  \&
  {Bessell}}{{Schneider} et~al.}{2012}]{Schneider}
{Schneider} A.,  {Song} I.,  {Melis} C.,  {Zuckerman} B.,   {Bessell} M.,
  2012, \mn@doi [\apj] {10.1088/0004-637X/757/2/163}, \href
  {https://ui.adsabs.harvard.edu/abs/2012ApJ...757..163S} {757, 163}

\bibitem[\protect\citeauthoryear{{Schweitzer} et~al.,}{{Schweitzer}
  et~al.}{2019}]{S19}
{Schweitzer} A.,  et~al., 2019, \mn@doi [\aap] {10.1051/0004-6361/201834965},
  \href {https://ui.adsabs.harvard.edu/abs/2019A&A...625A..68S} {625, A68}

\bibitem[\protect\citeauthoryear{{Schwenke}}{{Schwenke}}{1998}]{Schwenke98}
{Schwenke} D.~W.,  1998, \mn@doi [Faraday Discussions] {10.1039/a800070k},
  \href {https://ui.adsabs.harvard.edu/abs/1998FaDi..109..321S} {109, 321}

\bibitem[\protect\citeauthoryear{{Solanki} \& {Hammer}}{{Solanki} \&
  {Hammer}}{2002}]{solanki02}
{Solanki} S.~K.,  {Hammer} R.,  2002, in {Bleeker} J.~A.,  {Geiss} J.,
  {Huber} M. C.~E.,  eds, , The Century of Space Science, Volume I.
p.~1065

\bibitem[\protect\citeauthoryear{{Sprague} et~al.,}{{Sprague}
  et~al.}{2022}]{S22}
{Sprague} D.,  et~al., 2022, \mn@doi [\aj] {10.3847/1538-3881/ac4de7}, \href
  {https://ui.adsabs.harvard.edu/abs/2022AJ....163..152S} {163, 152}

\bibitem[\protect\citeauthoryear{{Tsuji}}{{Tsuji}}{1973}]{tsuji73}
{Tsuji} T.,  1973, \aap, \href
  {https://ui.adsabs.harvard.edu/abs/1973A&A....23..411T} {23, 411}

\bibitem[\protect\citeauthoryear{{Tsuji}, {Ohnaka}  \& {Aoki}}{{Tsuji}
  et~al.}{1996}]{Tsuji}
{Tsuji} T.,  {Ohnaka} K.,   {Aoki} W.,  1996, \aap, \href
  {https://ui.adsabs.harvard.edu/abs/1996A&A...305L...1T} {305, L1}

\bibitem[\protect\citeauthoryear{{Unsold}}{{Unsold}}{1955}]{unso55}
{Unsold} A.,  1955, {Physik der Sternatmospharen, MIT besonderer
  Berucksichtigung der Sonne.}

\bibitem[\protect\citeauthoryear{{Vardya} \& {B{\"o}hm}}{{Vardya} \&
  {B{\"o}hm}}{1965}]{vardya}
{Vardya} M.~S.,  {B{\"o}hm} K.~H.,  1965, \mn@doi [\mnras]
  {10.1093/mnras/131.1.89}, \href
  {https://ui.adsabs.harvard.edu/abs/1965MNRAS.131...89V} {131, 89}

\bibitem[\protect\citeauthoryear{{Verro} et~al.,}{{Verro}
  et~al.}{2022}]{xshoot}
{Verro} K.,  et~al., 2022, \mn@doi [\aap] {10.1051/0004-6361/202142388}, \href
  {https://ui.adsabs.harvard.edu/abs/2022A&A...660A..34V} {660, A34}

\bibitem[\protect\citeauthoryear{{Vyssotsky}}{{Vyssotsky}}{1943}]{1943}
{Vyssotsky} A.~N.,  1943, \mn@doi [\apj] {10.1086/144527}, \href
  {https://ui.adsabs.harvard.edu/abs/1943ApJ....97..381V} {97, 381}

\bibitem[\protect\citeauthoryear{{Wang} et~al.,}{{Wang} et~al.}{2022}]{wang}
{Wang} Y.-F.,  et~al., 2022, \mn@doi [\aap] {10.1051/0004-6361/202142009},
  \href {https://ui.adsabs.harvard.edu/abs/2022A&A...660A..38W} {660, A38}

\bibitem[\protect\citeauthoryear{{Warner} \& {McGraw}}{{Warner} \&
  {McGraw}}{1974}]{warner}
{Warner} B.,  {McGraw} J.~T.,  1974, The Observatory, \href
  {https://ui.adsabs.harvard.edu/abs/1974Obs....94..313W} {94, 313}

\bibitem[\protect\citeauthoryear{{Weniger}}{{Weniger}}{1967}]{weniger}
{Weniger} S.,  1967, in {Hack} M.,  ed., Late-Type Stars. p.~25

\bibitem[\protect\citeauthoryear{{Yurchenko}, {Williams}, {Leyland}, {Lodi}  \&
  {Tennyson}}{{Yurchenko} et~al.}{2018}]{yurchenko18}
{Yurchenko} S.~N.,  {Williams} H.,  {Leyland} P.~C.,  {Lodi} L.,   {Tennyson}
  J.,  2018, \mn@doi [\mnras] {10.1093/mnras/sty1524}, \href
  {https://ui.adsabs.harvard.edu/abs/2018MNRAS.479.1401Y} {479, 1401}

\bibitem[\protect\citeauthoryear{{Zhang}, {Zhang}, {Comte}, {Homeier}, {Wang},
  {Hejazi}, {Li}  \& {Luo}}{{Zhang} et~al.}{2023}]{zhang}
{Zhang} S.,  {Zhang} H.-W.,  {Comte} G.,  {Homeier} D.,  {Wang} R.,  {Hejazi}
  N.,  {Li} Y.-B.,   {Luo} A.~L.,  2023, \mn@doi [\apj]
  {10.3847/1538-4357/aca28d}, \href
  {https://ui.adsabs.harvard.edu/abs/2023ApJ...942...40Z} {942, 40}

\bibitem[\protect\citeauthoryear{{de Jong}}{{de Jong}}{2019}]{dejong}
{de Jong} R.,  2019, in Preparing for 4MOST. p.~1,
  \mn@doi{10.5281/zenodo.3244904}

\makeatother
\end{thebibliography}

%\begin{thebibliography}{99}
%\end{thebibliography}

% Don't change these lines
\bsp	% typesetting comment
\label{lastpage}
\end{document}